\documentclass[letterpaper]{JHEP3}
\usepackage{array}

\usepackage{nicefrac}
\usepackage{amsmath}

\usepackage{amssymb,amsfonts,mathrsfs}
\usepackage{graphicx}
\usepackage{epsfig}

\usepackage{amsfonts}

\newcommand{\Tr}{\mbox{Tr}}

\newcommand{\suup}{1}
\newcommand{\sudown}{2}

\def\s{\sigma}

\def\a{\alpha}

\def\h{\eta}

\def\CN{{\cal N}}

\def\half{{\frac12}}

\def\IC{\relax\hbox{$\inbar\kern-.3em{\rm C}$}}

\def\IC{{\bf C}}
\def\IP{{\bf P}}

\def\CN{{\cal N}}

\def\bea{\begin{eqnarray}}
\def\eea{\end{eqnarray}}
\def\be{\begin{equation}}
\def\ee{\end{equation}}
\def\ba{\begin{align}}
\def\ea{\end{align}}
\def\bse{\begin{subequations}}
\def\ese{\end{subequations}}
\def\1F1{{}_1\!F_1}
\def\2F0{{}_2\!F_0}

\def\a{\alpha}

\def\h3{$\textrm{H}_3^+$}

\def\IC{{\mathbb C}}

\def\Tr{{\rm Tr}}

\def\lbldef#1#2{\expandafter\gdef\csname #1\endcsname {#2}}

\def\href#1#2{#2}

\newcommand{\beq}{\begin{equation}}
\newcommand{\eeq}{\end{equation}}
\newcommand{\ber}{\begin{eqnarray}}
\newcommand{\eer}{\end{eqnarray}}

\def\be{\begin{eqnarray}}
\def\ee{\end{eqnarray}}

\newcommand{\cN}{{\cal N}}

\def\({\left(}
\def\){\right)}
\def\[{\left[}
\def\]{\right]}
\def\<{\langle}
\def\>{\rangle}

\def\Q{\widetilde {\mathcal Q}_{{\suup}\dot{-}}}

\title{ Bootstrapping the superconformal index with surface defects  }
\author{Davide Gaiotto$^{\clubsuit,\diamondsuit}$, 
Leonardo Rastelli$^\spadesuit$, 
 and Shlomo S. Razamat$^{\diamondsuit}$ 
\\
\\
$^\clubsuit$ Perimeter Institute for Theoretical Physics,
 Waterloo, Ontario N2L 2Y5, Canada\\
$^\diamondsuit$\it Institute for Advanced Study, Princeton, NJ 08540, USA\\
$^\spadesuit$\it C.N. Yang Institute for Theoretical Physics,
\it Stony Brook, NY 11794-3840, USA\\
}

\bigskip

\abstract{

\

The analytic properties of the  ${\cal N} = 2$ superconformal index are given a physical interpretation
in terms of certain BPS surface defects, which
arise as the IR limit  of supersymmetric vortices. 
The residue of the  index 
at a pole in flavor fugacity is interpreted as
the index  of  a superconformal field theory without this flavor symmetry, but endowed with an additional
surface defect. The residue can be efficiently extracted
by acting on the index with a difference operator of Ruijsenaars-Schneider  type.
By imposing the associativity constraints of S-duality, we are then able to evaluate the  index 
of all   generalized
quiver theories of type $A$, for generic values of the three superconformal fugacities, with or without surface defects.
}

\begin{document}

\section{Introduction}

In recent years  much has been learned about rigid supersymmetric quantum field theories 
 by  placing them on curved compact manifolds, notably (squashed) spheres.\footnote{More general susy-preserving backgrounds are beginning to be explored \cite{Festuccia:2011ws,Dumitrescu:2012ha}.} 
 In several cases localization techniques have facilitated  the exact calculation of the 
 partition functions on these backgrounds, possibly in the presence of supersymmetric defects.
Four important classes of examples are   expectation values of circular Wilson loops in $4d$ ${\cal N}=2$ theories on $S^4$ \cite{Pestun:2007rz},
  partition functions of   $3d$ ${\cal N}=2$ theories on $S^3$ \cite{Kapustin:2009kz,Kapustin:2010xq,Hama:2010av},  partition functions of  ${\cal N}=1$ and ${\cal N}=2$ theories 
on $S^3 \times S^1$~\cite{Kinney:2005ej,Romelsberger:2005eg} and partition functions of ${\cal N}=2$ theories on $S^2 \times S^1$~\cite{Bhattacharya:2008zy,Kim:2009wb,Imamura:2011su,Krattenthaler:2011da,Kapustin:2011jm,Benini:2011nc}
  (the latter are usually referred to as ``superconformal indices'' if the theory is conformal).

These observables have been  indispensable tools in the exploration of the  new classes of $4d$ and $3d$ theories, dubbed respectively
class ${\cal S}$~\cite{Gaiotto:2009we, Gaiotto:2009hg} and class ${\cal R}$~\cite{DGG}\footnote{More precisely, the class ${\cal R}$
consists of ${\cal N}=2$ SCFTs which admit an abelian
Chern-Simons-Matter description, deformed by a certain class of
superpotentials.
Conjecturally, it includes the $3d$ theories defined by the
compactification of the (2,0) theories on $3d$ manifolds which admit an
ideal triangulation.
See also~\cite{Terashima:2011qi,Cecotti:2011iy} for related $3d$ discussions.}, 
which arise by twisted compactifications of the $6d$ (2,0) theory on Riemann surfaces and on hyperbolic three-manifolds. This line of enquiry
 has led to the discovery of remarkable $4d/2d$ and $3d/3d$ relations. 
For example, according to the AGT conjecture~\cite{Alday:2009aq}, the $S^4$ partition function of a class ${\cal S}$ theory is computed by a Liouville/Toda correlator
 on the associated Riemann surface, whose complex structure moduli correspond the exactly marginal $4d$ gauge couplings. 
 In a similar spirit, the $S^3 \times S^1$ partition function 
 of a class ${\cal S}$ theory, being independent of the exactly marginal gauge theory couplings,
is  computed by a topological quantum field theory correlator on the Riemann surface~\cite{Gadde:2009kb}.  

In this paper we present a complete characterization of this $2d$ TQFT,
and thereby provide an algorithm to evaluate the superconformal index of the general theory of class ${\cal S}$.\footnote{More accurately
we focus on class ${\cal S}$ theories of type A, that is, the $SU(N)$ generalized quivers. The generalization to  theories of type-D and E is in progress.}
The superconformal index is superficially a simpler observable than the $S^4$ partition function, 
since it can be evaluated by a free field calculation in any theory that admits a Lagrangian description. 
It is still however a very non-trivial function of the fugacities associated to the flavor symmetries and of  three additional superconformal fugacities $(p,q,t)$.
Its expression for a generic class ${\cal S}$ theory, which is non-Lagrangian, is a priori unknown.
For a two-dimensional slice $(q, t)$ of the three-dimensional superconformal fugacity space
an explicit description of the $2d$ TQFT was proposed in~\cite{Gadde:2011ik,Gadde:2011uv} and shown to pass many checks.
The key  was to find a complete basis of functions of the flavor fugacity variables
 that diagonalizes the topological structure constants. The diagonalizing functions turn out to be closely
 related to Macdonald polynomials.  Here we obtain the general answer, valid for arbitrary $(p,q,t)$,
 by a completely different strategy. The new method sheds light on the appearance of the Macdonald polynomials
 and yields naturally their expected ``elliptic'' generalization,  which would have been difficult to obtain in the previous approach.

The route that will lead us to the derivation of the general index is somewhat indirect, but it is guided
by some simple physical ideas.
A first source of inspiration is the AGT correspondence, where normalizable Liouville vertex operators
are associated with flavor symmetries of the $4d$ gauge theory, while degenerate vertex operators correspond to 
inserting extra surface defects in the $4d$ theory.  The degenerate operators are the key to the solution of Liouville  theory by the conformal bootstrap~\cite{Teschner:1995yf}:
 considering their fusion with normalizable vertex operators one can derive functional equations that admit a unique solution.
By analogy, we expect that adding surface defects to the $S^3 \times S^1$ partition function should correspond to inserting special ``degenerate'' punctures
in the $2d$ TQFT correlator, and that their fusion with the ordinary flavor punctures will lead to  ``topological'' bootstrap equations. This is indeed what we find.
Another useful heuristic  principle is that since divergences in a partition function must be related to flat bosonic directions,  it should be possible
to interpret  the residue of the index at any of its poles   in terms of the behavior of the $4d$ field theory ``far away'' in moduli space.

Guided by this intuition, we set out to evaluate the superconformal index of class ${\cal S}$ theories endowed with BPS surface defects.\footnote{See
~\cite{Nakayama:2011pa} for a previous attempt to incorporate surface defects in the superconformal index.
The index of some ${\cal N}=2$ theories
in the presence of {\it loop} operators has been evaluated in~\cite{Gang:2012yr}, while~\cite{Gang:2012ff} studied the index in presence of duality domain walls.
}
This is a very interesting observable in its own right. Surface defects are among the least studied objects in four-dimensional
quantum field theory and only recently they have
started to receive proper attention (see {\it e.g.} ~\cite{Gukov:2006jk,Gukov:2008sn,Gaiotto:2009fs}).
Surface defects are in fact the only defects compatible with localization with minimal (${\cal N}=1$)  supersymmetry in four dimensions.
While in this paper we focus on ${\cal N}=2$ theories, 
 we hope that our results will  serve as a stepping stone towards 
the study of surface defects in ${\cal N}=1$ theories.

 In theories that admit a Lagrangian description, such as the class ${\cal S}$ theories of type $A_1$, it should be possible
 to evaluate the index and  the $S^4$ partition function in the presence of surface defects by  localization techniques. This is a very interesting
 direction for future work. Here we resort instead to a less direct construction that however also applies  to the non-Lagrangian higher-rank theories.
 We start with the physical picture of a surface defect as the IR end point of a BPS vortex solution. 
The construction proceeds by embedding a given SCFT ${\cal T}_{IR}$ into a larger theory ${\cal T}_{UV}$ such that
turning on a  spatially constant Higgs branch vacuum expectation value (vev) one
flows back to the original  ${\cal T}_{IR}$.  If one  then modifies the UV boundary conditions of the RG flow by 
considering instead a position-dependent vev, the infrared endpoint  is
 ${\cal T}_{IR}$ endowed with an additional BPS surface defect.  The theory ${\cal T}_{UV}$
 has an additional $U(1)_f$ flavor symmetry with respect to ${\cal T}_{IR}$. We  argue  that
 the residues of the index of ${\cal T}_{UV}$ at some special poles in the $U(1)_f$ flavor fugacity
 capture the index of  ${\cal T}_{IR}$ in the presence of surface defects. Although ${\cal T}_{UV}$ does not contain
 any surface defects, the analytic structure of its index encodes the possibility to generate them by RG flows.
 By this route we are led to formulate a precise prescription
to evaluate the index of  ${\cal T}_{IR}$ with extra surface defects. 
 
 The prescription can be formulated entirely in terms of the index of  ${\cal T}_{IR}$. Remarkably, adding
  surface defects to  ${\cal T}_{IR}$ amounts to acting on its index with certain difference operators
  ${\frak S}_{(r,s)}$,  closely related to   the Hamiltonians of the elliptic Ruijsenaars-Schneider (RS)
model~\cite{Ruijsenaars:1986vq,Ruijsenaars:1986pp,R}. The difference operators act as shifts of one of the $SU(N)$ flavor
fugacities.  The analogy with AGT is compelling: the action of the difference operator on one of the flavor punctures  
corresponds to the fusion of a degenerate Liouville primary with one of the normalizable primaries.
Generalized S-duality predicts that one should get the same result independently of which  flavor puncture
 the difference operator is acting on. This immediately leads to the conclusion
 that the functions that diagonalize the topological structure constants must be eigenfunctions
 of the difference operators (under the assumption that the spectrum is non-degenerate). By this route we are led to a complete determination of the $2d$ TQFT
 and thus of the index of a general class ${\cal S}$ theory of type $A$, with or without surface defects.
 For the two-dimensional slice $(q,t)$ of superconformal fugacity space the eigenfunctions
 turn out to be proportional to Macdonald polynomials and we re-derive the results of ~\cite{Gadde:2011uv}. 
 For arbitrary values $(p,q,t)$ of the superconformal fugacities the eigenfunctions are not known in closed analytic form.
 As a demonstration of principle that they exist, that they have a non-degenerate spectrum
 and that they are in fact calculable, we discuss a perturbative scheme to determine the eigenfunctions for small $q$ or $p$ --
 an approach that may have independent mathematical interest.

The structure of this paper is as follows. In section~\ref{motsec} we present
 an RG construction of certain BPS surface defects.
 In section~\ref{prescsec} we give a physical interpretation of some special poles of the index 
  in terms of such surface defects. We
 spell out a precise prescription for evaluating the index of a theory in the presence
 of these defects.
In section~\ref{su2sec} we apply our prescription to the case of $A_1$ quivers of class ${\cal S}$
and recast it in terms of the action of difference operators of RS type. In section~\ref{propsec} we study
the properties of these difference operators and interpret them physically.
In section \ref{bootsec} we combine the consistency conditions of generalized S-duality and the explicit
form of the difference operators to ``bootstrap'' the index of a a general $A_1$ quiver.
In section \ref{$3d$sec} we dimensionally reduce our results to $S^3$ and interpret them in the context
of $3d$ gauge theories.
In section~\ref{sunsec} we present the extensions to the higher-rank theories.
We close  in section~\ref{sumsec} with a discussion and a list of open problems.
We collect in
three appendices some additional material not needed on a first reading.  In appendix A
we describe the embedding of the $2d$ superalgebra that lives at the location of the defect
into the $4d$ superalgebra, and recall some results on  $2d$ partition functions.
In appendix B we present a perturbative approach to the calculation of the elliptic RS
wavefunctions. Finally in appendix C we describe the generalization of our results to the case
of flavor punctures with reduced symmetry.

\

\

\section{An RG construction of supersymmetric surface defects }\label{motsec}

We begin by discussing a general construction 
of BPS surface defects, applicable to
a large class of  ${\cal N}=2$ superconformal field theories.
The construction
proceeds by embedding a given SCFT ${\cal T}_{IR}$ into a larger theory ${\cal T}_{UV}$, such that
turning on a  spatially constant Higgs branch vacuum expectation value (vev) one
flows back to the original  ${\cal T}_{IR}$.  If one  then modifies the UV boundary conditions of the RG flow by 
considering instead a position-dependent vev, the infrared endpoint  is
 ${\cal T}_{IR}$ endowed with an additional BPS surface defect.

In the rest of the paper we will explain how the pole structure
of the superconformal index can be understood physically in terms of these surface defects.
We will need to 
 make some assumptions about the flavor charges of the Higgs branch vevs which initiate the RG flow.
These assumptions, and many of our calculations, can be elegantly stated in terms of gauging a $U(1)_f$ flavor symmetry of $ {\cal T}_{UV}$.
There are reasons for which gauging $U(1)_f$  should be a bad idea: the gauged theory is not UV complete, and has an anomalous $U(1)_r$ R-symmetry. 
We believe that these problems are less serious than they appear, and we will sketch how one could overcome them. 
In any case, we  also present an alternative analysis that does not rely on  gauging $U(1)_f$ and leads to the same conclusions.

\

\subsection{RG flow by a constant Higgs branch vev: the gauged perspective}\label{gaugedconstsec}

\begin{figure}
\begin{center}
\begin{tabular}{ccc}
\includegraphics[scale=0.4]{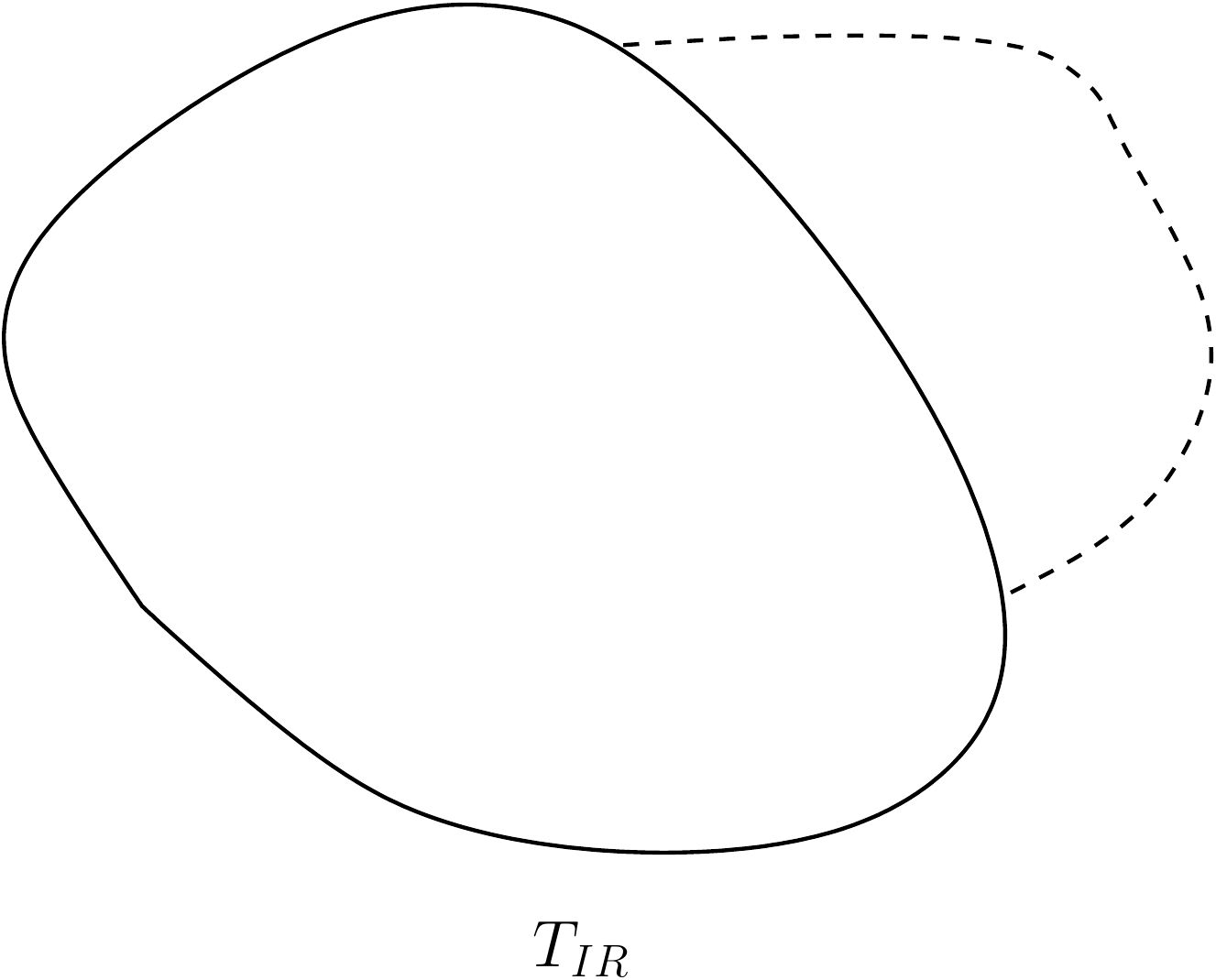}&\;\;\;\;&
\includegraphics[scale=0.4]{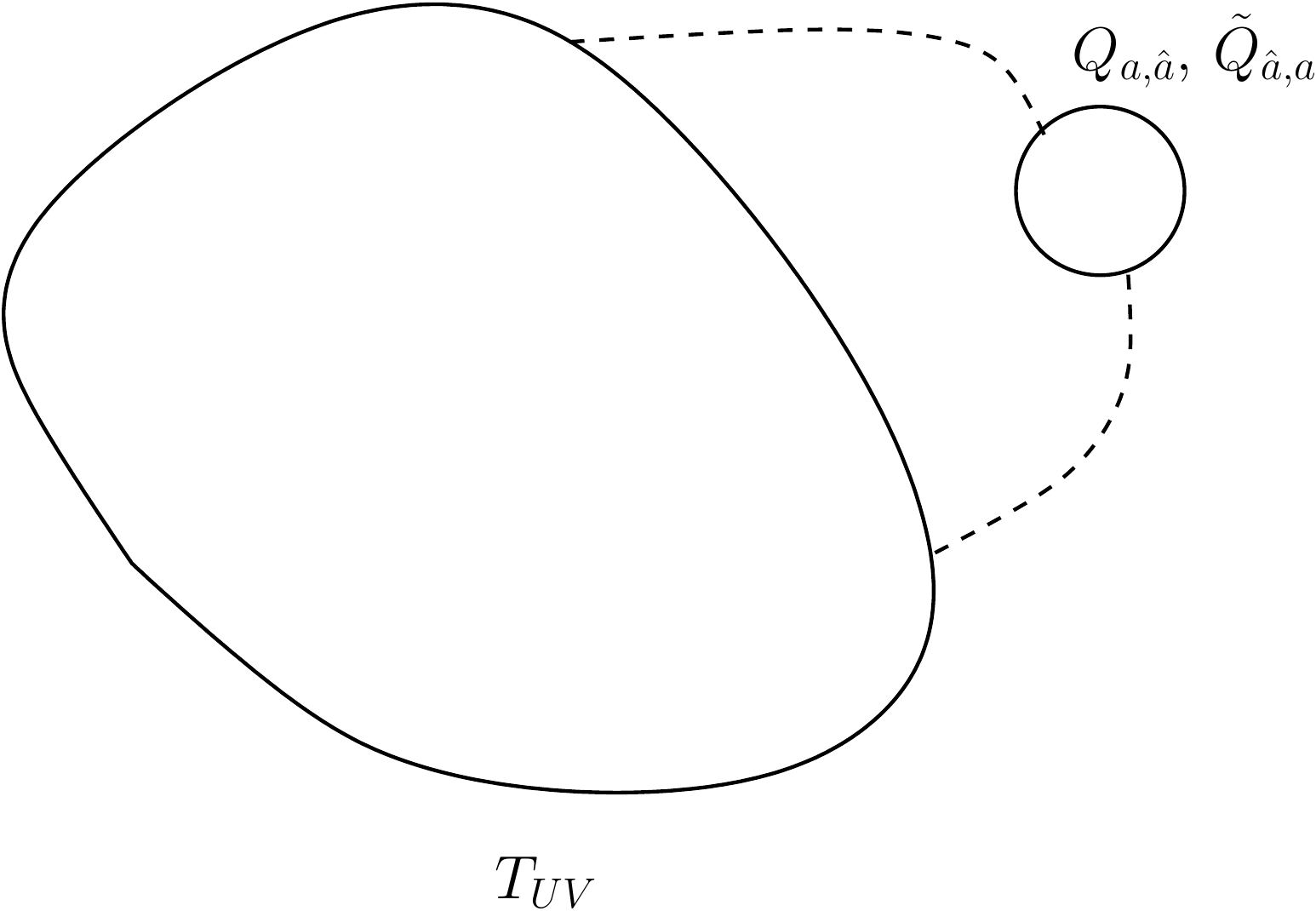}
\end{tabular}
\end{center}
\caption{Our first RG example. On the left we  indicate schematically the quiver for the SCFT ${\cal T}_{IR}$,
and on the right the quiver for  ${\cal T}_{UV}$.
The dashed lines represent 
$SU(N)$ gauge groups.
\label{T0fig}}
\end{figure}

As our main example, we take ${\cal T}_{IR}$ to be a generalized superconformal quiver of type $A_{N-1}$.
We focus on a link of the quiver, corresponding to an $SU(N)$ gauge group. We cut the link
and insert an extra node corresponding to a free hypermultiplet in the bifundamental representation of $SU(N) \times SU(N)$,
see figure~\ref{T0fig}. The resulting superconformal field theory is what we call ${\cal T}_{UV}$. Relative to ${\cal T}_{IR}$,
${\cal T}_{UV}$ has an extra $U(1)_f$ flavor symmetry acting on the bifundamental hypermultiplet only. 
We now describe a supersymmetric RG flow that connects  ${\cal T}_{UV}$ and ${\cal T}_{IR}$. The flow is initiated
by gauging the $U(1)_f$  symmetry in the presence of a Fayet-Iliopolous (FI) parameter, which introduces a scale
and forces the scalars in the  hypermultiplet associated to the extra node to acquire a vev.  
We are interested in the symmetry-breaking pattern that preserves
the diagonal $SU(N)$ of the original $SU(N) \times SU(N)$ non-abelian flavor symmetry.
This is achieved by choosing the vevs to be proportional to the unit matrix,
\be
Q_{a  \hat a} = q \delta_{a \hat a} \, , \qquad \tilde Q_{a \hat a} = \tilde q \delta_{a \hat a} \, .
\ee
The non-abelian D-term constraints are then automatically satisfied.
The $U(1)_f$ moment maps are 
\begin{equation}
\mu_3 = \mathrm{Tr}\,( |Q|^2 - |\tilde Q|^2)    \qquad \qquad \mu_1 + i \mu_2 =  \mathrm{Tr} Q \tilde Q \,.
\end{equation}
By an $SU(2)_R$ rotation ({\it i.e.} a choice of an ${\cal N}=1$ subalgebra) 
 we  align the FI parameters $v_i$ along $\mu_3$, so that the $U(1)_f$ D-term constraints read
\be
\mu_3 =  N (|q|^2 - |\tilde q|^2) =  v   \, , \quad  \mu_1 + i \mu_2  =  N q \tilde q = 0  \, ,
\ee
which  have a unique solution up to $U(1)_f$ gauge rotations. Taking for definiteness $v>0$, the solution is $|q|^2 =v/N$, $\tilde q = 0$.

The $U(1)_f \times SU(N)_1 \times SU(N)_2$ gauge symmetry is higgsed  to a diagonal $SU(N)$, 
and the bifundamental hypermultiplets are eaten up in the process. In other terms, the extra node that we added
to the generalized quiver is removed by the RG flow and in the IR we recover the original theory
 ${\cal T}_{IR}$.  A comment about R-symmetry breaking and its restoration is in order. The FI parameter
 breaks explicitly the $SU(2)_R$ symmetry of the UV theory to an $SO(2)_R$ subgroup.\footnote{However, the theory has still exact ${\cal N}=2$ supersymmetry,
 as explained in \cite{Hanany:1997hr,Vainshtein:2000hu}.} 
  The 
 vev further breaks this $SO(2)_R$ spontaneously. There is however a linear combination $SO(2)_{\bar R}$
 of $SO(2)_R$ and $U(1)_f$ which is preserved in the new vacuum. In our conventions
 we assign 
 \be
 R_Q = \frac{1}{2}  \, , \qquad f_Q=- 1 \,,
 \ee 
 so that the linear combination
 \be
 \bar R \equiv R + \frac{f}{2} \, 
 \ee 
leaves $Q$ invariant. In the IR, one must recover the full $SU(2)_{\bar R} \times U(1)_{\bar r}$ R-symmetry 
of the ${\cal N}=2$ superconformal algebra.
We identify the $SO(2)_{\bar R}$ symmetry, which is preserved all along the flow, with the Cartan subalgebra of the infrared  $SU(2)_{\bar R}$.

We can also apply the same construction to the external leg of  the generalized quiver that defines ${\cal T}_{IR}$.
The external leg is associated to an $SU(N)$ flavor symmetry.  To obtain 
 ${\cal T}_{UV}$, we break the leg and insert again a free bifundamental hypermultiplet. 
Again  ${\cal T}_{UV}$ has an extra $U(1)_f$ symmetry.
 One can repeat
 exactly the same steps as above and define an RG flow between ${\cal T}_{UV}$ with higgsed $U(1)_f$
 and  ${\cal T}_{IR}$. \footnote{There is a small caveat here concerning the difference between $SU(N)$ and $PSU(N)$ flavor symmetry. We will come back to it in sec \ref{imppoles}}
\begin{figure}
\begin{center}
\begin{tabular}{ccc}
\includegraphics[scale=0.4]{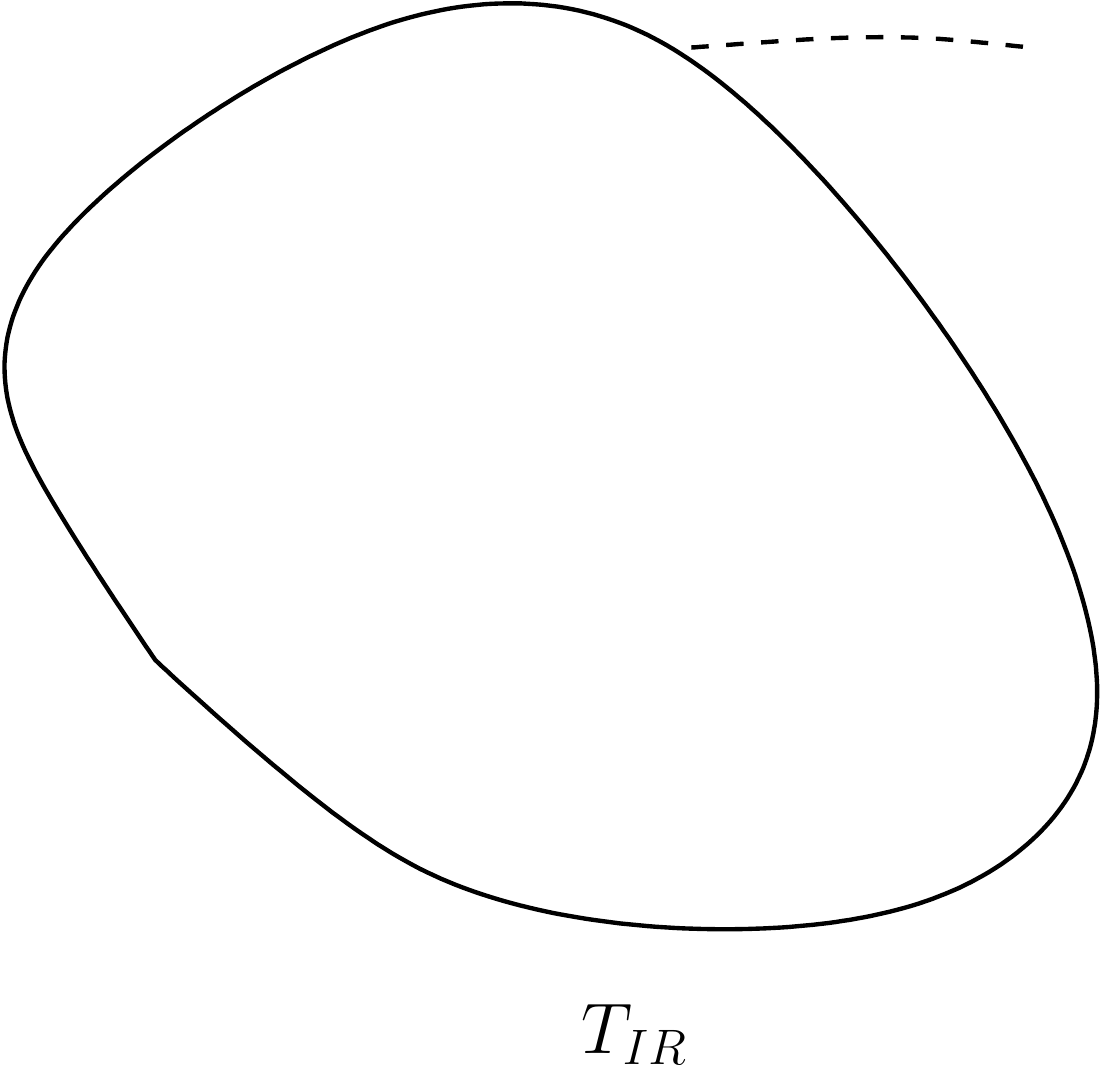}&\;\;\;\;&
\includegraphics[scale=0.4]{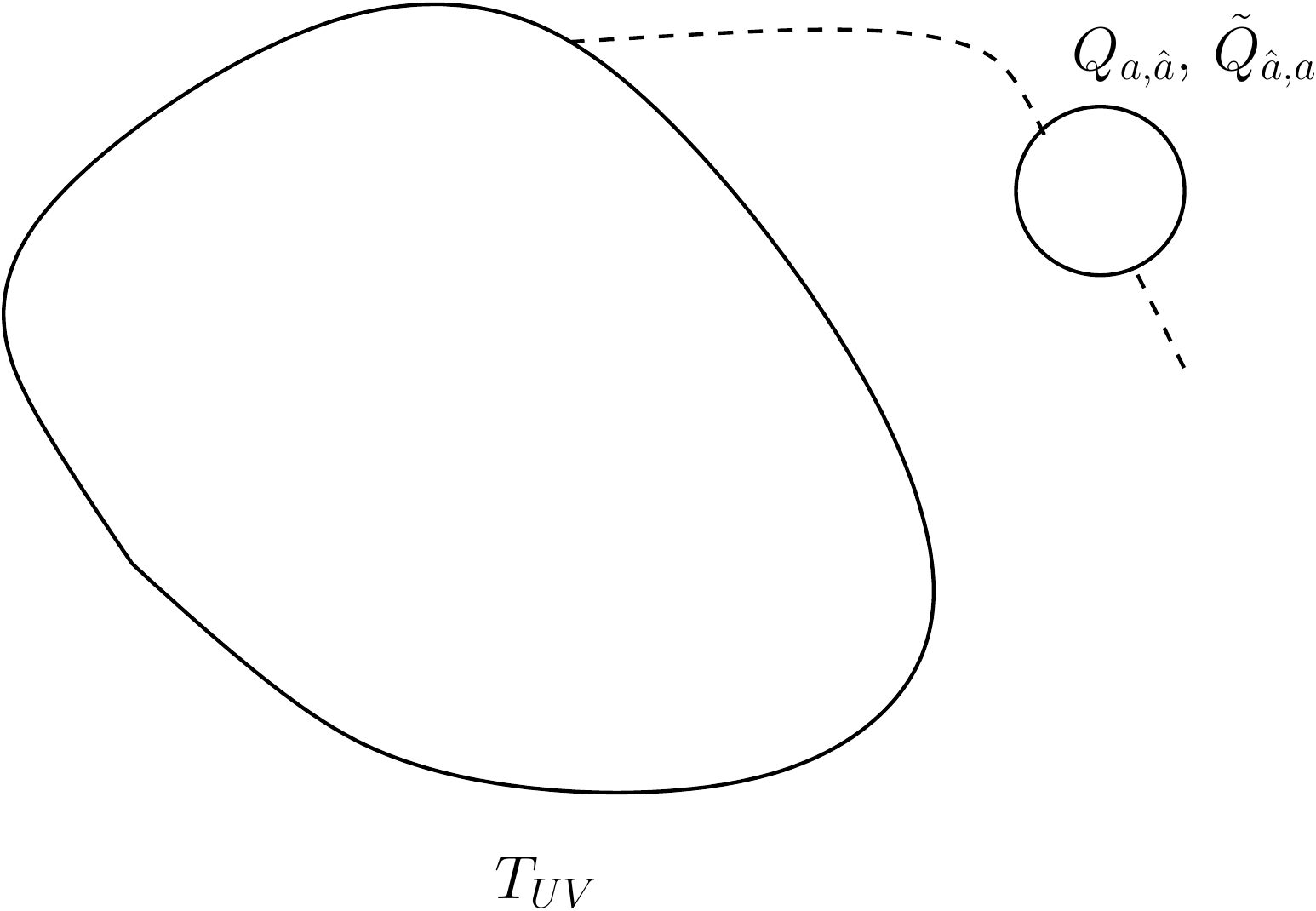}
\end{tabular}
\end{center}
\caption{Our second RG example. \label{T0fig2}}
\end{figure}

While we have focused on two very concrete examples,
the basic idea  is more general. Given an ${\cal N}=2$ SCFT with a $U(1)_f$ flavor symmetry,
we can gauge the $U(1)_f$ and turn on an FI parameter.
The $U(1)_f$ symmetry is then higgsed,  four real directions in the Higgs branch are eaten, and 
 at much lower energies the theory flows to a new $\cN=2$ SCFT ${\cal T}_{IR}$. 
In geometric terms, the IR Higgs branch  is the hyperkahler quotient of the UV Higgs branch by the tri-holomorphic 
$U(1)_f$ isometry which we gauged: we set the moment maps equal to the FI parameters, and remove the orbits of the $U(1)_f$ action. 
In order to define the RG flow properly, we need to pick a vacuum for the theory, {\it i.e.} a  choice of vev for the Higgs branch fields. 
With no loss of generality, we can pick a complex structure, or $\CN=1$ subalgebra, and only turn on a real FI parameter, 
so that a Cartan subgroup $SO(2)_R$ of $SU(2)_R$ is not explicitly broken.  One can then show that  there is always a choice of vacuum such that
$SO(2)_R$ is not spontaneously broken, {\it i.e.} a point on the Higgs branch where an $SO(2)_R$ rotation can be compensated by a 
$U(1)_f$ gauge rotation.  Indeed, the Higgs branch of the UV SCFT is a hyperkahler cone, with a  radial coordinate $\rho$ which is the moment map for $SO(2)_R$. 
We can look for the point of minimum $\rho$ among solutions of the moment map constraints. At that point, the $SO(2)_R$
isometry is aligned with the $U(1)_f$ isometry, and there is a linear combination $SO(2)_{\bar R}$ which is unbroken. 
In general one expects that $SO(2)_{\bar R}$ should become the Cartan subgroup of the IR $SU(2)_{\bar R}$ R-symmetry group.

\
   
\subsection{RG flow by a variable Higgs branch vev: the gauged perspective}\label{varhiggssec}

We now consider the same setup of figure \ref{T0fig}, but with a more interesting choice of UV boundary conditions.
We still embed ${\cal T}_{IR}$ 
into  the larger theory ${\cal T}_{UV}$,  gauge the extra $U(1)_f$ flavor symmetry
and turn on an FI parameter, but  rather than choosing the trivial background that brings
us back to  ${\cal T}_{IR}$, we consider  a position-dependent profile for the fields.
A very interesting class of backgrounds are  the non-abelian vortex solutions that preserve $(2,2)$ $2d$ supersymmetry.
 Vortices are solutions that depend on two of the four Cartesian coordinates\footnote{At this stage we consider
gauge theories in Minkowski space.}, which we parametrize by a complex coordinate $z$.

Vortex solutions localized in the $z$ plane exist in general only if the FI parameter $v$ is turned on. 
Indeed, the tension of a BPS vortex solution is $2 \pi r v$, where the integer $r$ counts the units of magnetic
$U(1)_f$ flux on the $z$ plane. Away from the location of the vortices 
the scalar fields asymptote to constant vevs selected by the choice of vacuum at infinity. This implies that
 the endpoint of the RG flow is still ${\cal T}_{IR}$ away from the location of the vortices. 
As we flow to the IR, the tension of the vortices goes to infinity, and they become
BPS surface defects in ${\cal T}_{IR}$.

The BPS equations for a vortex solution have a simple structure, and can be organized into a set of F-term constraints and some D-term constraints. 
The F-term constraints  force the complex scalars in the hypermultiplets 
to be covariantly holomorphic, and to satisfy the usual condition $\mu_1 + i \mu_2=0$ for all gauge groups. 
The D-term equation is deformed to
\begin{equation}
\mu_3 = F_{z, \bar z} + v\,
\end{equation}
for all gauge groups, where the FI term $v$ lives in the $U(1)_f$ subgroup.

The first consequence of these equations is that the vev of gauge-invariant chiral operators, {\it i.e.}
chiral gauge-invariant operators in ${\cal T}_{UV}$ with zero $U(1)_f$ gauge charge, is holomorphic on the $z$ plane. 
As it goes to a constant vev to infinity, the vev is constant everywhere. 
The moduli space of BPS solutions can be 
 identified 
with the space of solutions to the F-term constraints modulo complexified gauge transformations,
dropping the D-term equations. One can gauge away the anti-holomorphic component of the gauge connection, $A_{\bar z} \equiv 0$,
so that the scalar vevs  can be taken to be holomorphic, and actually polynomial in $z$. One is left with a quotient by polynomial 
complex gauge transformations. 

The moduli space of BPS solutions with a given vortex number $r$ can be rather intricate. 
We wish to focus on backgrounds where the only scalars that receive a vev at infinity belong to hypermultiplets of the extra node in figure 1.
The baryon operator $B = \det Q$ only carries $U(1)_f$ gauge charge.
The only invertible holomorphic $U(1)_f$ transformations are constant rescalings, so we can write 
\be
B(z) = P_r(z) = \prod_{i=1}^r (z- z_i)\, ,
\ee
up to an overall rescaling. The degree $r$ is the vortex number, and the  $z_i$ are gauge-invariant parameters identified with the position of the $r$ vortices. 

Although we cannot generally describe the full moduli space of vortex solutions for a generic ${\cal T}_{UV}$, we can describe a 
universal subspace, consisting of solutions where only $Q$ receives a vev. 
Then we are effectively considering solutions in ${\cal N}=2$ SQCD with $N_f = N_c$ flavors. Supersymmetric vortices 
in this model have been extensively studied. We refer  to \cite{Shifman:2007ce, Tong:2005un, Eto:2005yh}
 for reviews  and only recall some of the basic facts here.

In addition to the $2r$ (real) position moduli, there are $ 2 (N-1) r$ internal orientational moduli, encoded in the holomorphic matrix
$Q(z)$ modulo gauge transformations that leave  ${\rm det}\, Q (z) = P_r(z)$ fixed.  For example, for $r=1$ there is a
 $\IC \times \IC \IP^{N-1}$ worth of vortex solutions: the $\IC$ factor describes the position, while 
the $\IC \IP^{N-1}$ factor parametrizes the breaking of the diagonal $SU(N)$ gauge group of ${\cal T}_{IR}$ down to $U(1) \times SU(N-1)$. 
For general $r$, 
the vortex moduli space can be described by a $U(r)$ gauged linear sigma model coupled to $N$ fundamentals and one adjoint chiral multiplet (which contains the center
 of mass degrees of freedom) \cite{Hanany:1997vm}.

As we flow to the IR, we can keep all the $r$ vortices at the origin, and set $P_r(z) = z^r$.
The background preserves a diagonal combination $\bar j$ of the angular momentum $j$ in the $z$ plane and of the flavor symmetry $f$,
\be
\bar j  \equiv j  + \frac{r}{N}\,f\, ,
\ee 
which should be identified with the angular momentum of the IR theory. There is also  (for any choice of $P_r(z)$, in fact)
 a preserved combination of the R and flavor symmetries,
\be
\bar R \equiv R + \frac{f}{2}\, \, ,
\ee
which we identify with the IR R-symmetry.
As the vortices have a tension which is proportional to the FI parameter, they become infinitely heavy surface defects in the 
IR SCFT ${\cal T}_{IR}$. We have just described the construction of an infinite family of surface defects $ {\frak S}_{(r,0)}$ in ${\cal T}_{IR}$, labeled by the number of vortices $r$.
We can repeat the construction by considering $s$ vortices in the plane $w$, orthogonal to the $z$ plane in $\mathbb{R}^4$; we consider the background with  $B(w) = w^s$,
and  obtain surface defects in ${\cal T}_{IR}$ which we label as  $ {\frak S}_{(0,s)}$.
Finally we can consider surface defects on both planes at the same time,  labelled as  $ {\frak S}_{(r,s)}$. 

We expect that the flat directions corresponding to the transverse positions of the vortices
(the coefficients  $\{ z_i \}$ of the polynomial $P_r(z)$) will flow to $r$ $2d$ free chiral multiplets $\phi_i$  in the IR. 
 It should be possible to strip off these free chiral multiplets, and define reduced surface defects $\bar {\frak S}_{(r,0)}$ in ${\cal T}_{IR}$.

While the difference in the holomorphic gauge couplings of the two $SU(N)$ gauge factors coupled to the extra node of figure~\ref{T0fig}
 does not affect the bulk IR SCFT, it is a marginal deformation parameter of the $\bar {\frak S}_{(r,0)}$ defects,
which behaves like a Kahler parameter for the $2d$ $(2,2)$ theory. Thus the $\bar {\frak S}_{(r,0)}$ defects should support an exactly marginal twisted chiral operator.\footnote{
We could also construct more intricate variants of the $\bar {\frak S}_{(r,0)}$ surface defects,  starting from a general linear quiver with $k$ $SU(N)$ nodes, and applying our construction 
to each of the $U(1)$ flavor symmetry groups of the bifundamental hypermultiplets. We expect to obtain  surface defects with $k-1$ exactly marginal twisted chiral operators. We leave the exploration of this construction for future work.}

\

\subsection{RG flow by Higgs branch vevs: the ungauged perspective}

In order to define our RG flow, we do not really need to gauge $U(1)_f$. We may simply turn on the appropriate Higgs branch vev in ${\cal T}_{UV}$, and look at physics below the scale set by the vev. 
Let us first consider  the constant-vev background.
Gauging the $U(1)_f$ ate up  three scalar fluctuations transverse to the moment map level set, and one along the $U(1)_f$ orbit. Thus we expect to see in the IR the same ${\cal T}_{IR}$ 
we encountered in the previous section, together with a hypermultiplet which captures these four extra directions. This hypermultiplet transforms with unusual quantum numbers
under the $SO(2)_{\bar R}$ symmetry   $\bar R = R + f/2$:
 it has two components of charge $0$, and components of charge $\pm 1$.  The $SO(2)_{\tilde R}$ is promoted in the IR to an $SU(2)_{\bar R}$ which acts as the standard R-symmetry of ${\cal T}_{IR}$.
 Under  $SU(2)_{\bar R}$  the hypermultiplet scalars transform as a singlet plus a triplet.\footnote{
  This is not in contradiction with the standard R-charge assignment for a hypermultiplet, since
this hypermultiplet, being  free and decoupled from ${\cal T}_{IR}$,
 has an accidental $Sp(1)$ flavor symmetry. The standard R-symmetry for the hypermultiplet is recovered
 as a diagonal combination of  $SU(2)_{\bar R}$ and $Sp(1)$.}

In the absence of the $U(1)_f$ gauge field, there are no dynamical vortex solutions.
But we can still consider the RG flow triggered by turning on
a position-dependent background with $B(z) = P_r(z) = z^r$ in ${\cal T}_{UV}$.
Away from $z=0$, the theory still flows to
 ${\cal T}_{IR}$ times the free hypermultiplet. 
The bulk theory is however modified at the origin by surface defects, which we are led to identify with the 
surface defects defined in the previous subsection. More precisely, we expect them to correspond 
to $\bar {\frak S}_{(r,0)}$ (as opposed to  ${\frak S}_{(r,0)}$), since
 the coefficients of $P_r$ are not fluctuating degrees of freedom in this setup.
 It is less obvious to describe the fate of the 
free hypermultiplet near the origin. The size of the gauge orbits now scales to zero as we approach the origin, as the Higgs branch vev goes to zero there. 
This suggests that the $\bar {\frak S}_{(r,0)}$ defects are coupled to a surface defect for the free hypermultiplet theory. Consistency
of this picture with the ``gauged'' picture of the previous subsection suggests that the degrees of freedom of the free hyper that survive
at the origin should match with the center-of-mass motion degrees of freedom of the vortices in the previous description.

\

\

\section{The prescription}\label{prescsec}

In this section, we 
  give a physical interpretation of a class of poles of the superconformal index
   in terms of the surface defects defined in the previous section. We will be led to a precise prescription
   to evaluate the index in the presence of surface defects.
 
Let us first review the definition of the superconformal index and fix our basic conventions.
The index of an ${\cal N}=2$ SCFT can be thought of as a trace over states of the theory in the radial quantization, i.e.
a partition function on $S^3\times S^1$~\cite{Kinney:2005ej},\footnote{In this paper we follow the notations of~\cite{Gadde:2011uv}.}
\begin{equation}\label{inddef}
{\cal I}=\mathrm{Tr} (-1)^F\,\left(\frac{t}{pq}\right)^r\,
 p^{j_{12}}\,
 q^{j_{34}}\,
 t^{R}\,
 \prod_i a_i^{f_i} \,.
\end{equation}
We denoted as $j_{12}$ as $j_{34}$ the rotation generators in two orthogonal planes:
$j_{12}=j_2+j_1$ and $j_{34}=j_2-j_1$ with $j_{1,2}$ being the Cartans of the Lorentz $SU(2)_1\times SU(2)_2$ isometry of $S^3$.
 $r$ is the $U(1)_r$ generator, and $R$ the $SU(2)_R$ generator of R-symmetries. 
The $a_i$ are fugacities for the flavor symmetry generators $f_i$.
We will always assume that
\be
|p|<1\,,\qquad
|q|<1\,,\qquad
|t|<1\,,\qquad 
|a_i|= 1\,,\qquad
\left|\frac{p\,q}{t}\right|<1\,.
\ee
The particular index~\eqref{inddef} counts states annihilated by supercharge 
$\widetilde {\cal Q}_{1\dot -}$ (and its 
Hermitian conjugate): this charge has $SU(2)_R$ charge $\half$, $r$-charge $-\half$,
and $SU(2)_1\times SU(2)_2$ charges $(0,\,-\half)$. Other choices of the supercharge 
will give an equivalent index for ${\cal N}=2$ theories~\cite{Kinney:2005ej}.
Thus the states which contribute to the index~\eqref{inddef}   satisfy~\cite{Gadde:2011uv}
\be\label{BPS}
2\left\{\widetilde {\cal Q}_{1\dot -},\,\left(\widetilde {\cal Q}_{1\dot -}\right)^\dagger\right\}=
E-2j_2-2R+r=0\,.
\ee
Let us also mention here the single letter indices (partition functions) of the basic ingredients
of ${\mathcal N}=2$ field theories, the hypermultiplet and the vector multiplet,
\be
&&{\cal I}^{s.l.}_{H}(p,q,t,a)=\frac{\sqrt{t}-\frac{pq}{\sqrt{t}}}{(1-p)(1-q)}(a+a^{-1})\,,\\
&&{\cal I}^{s.l.}_V(p,q,t)=-\frac{p}{1-p}-\frac{q}{1-q}+\frac{\frac{pq}{t}-t}{(1-p)(1-q)}\,.\nonumber
\ee Here $a$ is the fugacity for the $U(1)$ charge of the half-hypers. To obtain the multi-particle 
index one computes the plethystic exponent the single-letter partition function. For example,
the full, multi-particle, index of the vector multiplet is
\be
{\cal I}_V={\text{ PE}}\left[{\cal I}^{s.l.}_V(p,q,t)\right]=
\exp\left\{
\sum_{n=1}^\infty \frac{1}{n}\,{\cal I}^{s.l.}_V(p^n,q^n,t^n)
 \right\}\,.
\ee
Given an index ${\cal I}({\bf a},\dots)$ of a theory with $SU(N)$ flavor symmetry labeled by fugacities ${\bf a}$ the index 
of a theory with this flavor symmetry gauged is given by
\be
\oint \left[\prod_{i=1}^{N-1}\frac{da_i}{2\pi i a_i}\right]\;\Delta({\bf a})\;
{\cal I}_V({\mathbf a})\;
{\cal I}({\bf a},\dots)\, ,
\ee where
\be
\Delta({\bf a})=\frac{1}{N!}\prod_{i\neq j}(1-a_i/a_j)\,,
\ee is the familiar  $SU(N)$ Haar measure, and 
\be
{\cal I}_V({\mathbf a})\equiv {\text{ PE}}\left[{\cal I}^{s.l.}_V(p,q,t)\left(
\sum_{i,\,j=1}^Na_i/a_j-1\right)
\right]\,,
\ee is the index of an $SU(N)$ vector multiplet.

\

\subsection{Some important poles}\label{imppoles}

The superconformal index has  a rich analytical structure, in particular it exhibits many poles in the flavor fugacities.
A divergence in the index is associated to the integration over a bosonic zero mode in the
$S^3 \times S^1$ partition function.
The introduction of a maximal set of superconformal and flavor fugacities is in fact motivated by the desire
to regulate these divergences. 

We wish to give a physical interpretation of the simple poles of the index and characterize their residues.
Not to obscure the main point with heavy notations,  consider a schematic definition of the  index,
 \be
{\cal I}(a,b)=\Tr(-1)^F\,a^f\,b^{g}\,,
\ee where $f$ and $g$ are some charges. 
Let us assume that ${\cal I}$ has a pole in fugacity $a$,
\be
{\cal I}=\frac{\widetilde {\cal I}(a,\, b)}{1-a^{f_{\cal O}}\, b^{g_{\cal O}}}\,.
\ee
The pole is naturally associated to a bosonic
operator, ${\cal O}$, with charges $f=f_{\cal O}$ and  $g=g_{\cal O}$.
The fact that we have the divergence means that  arbitrary high powers of the operator ${\cal O}$ contribute to the index.
\footnote{In the simplest case,
this operator generates a ring and all the its powers contribute. More generally, there can be relations,
which are taken into account by the numerator $\widetilde {\cal I}(a,\, b)$.}
Although for generic values of the fugacities  fluctuations  are massive and the $S^3 \times S^1$ partition function is finite, for $a^{f_{\cal O}}\, b^{g_{\cal O}} =1$
a flat direction opens up and the partition function diverges. 
 The residue  at the pole 
 is given by $\widetilde {\cal I} (
b^{-g_{\cal O}/f_{\cal O}}
,\,b)$, so it can be written
\be
\widetilde {\cal I}=\Tr(-1)^F\,\,b^{g'}\,.
\ee
where $g'$ denotes the shifted charge
\be
g'=g-\frac{g_{\cal O}}{f_{\cal O}}\, f\,.
\ee The shifted charge is preserved in the background where ${\cal O}$ has a non-zero vev.
To characterize the residue, we can use the understanding of the divergence 
 as arising from a zero-mode.
We then expect the divergence of the partition function to be controlled by the behavior of theory ``at infinity'' in the moduli space parametrized by the vev of ${\cal O}$,
that is, by the properties of  the  IR theory  reached at the endpoint of the 
 the RG flow triggered by giving   ${\cal O}$ a vev.  We should interpret  $\widetilde {\cal I}$ as the index of the IR fixed of this RG flow.
 
By computing the residue of fugacity $a$ we have lost the $U(1)$ symmetry corresponding to it in $\widetilde{\cal I}$ and thus the IR theory has smaller global symmetry, {\it i.e.}
giving vev to ${\cal O}$ explicitly breaks some of the symmetries of the UV theory. 
Note that the number of states contributing to $\widetilde {\cal I}$ is smaller (or equal) to the number of states contributing to ${\cal I}$ since 
after shifting the charges some states might cancel out but new states do not start contributing. This is similar in spirit to Romelsberger's prescription
for computing indices of IR fixed points in the ${\cal N}=1$ setups~\cite{Romelsberger:2005eg,Romelsberger:2007ec,Dolan:2008qi,Festuccia:2011ws}. 
\

Let us specialize this general framework to the flavor fugacities poles that arise from Higgs-branch zero modes.
Consider a general SCFT with a Higgs branch of vacua ${\cal M}$, parameterized by a ring of BPS operators $\{ O^\a \}$,
which sit in spin $R^\a$ representations of $SU(2)_R$,  have charges $f^\a_i$ 
under the flavor symmetries and carry no angular momenta or $r$-charge. 
The index receives a contribution from the highest weight component in the $SU(2)_R$ multiplet, complex chiral fields which we  still denote as  $O^\a$.
 The operator ring matches the ring of holomorphic functions 
on ${\cal M}$. 
For every generator $O^\a$, all the powers of operator $O^\a$ contribute to the index, and
summing the geometric series  this implies that the index has a pole\footnote{If we specialize to $p=q=0$, 
so that the chiral operators on the Higgs branch are the only operators which contribute to the 
index, the index takes in general the form of a polynomial in $t$ over the product of factors 
$(1- t^{R_\a} \prod_i a_i^{f^\a_i})$. In some cases, this specialization of the superconformal fugacities relates 
the index to the Hibert series (as defined {\it e.g.} in~\cite{Benvenuti:2010pq}) of the Higgs branch~\cite{Gadde:2011uv}.
This relation was explored recently in~\cite{DR,Keller:2012da,Hanany:2012dm}. 
} at $t^{R_\a} \prod_i a_i^{f^\a_i} =1$.

 We now focus on  the constant Higgs branch vevs that we have considered in section \ref{motsec},
where we have a clear understanding of how the R-charges of the UV theory match the R-charge of the effective IR description.
We can then hope to relate the residue of  UV index  at the pole with the index of the IR theory. 
Let us  consider the setup of figure 2, and study the divergence of the UV index associated
to the baryon operator $B$. 
The position of the pole can be written as $t^{R_B} a^{f_B}=1$, where  $a$ is the fugacity for $U(1)_f$.
The $SU(2)_{\bar R}$  Cartan generator is $\bar R= R - \frac{R_B}{f_B} f$, so
if we sit at the pole $a = t^{-R_B/f_B}$ we are simply replacing $t^R$ with $t^{\bar R}$, {\it i.e.} evaluating the index with the fugacities which are appropriate for 
the IR theory.

Using $R_B = N/2$ and $f_B = N$,  the equation $t^{N/2} a^N =1$ implies
 that the index has $N$ poles at the locations
\be
a=\exp\left[\frac{2\pi i }{N}\,\ell\right]\, t^\half\,,\qquad \ell=0,\cdots, N-1\,.
\ee The residues of all these poles quantify 
 the same physics: the IR fixed point of a theory with a vev for baryonic operator.
This baryonic operator breaks the $U(1)$ flavor symmetry parametrized by $a$ to ${\mathbb Z}_N$
symmetry, and looking at different poles of the index corresponds to the index of the IR theory
with one of the $N$ group elements of ${\mathbb Z}_N$ inserted in the trace formula.
Thus, it makes sense to either sum over the $N$ poles or consider the contribution 
of one of the poles multiplied by $N$. In our setup of figure 2 the former corresponds
to gauging the ${\mathbb Z}_N$ center of the $SU(N)$ flavor symmetry and thus leaving only 
$PSU(N)$ as the flavor symmetry in the IR: this scenario is the one naturally occuring in the gauged
perspective of section~\ref{gaugedconstsec}.
 The latter point of view however corresponds to keeping the full $SU(N)$ flavor 
symmetry which is more natural when considering flows initiated by a vev:
this will be the point of view we will adopt in the rest of the paper. 

In the IR we also expect to have the free hypermultiplet with scalars transforming in the $1+3$ irreps of $SU(2)_{\bar R}$. The index of such a hypermultiplet captures exactly the 
divergent part of the UV index at $t^{N/2} a^{-N}=1$, as it contains exactly the massless fluctuation which causes the divergence. 
If we start from the index of a standard free hypermultiplet, we can get the index of the $1+3$ hypermultiplet by setting $a=t^{1/2}$.
Up to the divergent zero mode, we get precisely ${\cal I}^{-1}_V$, the inverse of the index of a $U(1)$ vector multiplet.
This motivates our final prescription:
\begin{equation}\label{prescC}
{\cal I}[{\cal T}_{IR}] = N\,{\cal I}_V \,\mathrm{Res}_{a = t^{1/2}}\;\frac{1}{a}\, {\cal I}[{\cal T}_{UV}] \,.
\end{equation}
The origin of the factor $N$ is explained in the paragraph above.
In summary, extracting the residue  at $a = t^{1/2}$ in the index of ${\cal T}_{UV}$ simply amounts to removing the extra node in the quiver (see figure 2)
and recovering the original theory ${\cal T}_{IR}$. This is in perfect agreement with the intuition coming from the RG with a constant vev for the baryon operator.

\

The index also receives contributions from holomorphic derivatives of the baryon operator
 in the $12$ and $34$ planes,  
 $\partial_{12}^r \partial_{34}^s B$, so  by the same logic as above 
 we expect poles at  $t^{R_B} p^r q^s  a^{f_B} =1$. 
 We wish to relate the residues at these poles to the theories obtained in the IR 
 by turning  on a space-dependent vev $B (z, w) = z^r w^s$. 
 As we have explained in section 2,
 the IR theory is ${\cal T}_{IR}$ in the presence of an additional surface defect  $\bar {\frak S}_{(r,s)}$,
 with some extra decoupled matter coming from the free hypermultiplet. 
  In the $S^3 \times S^1$ geometry,
 the surface defects  fill the ``temporal'' $S^1$ and (for $r, s \neq 0$) the two maximal circles inside the $S^3$ fixed by the $j_{12}$ and $j_{34}$ rotations,
 respectively (or only one of them for $r=0$ or $s=0$).

We are almost ready to compute the index of the theory ${\cal T}_{IR}$ in the presence of surface defects.
  What remains to be done is to figure out which contribution from the free hypermultiplet should we strip off. 
A simple choice is to just strip off the bulk contribution. We will later identify that with the index for the theory in the presence of the ${\frak S}_{(r,s)}$
 surface defect,
\begin{equation}
{\cal I}[{\cal T}_{IR},{\frak S}_{(r,s)}] =N\, {\cal I}_V\, \mathrm{Res}_{a = t^{1/2} p^{r/N} q^{s/N}} 
\;\frac{1}{a}\,
{\cal I}[{\cal T}_{UV}] \,.
\end{equation}
Again, when we set $a = t^{1/2} p^{r/N} q^{s/N}$ we are shifting the UV rotation generators to become the 
rotation generators appropriate to the IR SCFT, as explained in section \ref{motsec}. 

\

\subsection{A toy model: the free hypermultiplet}

To understand which terms should be stripped off from the index to get rid of the decoupled degrees of freedom, it is useful to 
look at a toy-model where the UV theory coincides with a single hypermultiplet. 
The index  of a free hypermultiplet is 
\begin{equation}\label{hyppee}
{\cal I}_{H}=\prod_{n,m \geq 0} \frac{1-t^{-1/2} p^{n+1} q^{m+1} a^{\pm1}}{1-t^{1/2} p^{n} q^{m} a^{\pm1}}\equiv\Gamma\left(t^\half\,a^{\pm1};q,p\right)\,. \end{equation}
Here and throughout the paper we use a shorthand notation where the $\pm$ exponent means
that we take the product over both choices of sign.
This index is thus given by a product of two elliptic Gamma functions.\footnote{
The relevance of these functions to the index computations was observed in~\cite{Dolan:2008qi}.
See also~\cite{Spiridonov2} for a nice review of the properties of elliptic Gamma functions.}
The denominator of~\eqref{hyppee} comes  from the  scalar
fields in the hypermultiplet and their derivatives. 
If we specialize to $a = t^{1/2} p^n q^m$ in the free hypermultiplet index, remove the zero mode and multiply it by ${\cal I}_V$, we get some very suggestive results. 
For example, at $a = t^{1/2} p$ 
\begin{equation}\label{res1}
R_{1,0} = \prod_{m \geq 0} \frac{(1-t^{-1} q^{m+1}) (1-t q^{m})}{(1-p^{-1}q^{m})(1-p q^{m+1})}=
\frac{\theta(t;q)}{\theta(p^{-1};q)}\,.
\end{equation} Here the theta-function is defined as
\be
\theta(z;q) \equiv \prod_{m=0}^\infty (1-z q^m)(1-z^{-1}q^{m+1}),\qquad \theta(qz;q)=\theta(z^{-1};q)\,. 
\ee We will also use the Pochhammer symbol,
\be
(a;\;b) \equiv \prod_{\ell=0}^\infty(1-a\;b^{\ell})\,.
\ee
The residue $R_{1,0}$ takes the form of a $2d$ index for a free $2d$ chiral multiplet, living on the circle fixed by the $j_{12}$ rotations.
We present the details of the precise relation between~\eqref{res1} and the $2d$ index 
in appendix~\ref{imbsec}.
The bosonic scalar field in the chiral multiplet has charge $1$ under rotations in the $12$ plane: it is exactly the 
$2d$ chiral field which we planned to strip off from ${\frak S}_{(1, 0)}$ to get $\bar {\frak S}_{(1, 0)}$. 

In order to read off the residue at $a = t^{1/2} p^r$, we can simplify some manipulations by looking at the single-letter partition function. We specialize to $a = t^{1/2} p^r$, subtract $1$ to remove the pole
\begin{equation}
-\frac{p}{1-p} -\frac{q}{1-q} + \frac{pq/t-t}{(1-p)(1-q)} + \frac{t p^r - p^{r+1} q +p^{-r} - t^{-1} p^{1-r} q}{(1-p)(1-q)} - 1
\end{equation} 
or
\begin{equation}
\frac{1-p^r}{1-p} \frac{p q - t}{1-q}+  \frac{1-p^{-r}}{1-p^{-1}} \frac{p^{-1}- q t^{-1}}{1-q}
\end{equation} 
which gives the residue 
\begin{equation} \label{Rr0}
R_{r,0} = \prod_{u=0}^{r-1} \prod_{m \geq 0} \frac{(1-t^{-1} p^{-u} q^{m+1}) (1-t p^u q^{m})}{(1-p^{-u-1}q^{m})(1-p^{u+1} q^{m+1})}=\prod_{u=0}^{r-1} \frac{\theta(tp^u;q)}{\theta(p^{-1-u};q)}\,.\end{equation}
This has the form of a $2d$ index for $r$ free chiral multiplets, living on the circle fixed by the $j_{12}$ rotations
(see appendix~\ref{imbsec} for the definition of the $2d$ index).
The $2d$ chiral fields we see in $R_{r,0}$  carry charges $1, \cdots r$ under the $12$ rotations, 
and so look exactly like the $2d$ chirals which we  planned to strip off from ${\frak S}_{(r,0)}$ to get $\bar{\frak  S}_{(r, 0)}$.

Finally, in order to read off the residue at $a = t^{1/2} p^r q^s$ we can use again the single-letter partition function
\begin{equation}
-\frac{p}{1-p} -\frac{q}{1-q} + \frac{pq/t-t}{(1-p)(1-q)} + \frac{t p^r q^s- p^{r+1} q^{s+1} +p^{-r} q^{-s}- t^{-1} p^{1-r} q^{1-s}}{(1-p)(1-q)} - 1
\end{equation} 
We can write that as a sum of three terms. The first term corresponds to $R_{r,0}$. The second to $R_{0,s}$, which is the same as $R_{s,0}$ with $p$ and $q$ exchanged,
\begin{equation} \label{R0s}
R_{0, s} = \prod_{u=0}^{s-1} \prod_{m \geq 0} \frac{(1-t^{-1} q^{-u} p^{m+1}) (1-t q^u p^{m})}{(1-q^{-u-1}p^{m})(1-q^{u+1} p^{m+1})}=\prod_{u=0}^{s-1} \frac{\theta(t q^u;p)}{\theta(q^{-1-u};p)}\,.\end{equation}
The third piece is 
\begin{equation}
(t-p q) \frac{1-p^r}{1-p}\frac{1-q^s}{1-q} -(t^{-1}-p^{-1} q^{-1}) \frac{1-p^{-r}}{1-p^{-1}}\frac{1-q^{-s}}{1-q^{-1}}
\end{equation}
This contributes to the residue through $r s$ ratios of the form 
\begin{equation}
\frac{(1-p^{n+1} q^{m+1})(1-t^{-1}p^{-n} q^{-m})}{(1-p^{-n-1} q^{-m-1})(1-t p^{n} q^{m})} = \frac{p q}{t}
\end{equation}
Hence the total residue is 
\begin{equation}
R_{r,s} = \left( \frac{p q}{t} \right)^{r s} R_{r,0} R_{0,s}
\end{equation}
Up to the prefactor, this has the form of the product of two $2d$ indices, one for a $2d$ theory on the circle fixed by $j_{12}$,
the other living on the circle fixed by $j_{34}$.  The prefactor
corresponds to a re-definition of the $U(1)_r$ 
charge by a constant shift. Such shifts of an abelian symmetry generator are common in the presence of defects, unless the generator 
sits in a non-abelian factor of the subgroup of the superconformal group preserved by the defect. Here we have two distinct surface defects,
in a quarter-BPS configuration. 

\

\

Inspired by this calculation, we propose our final recipe to completely strip off the free hypermultiplet contributions:
\begin{equation}\label{finpresc}
\boxed{{\cal I}[{\cal T}_{IR}, \bar {\frak S}_{(r,s)}] =
N\; R_{r,s}^{-1}\;{\cal I}_V\;\mathrm{Res}_{a = t^{1/2} p^{r/N} \; q^{s/N}} 
\;\frac{1}{a}\,
{\cal I}[{\cal T}_{UV}]}\,.
\end{equation}

\

\subsection{Gauging $U(1)_f$ and the index}\label{u1gauging}

Our basic prescription (\ref{finpresc}) has a suggestive interpretation in terms of the index of the theory where the $U(1)_f$
symmetry is gauged,
\begin{equation} \label{gaugedindex}
\oint \frac{da}{2 \pi i a} \, a^\xi   {\cal I}_V \, {\cal I} \, [{\cal T}_{UV}] \, .
\end{equation}
Here the contour integral is on the unit circle $|a|=1$ and we added a FI parameter $\xi$ (in units where the $S^3$ has unit radius), which can
be turned on in the index.\footnote{We thank N.~Seiberg for comments on this point.}

Notice that  (\ref{gaugedindex}) only makes sense if $\xi$ is quantized. 
If $\xi>0$, it is natural to try and evaluate the contour integral by picking the residues at $|a|<1$. 
In particular, we will pick the residues at $a= t^{1/2} p^{r/N} q^{s/N}$. In general there could be other poles.
The dominant contribution at large $\xi$ will be 
the residue at $a = t^{1/2}$, which is precisely our prescription for the index of ${\cal T}_{IR}$, while  the subleading contributions (or a subset of them)
correspond to the index of  ${\cal T}_{IR}$ with the insertions of ${\frak S}_{(r,s)}$.
The UV index can then be written as a sum over the indices of the IR theories 
associated to the different supersymmetric backgrounds. 

One may worry  that the  $U(1)_f$ gauged theory is not well defined since it has both a Landau pole and an anomalous $U(1)_r$ symmetry.
The presence of a Landau pole may not be a serious problem, as the index is independent of the gauge coupling, and we can suppress the Landau pole 
arbitrarily by making the gauge coupling smaller and smaller. The anomaly in $U(1)_r$ would appear to be a more serious obstruction. 
In principle, one can reabsorb the $\mathrm{Tr} F \wedge F$ anomaly for the $U(1)_r$ symmetry 
by redefining the $U(1)_r$ charge as 
\begin{equation}
r \to r + b_0 \int_{S^3} A \wedge F\,,
\end{equation} where $b_0$ is some proportionality constant.
The Chern Simons action is usually not gauge invariant, but for an Abelian gauge theory ($F=dA$) $U(1)_f$, on a three manifold with no two-cycles, such as $S^3$, 
the gauge variation 
\begin{equation}
 \int_{S^3} d\Lambda \wedge F
\end{equation}
is actually zero. 

The redefinition of the $r$ charge will place an overall power of $p q/t$ in front of the contribution to the index of interesting topological sectors
for $U(1)_f$. A pair of linked vortices may exactly fit the bill. Each vortex contains a unit of flux in the core, {\it i.e.} the integral of $A$ around the vortex should be one. 
If we have linked vortices, at the core of each we have flux $F$ from that vortex, and potential $A$ from the other vortex, which conspire to give a unit contribution to $\int_{S^3} A \wedge F$.
This may neatly explain  the prefactor $\left( \frac{p q}{t} \right)^{r s}$ we encountered in $R_{r,s}$.

\

\

\section{Residues and Ruijsenaars-Schneider models:  $A_1$ theories }\label{su2sec}

We now turn to a detailed analysis of the analytic
properties of the index for the $A_1$ theories of class ${\cal S}$, the $A_1$ generalized quivers of ~\cite{Gaiotto:2009we}.
The generalization to higher rank is conceptually straightforward
and will be presented in section~\ref{sunsec}.
The $A_1$ case is technically the simplest as there is only one type of flavor puncture,
carrying an $SU(2)$ flavor symmetry. The basic building block is the free half-hypermultiplet
in the trifundamental representation of $SU(2)^3$,  associated to the three-punctured sphere.
The general $A_1$ theory of class ${\cal S}$ is associated to a pair-of-pants decomposition
of a genus ${\frak g}$ surface with $p$ punctures.

\begin{figure}
\begin{center}
\includegraphics[scale=0.35]{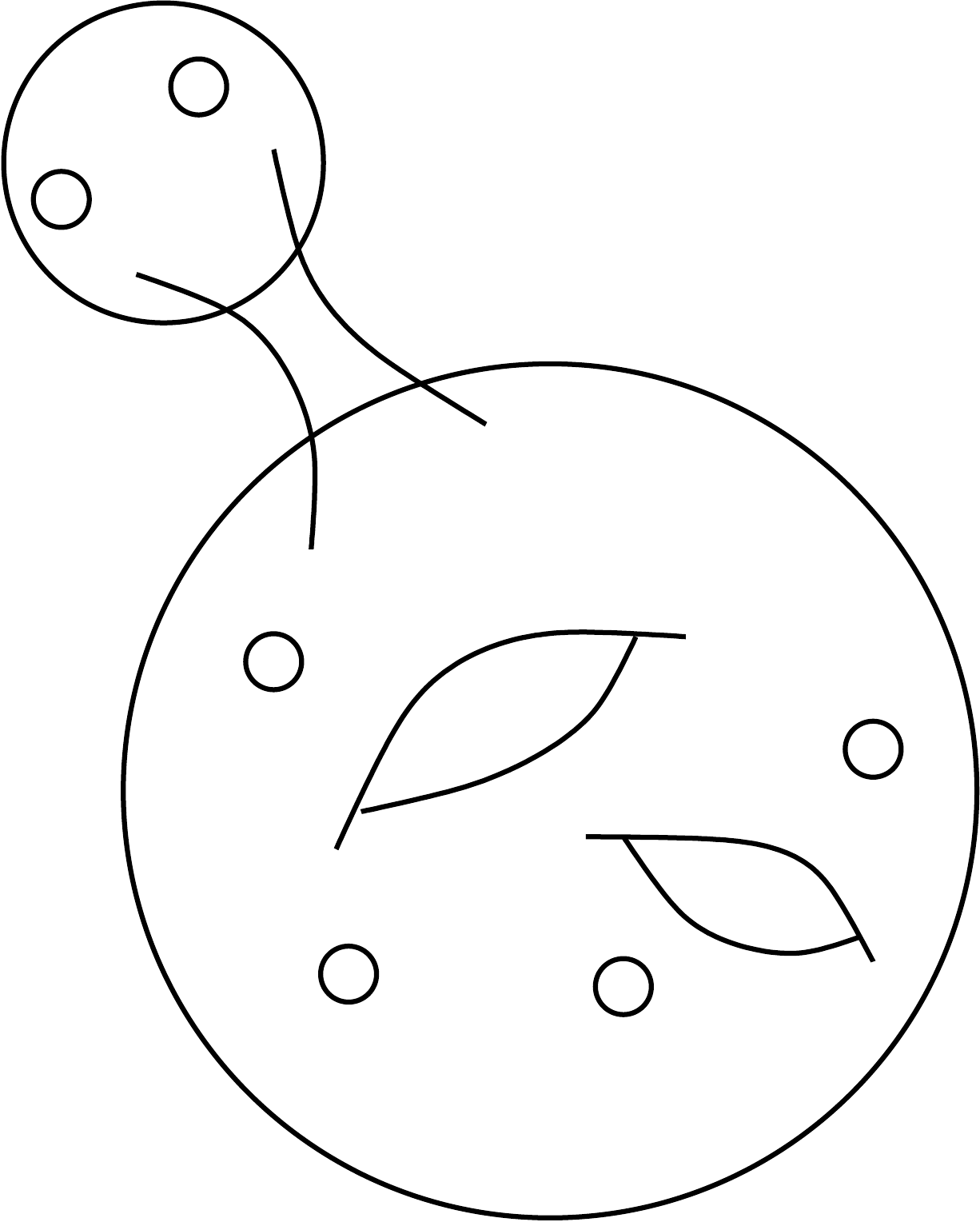}
\end{center}
\caption{The setup of figure 2  for  the $A_1$ generalized 
quivers. The SCFT ${\cal T}_{IR}$, associated
to a Riemann surface ${\cal C}$ with $s$ punctures, is coupled
to the SCFT of free trifundamental half-hyper, associated to a  sphere with three punctures.
The two SCFTs are coupled by choosing an $SU(2)$ flavor symmetry of each theory,
    and weakly gauging a diagonal $SU(2)$ subgroup. The resulting SCFT is denoted by 
 ${\cal T}_{UV}$, and corresponds to the degeneration limit of a surface with $s+1$ punctures
 shown in the figure.
\label{figdeg}}
\end{figure}

We consider  the setup of figure 2. The SCFT  ${\cal T}_{UV}$ is taken to be an 
 $A_1$ class ${\cal S}$ theory with $p=s+1$ punctures, $s \geq 1$.
We are interested in  the degeneration limit  where 
a three-punctured sphere is connected by a long cylinder to the rest of the surface,
an $s$-punctured Riemann surface which we denote by ${\cal C}$,  see   figure~\ref{figdeg}. 
The field theory interpretation of this geometry
is familiar: the SCFT associated to ${\cal C}$ is coupled to a  trifundamental half-hyper,
by weakly gauging a diagonal $SU(2)$ group. In the notation of figure 2, the theory associated to ${\cal C}$ is ${\cal T}_{IR}$.

We then focus on the functional dependence of the superconformal index of ${\cal T}_{UV}$ on either
 one of the flavor fugacities of the weakly-coupled hyper. As outlined in sections 3 and 4,
 we expect to find poles at certain values of the flavor fugacity, and to be able to interpret
 the residue at each of these poles  as the index of ${\cal T}_{IR}$ in the presence
 of additional surface defects.  The analysis will be somewhat technical
 but the main idea is simple. The basic point is that since the index of ${\cal T}_{UV}$
 is a contour integral in the fugacity of the gauged $SU(2)$ associated to the long cylinder,
 a pole in the flavor fugacity arises
 when the integration contour is ``pinched'' by two singularities.
  A little calculation reveals that the residue 
 takes the form of a difference operator acting on the index of ${\cal T}_{IR}$.
 Our main claim is that the action of this difference operator introduces a surface defect
 in ${\cal T}_{IR}$, equation (\ref{Sr0}).
 Remarkably, the difference operator turns out to be the Hamiltonian
 of the relativistic Calogero-Moser model, also known as the elliptic Ruijsenaars-Schneider (RS) 
model.\footnote{The RS integrable models appear in 
several contexts in the study of ${\cal N}=2$ gauge theories, see {\it e.g.}~\cite{Nekrasov:2009rc,Nekrasov:2010ka,Chen:2011sj,Bulycheva:2012ct}.
The same quantum mechanical integrable models are also related to gauge theories in lower dimensions (see {\it e.g}.~\cite{Gorsky:1993pe,Minahan:1993mv,
Gorsky:1993dq,Gorsky:1994dj,gorsky,Gorsky:2000px}).  It would be interesting to understand the connection
between these lower-dimensional gauge theories and the $2d$ TQFT that computes the index, which for $q=t$ 
is known to reduce to $2d$ $q$-deformed Yang-Mills in the zero area limit \cite{Gadde:2011ik}.
}

In the following sections we will employ these difference operators to extract several physical statements about the index
of theories of class ${\cal S}$. In particular we will gather evidence for the interpretation of the difference operators as
operators introducing surface defects into the index computation.
The difference operator can act on any one of the $s$ flavor  fugacities  of ${\cal T}_{IR}$,
and by generalized S-duality we expect to obtain the same
result regardless of which fugacity we decide to act on.  This constraint
is so powerful that it allows 
to completely fix the index for any $A_{1}$ generalized quiver, with or without surface defects.
Of course, in the $A_1$ case all theories have an explicit Lagrangian description,
and  the index of any quiver admits an explicit representation as a matrix integral.
What this new logic buys us is a  representation of the index where
the structure constants of the associated $2d$ TQFT have been diagonalized. 
More importantly, this algorithm will generalize easily  to the higher-rank quivers,
which generically do not admit a direct Lagrangian description.

For general values of the three superconformal fugacities $(p, q, t)$, the answer
is given in terms of  eigenfunctions and eigenvalues of the elliptic RS model, whose
analytic form is not fully known. We construct however approximate eigenfunctions in a certain limit,
which serves as a proof of concept of their existence and uniqueness (see appendix~\ref{eigenfuncsec}). For $p \to 0$ (or equivalently $q \to 0$),
the elliptic RS  model degenerates to the much better studied trigonometric RS model, whose eigenfunctions are
the  Macdonald's polynomials. By this route we recover from ``first principles'' the results
of \cite{Gadde:2011uv}.  
Finally we consider the dimensional reduction of the $4d$ superconformal index to the $3d$ partition function.

 \

\

Let us introduce the basic ingredients.
The index associated to the  three-punctured sphere, the index of the free trifundamental 
half-hypermultiplet, is given by
\be\label{hyp}
{\cal I}_{hyp}(a,b,c)=\prod_{m,n\geq0}
\frac{1-p^{n+1}q^{m+1}t^{-\half}\,a^{\pm1} b^{\pm1} c^{\pm 1}}{1-p^{n}q^{m}t^{\half}a^{\pm1} b^{\pm1} c^{\pm1} }=
\Gamma\left(t^\half\,a^{\pm1} b^{\pm1} c^{\pm 1};\,q,\,p\right)
\,.
\ee
We couple this free SCFT to a general $A_{1}$ quiver SCFT, which we call  ${\cal T}_{IR}$,
associated to a Riemann surface ${\cal C}$, by gauging fugacity $c$. The result is the SCFT that we call ${\cal T}_{UV}$,
see figures~2 and 3.  
 The index ${\cal I}[{\cal T}_{UV}]$ ($\equiv {\cal I}$ for short) reads
\be\label{inda}
{\cal I}(a, \dots)=\oint\frac{dc}{2\pi ic}\Delta(c)\,{\cal I}_V(c)\, {\cal I}_{hyp}(a,b,c)\, 
{\cal I}_{\cal C}(c^{-1},\dots)\, ,
\ee
where ${\cal I}_{\cal C} \equiv {\cal I}[{\cal T}_{IR}]$. 
From the physical considerations of sections 2 and 3, we expect  ${\cal I}(a, \dots)$ to have  simple poles
at $a = t^{1/2} p^{r/2} q^{s/2}$, for $r$ and $s$  non-negative integers.\footnote{We do not claim these are the only poles in general. 
The poles we discuss here  are the ones for which we found a simple interpretation
in terms of the behavior at infinity of the Higgs branch. 
Other poles may have different interpretations: see {\it e.g.}~\cite{DR} for some special examples.
}

Let us start with the simplest case, the pole at
\be\label{cpole}
a=t^\half\,.
\ee
To verify that  ${\cal I}$ has indeed a pole at this value of the fugacity,
we proceed as follows. The integrand has (among others) simple poles in $c$ at 
\be\label{poles1}
c=t^\half b^{\pm1} a^{-1},\qquad
c=t^{-\half} b^{\pm1} a^{}\,.
\ee
Setting $a=t^\half$ these two poles collide and pinch the $c$ integration contour 
producing a simple pole in $a$.
To compute the residue at $a=t^\half$ of ${\cal I}$  we can pick up the residues at
the  poles (\ref{poles1}) in $c$ inside the integration
contour. The other  poles in $c$ do not contribute
to the residue in $a$. Explicit evaluation gives
\be
{\text{Res}}_{a=t^\half}\;\frac{1}{a}\,\,\left[{\cal I}\right]&&=
\sum_{\pm}{\text{Res}}_{a=t^\half;c=t^\half b^{\pm1}a^{-1} }\;\frac{1}{a}\,\,\left[\Delta(c)\,{\cal I}_V(c)\, {\cal I}_{hyp}(a,b,c)\,
{\cal I}_{\cal C}(c^{-1},\dots)\right]\nonumber\\
\qquad&&=\half\,\half\frac{2\;{\cal I}_{\cal C}(b,\dots)\Delta(b)\,{\cal I}_V(b)}
{\prod_{\pm}(1-b^{\pm2})}
\,PE\left[\frac{p+q-2pq+t-\frac{pq}{t}}{(1-p)(1-q)}(b^2+b^{-2}+2\right]\\
\qquad&&=\half{\cal I}_V^{-1}\;{\cal I}_{\cal C}(b,\dots)\,.\nonumber
\ee The origin of the different numeric factors is as follows. The factor of $2$ comes
from summing over the two poles in~\eqref{poles1}. One factor of $\half$ comes from the Haar measure of $SU(2)$ and the
second factor of $\half$ comes from evaluating the hypermultiplet index at the pole. 

There is also a pole at $a=-t^{\half}$,
whose residue is $\half {\cal I}_V^{-1}\,{\cal I}_{\cal C}(-b,\dots)$. 
As we discussed in section 3 we need only consider the pole ~\eqref{cpole}
since with either sign the poles describe the same physics.\footnote{Let us reiterate in this 
concrete context the logic of section~\ref{imppoles}: in the gauged perspective, 
as the contour integral corresponding to the gauging of $U(1)_f$, 
is around the unit circle  both poles at $\pm t^\half$ contribute.
The index of the IR theory is invariant under negating any of the flavor fugacities.
 In other words the flavor symmetry 
has additional discrete ${\mathbb Z}_2$ component which is gauged when we compute the residues.
Thus, after gauging the $U(1)_f$ flavor symmetry (which for $A_1$ quivers 
is enhanced to $SU(2)$), we only have $PSU(2)$ flavor symmetry, as opposed to $SU(2)$, associated with one of the punctures. In the un-gauged perspective it is more natural
to consider one of the poles, say at $a=t^\half$, taken twice and thus keeping the center of $SU(2)$ 
not gauged.
}

All in all, we can summarize our calculation by the simple relation\footnote{Recall the symbols ${\cal I}[{\cal T}_{IR}] \equiv {\cal I}_{ {\cal C}}$ are used interchangeably.}
\be
{\cal I}[{\cal T}_{IR}]=2\;{\cal I}_V\,{\text{Res}}_{a=t^\half}\;\frac{1}{a}\,\,\left[{\cal I}[ {\cal T}_{UV} ] \right] \, ,
\ee
which precisely confirms~\eqref{finpresc} for $r=s=0$.
Taking the residue of the UV index at the ``extra'' $U(1)_f$ flavor fugacity  $a=t^\half$ 
we have ``completely closed'' the $U(1)_f$ puncture and got back the IR index.

\

Let us now 
consider the poles with non-trivial $p$ and $q$ dependence,
\be\label{pole}
a=\pm t^\half \, p^{r/2}\, q^{s/2}\,,
\ee 
with non-negative integers $r$ and $s$. 
We need only focus  on the positive sign. We give the details of the calculation for $s >0$, $r = 0$,
as the generalization to non-zero $r$ is straightforward we will simply quote the result at the end of this subsection.
We start by considering the poles of the integrand of~\eqref{inda} at 
\be\label{twolines}
&&c=t^\half q^{m_1}\frac{1}{a\,b}\,,\qquad c^{-1}=t^\half q^{m_2}\frac{b}{a}\,,\\
&&c=t^\half q^{m_1}\frac{b}{a}\,,\qquad c^{-1}=t^\half q^{m_2}\frac{1}{a\,b}\,.\nonumber
\ee If the non-negative integers $m_i$ are chosen such that
\be
s=m_1+m_2\,,
\ee then the pairs of poles in each line of (\ref{twolines}) collide when $a=t^\half \, q^{\half s}$.
Further, since  all expressions are symmetric under $c\to c^{-1}$ the residues at the poles
corresponding to the two lines in~\eqref{twolines} are equal under exchange of $m_1$ with $m_2$.
It follows that to compute  the residue at~\eqref{pole}, we need to keep track of the
 terms in the $c$ contour integral  classified by a partition of $s$
into two non-negative parts. 
{\it E.g.} for $s=1$ we get two different choices, $m_1 =1$, $m_2=0$ and $m_1 =0$, $m_2=1$.
Let us then compute the residues. The contribution of the 
hyper is given by 
\be \label{Afactor}
&&{\cal A}_{\{m_i\}}=2   \, \half {\cal I}_{\cal C}(q^{\frac{s}{2}-m_i}b^{}_{i})\,\prod_{i,j=1}^2\prod'_{m,n\geq0}
\frac{1-p^{n+1}q^{m-m_j+1}t^{-1}b_j/ b_{i} }{1-p^{n}q^{m+m_j}t^{} b_i/ b_{j} }
\frac{1-p^{n+1}q^{m+m_j+1} b_i/ b_{j} }{1-p^{n}q^{m-m_j} b_j/ b_{i} }\, ,\nonumber\\
\ee  
where we have defined $b_1 \equiv b$, $b_2 \equiv b^{-1}$ and  the prime over the second product indicates that we are  omitting the diverging term.
The factor of $2$ in front is coming from the two lines in~\eqref{twolines}, while the factor of half arises since the pole in $a$ appears as $1/(1-t\,q^sa^{-2})$.

We have to  multiply each of these factors by the index of the vector multiplet and the Haar measure
evaluated at the pole and sum over all the partitions of $s$,  which we will denote by 
$\pi(s)$. The index of the vector multiplet together with the Haar measure evaluated at the pole is given by
\be \label{Bfactor}
{\cal B}_{\{m_i\}}=\frac{{\cal I}^{-1}_V}{2!}\prod_{i,j=1}^2
\prod_{m,n\geq0}
\frac{1-p^{n}q^{m+m_j-m_i}t^{} b_i/ b_{j} }{1-p^{n+1}q^{m+m_i-m_j+1}t^{-1}b_j/ b_{i} }
\frac{1-p^{n}q^{m+m_i-m_j} b_j/ b_i }{1-p^{n+1}q^{m+m_j-m_i+1} b_i/ b_{j} }\,.\nonumber\\
\ee
Combining the two factors (\ref{Afactor}) and (\ref{Bfactor}),
and further multiplying by 
twice the index of free vector  multiplet, we finally have
\be 
 2\;{\cal I}_V\,{\text{Res}}_{a=t^\half  q^{s/2}}\;\frac{1}{a}\,\,\left[{\cal I}  \right]    =\sum_{\{m_i\}\in\pi(r)}{\cal I}_V\,{\cal A}_{\{m_i\}}{\cal B}_{\{m_i\}}=\sum_{\{m_i\}\in\pi(r)}
f^{(r)}_{\{m_i\}}(b)\,{\cal I}_{\cal C}(q^{\half r-m_i}b^{}_{i})\,.\nonumber\\
\ee 
We see that
 the computation of the residue at the pole~\eqref{pole} amounts to applying a difference
operator ${\frak S}_{(0, s)}$ to the index,
\be
{\frak S}_{(0,s)}\,\,{\cal I}_{\cal C}\equiv  2\;{\cal I}_V\,{\text{Res}}_{a=t^\half  q^{s/2}}
\;\frac{1}{a}\,
\left[{\cal I}  \right]  \,.
\ee 

The generalizations of this calculation to non-zero $r$ in~\eqref{pole} is immediate.
The residue is given  by a sum over terms coming from partitions
of both $s$ and $r$,
\be
s=\sum_{i=1}^2 m_i,\qquad r=\sum_{i=1}^2 m'_i\,.
\ee  We will denote the operator which computes the residue at~\eqref{pole} by
${\frak S}_{(r,s)}$.

\

\noindent

 Let us now compute explicitly the basic operator ${\frak S_{(0,1)}}$.
First the  factors $f^{(s=1)}_{\{m_i\}}$ are given by
\be
f^{(s=1)}_{\{0, 1\}}&=&\prod_{m\geq 0}
\frac{(1-p^mt)(1-p^{m+1}t^{-1})}{(1-p^{m+1}q)(1-p^mq^{-1})}
\,\frac{(1-p^{m+1}q t^{-1} b_2/b_1)(1-p^mq^{-1}tb_1/b_2)}{(1-p^{m+1}b_1/b_2)(1-p^mb_2/b_1)}\nonumber\\
&=&\frac{\theta(t;p)}{\theta(q^{-1};p)}\,
\frac{\theta(\frac{t}{q}b_1/b_2;p)}{\theta(b_2/b_1;p)}\,.
\ee Thus the residue at $a=t^\half q^{\frac{1}{2}}$ is given by the action of the following operator
\be\label{opdef}
{\frak S}_{(0,1)}\, {\cal I}_{\cal C}&=&\frac{\theta(t;p)}{\theta(q^{-1};p)}\,\sum_{i=1}^2
\prod_{j\neq i}\frac{\theta(\frac{t}{q}b_i/b_j;p)}{\theta(b_j/b_i;p)}\,
{\cal I}_{\cal C}(b_i\to q^{-\frac{1}{2}}b^{}_i,\, b_{j\neq i}\to q^{\frac{1}{2}}b_j^{})\,,\\
&=&\frac{\theta(t;p)}{\theta(q^{-1};p)}\,\left[
\frac{\theta(\frac{t}{q}b^{-2};p)}{\theta(b^{2};p)}{\cal I}(b\,q^{1/2},\dots)+
\frac{\theta(\frac{t}{q}b^{2};p)}{\theta(b^{-2};p)}{\cal I}(b\,q^{-1/2},\dots)
\right]\,.\nonumber
\ee 
The operator ${\frak S}_{(0,1)}$ is, up to a conjugation by a simple function (see appendix~\ref{eigenfuncsec}),
the basic difference operator of the $A_1$ RS model!
We recognize the prefactor 
$
\frac{\theta(t;p)}{\theta(q^{-1};p)}
$
as the function $R_{0,1}$ introduced in (\ref{R0s}). This 
is  the $2d$ index of one free chiral field in two dimensions (see
appendix~\ref{imbsec}) and  we interpret it as the center of mass degree of freedom of the surface defect.
As we have discussed in section~\ref{prescsec}, it is natural to strip off this factor from the definition
of the defect. We can then define the difference operator associated to the ``bare'' defect, $\bar {\frak S}_{(0,1)} \equiv R_{0,1}^{-1}  {\frak S}_{(0,1)}$.
However,  we will mostly write equations for the  ``full'' operators ${\frak S}_{(r,s)}$, since they are somewhat easier to manipulate algebraically.

The operator corresponding to residue at $a=t^\half \, p^\half$,   ${\frak S}_{(1,0)}$, 
is simply obtained by exchanging
$q$ and $p$ in  (\ref{opdef}).
It is not difficult to write down the operators for general $s$,
\be\label{Sr0r}
&&{\frak S}_{(0,s)}\, {\cal I}_{\cal C}=\sum_{m=0}^s
\frac{\prod_{n=0}^{m-1}\theta(tq^{n};p)\prod_{n=0}^{s-m-1}\theta(tq^{n};p)}
{\prod_{n=1}^{m}\theta(q^{-n};p)\prod_{n=1}^{s-m}\theta(q^{-n};p)}\times\\
&&\qquad
\left[\prod_{n=0}^{m-1}\frac{\theta(t\,q^{s-2m+n}b^{2};p)}{\theta(q^{-s+m+n}b^{-2};p)}\right]\,
\left[\prod_{n=0}^{r-m-1}\frac{\theta(t\,q^{2m-s+n}b^{-2};p)}{\theta(q^{-m+n}b^{2};p)}\right]\,
{\cal I}_{\cal C}(b\to q^{\half s-m}b^{})\,.\nonumber
\ee
Finally we quote the  general  result for all $r$ and $s$,
\begin{equation}\label{Sr0}
\boxed{
\begin{array}{l}
{\cal I}[{\cal T}_{IR},{\frak S}_{(r,s)}]=2\;{\cal I}_V\,{\text{Res}}_{a=t^\half\,  p^{r/2}\,q^{s/2}  }
\;\frac{1}{a}\,
\left[{\cal I} [ {\cal T}_{UV}]  \right]  =  {\frak S}_{(r,s)}\;{\cal I}[{\cal T}_{IR}]=\\
\\
\qquad
\displaystyle\sum_{\sum_{i=1}^2 n_i=r}
\displaystyle\sum_{\sum_{i=1}^2 m_i=s} {\cal I}_{\cal C}(p^{\frac{r}{2}-n_i}\,q^{\frac{s}{2}-m_i}
\,b^{}_{i})\,\times\\
\\
\qquad\qquad\displaystyle\prod_{i,j=1}^2
\left[{{\displaystyle\prod_{m=0}^{m_i-1}}}'\frac{\theta(p^{n_j}q^{m+m_j-m_i}tb_i/b_j;p)}
{\theta(p^{-n_j}q^{m-m_j}b_j/b_i;p)}\right]
\left[{\displaystyle\prod_{n=0}^{n_i-1}}'\frac{\theta(q^{m_j-m_i}p^{n+n_j-n_i}tb_i/b_j;q)}
{\theta(q^{m_i-m_j}p^{n-n_j}b_j/b_i;q)}\right]\,,
\end{array}
}
\end{equation}
where the prime on the product means as always that we omit the divergent term.
Let us stress here that although in the analysis of this section we have used the jargon of theories of class ${\cal S}$ the final result~\eqref{Sr0}
applies to {\it any} theory of the setup of figure~2.

\

\

\section{Properties of the operators ${\frak S}_{(r,s)}$}\label{propsec}

We now proceed to investigate the algebraic properties of the difference operators  ${\frak S}_{(r,s)}$
obtained in the previous section. These operators have many beautiful mathematical properties. Some of
these properties are to be anticipated on purely physical grounds following the fact that ${\frak S}_{(r,s)}$ 
compute resiudes for theories enjoying generalized S-duality.

We will find further evidence for our main claim, that
${\frak S}_{(r,s)}\,{\cal I}[{\cal T}_{IR}]$ is the index of a ${\cal T}_{IR}$ in the presence 
of two surface defects along the two orthogonal circles of $S^3$ fixed by $j_{12}$ and $j_{34}$.
We will see that the two defects are naturally associated to the spin $r/2$ and spin $s/2$  representations of $SU(2)$. 

\

\subsubsection*{Factorization}

We claim
\be \label{factorizationclaim}
{\frak S}_{(r,s)}=\left(\frac{p\,q}{t}\right)^{rs}{\frak S}_{(r,0)}{\frak S}_{(0,s)}=\left(\frac{p\,q}{t}\right)^{rs}{\frak S}_{(0,s)}{\frak S}_{(r,0)}\,.
\ee
To show this we write
\be
&&{\frak S}_{(0,s)}{\frak S}_{(r,0)}{\cal I}=\sum_{\sum_{i=1}^2 n_i=r}
\sum_{\sum_{i=1}^2 m_i=s} {\cal I}(q^{\frac{s}{2}-m_i}
p^{\frac{r}{2}-n_i}b^{}_{i})\,\prod_{i,j=1}^2\\
&&\;\left[\prod_{n=0}^{n_i-1}\frac{\theta(p^{n+n_j-n_i}tb_i/b_j;q)}
{\theta(p^{n-n_j}b_j/b_i;q)}\right]
\left[\prod_{m=0}^{m_i-1}\frac{\theta(p^{n_j-n_i}q^{m+m_j-m_i}tb_i/b_j;p)}
{\theta(p^{n_i-n_j}q^{m-m_j}b_j/b_i;p)}\right]
\,,\nonumber
\ee 
and then using 
\be
\theta(x;q)=(-1)^nx^nq^{\half n (n-1)}\theta(q^nx;q),
\ee we establish (\ref{factorizationclaim}). This factorization property has a natural physical interpretation. We have claimed that the operators ${\frak S}_{(r,s)}$
introduce two surface defects, labeled by integers $r$ and $s$, along the two orthogonal planes,
and thus can be introduced (almost) without interfering with each other. The two defects feel each other's presence only through
the proportionality factor of $\left(\frac{p\,q}{t}\right)^{rs}$,   which as explained in section~\ref{u1gauging}
amounts to a redefinition of the IR r-charge.

\

\subsubsection*{Commutativity of the operators}

\begin{figure}
\begin{center}
\includegraphics[scale=0.60]{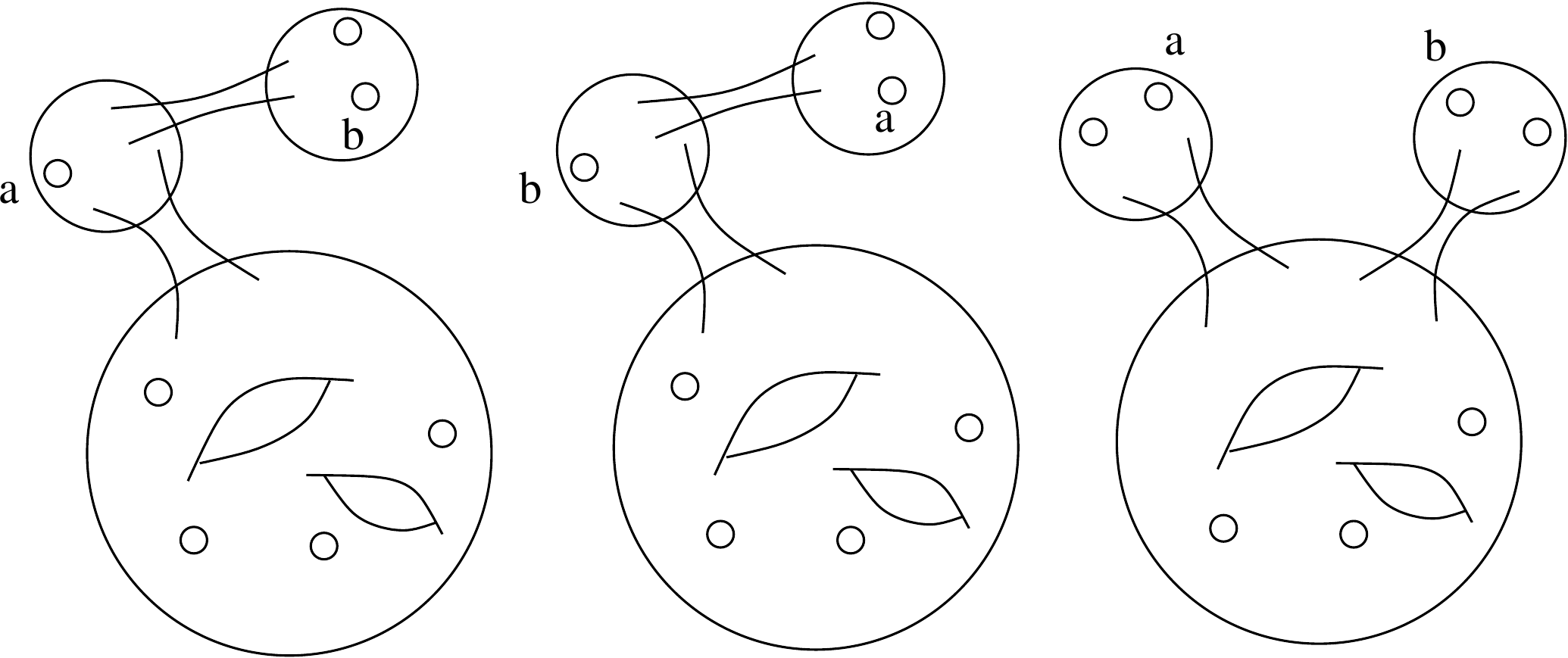}
\end{center}
\caption{By generalized S-duality, the order in which one extracts the residues in fugacities $a$ and $b$ is expected
to be immaterial, indeed the three different decompositions of the surface shown in the figure are topologically equivalent.
This implies  $[{\frak S}_{(r,s)} \, ,{\frak S}_{(r',s')}] = 0$.
\label{comm}}
\end{figure}

The two operators ${\frak S}_{(1,0)}$ and  ${\frak S}_{(0,1)}$ commute,
\be
\left[{\frak S}_{(1,0)},\,{\frak S}_{(0,1)}\right]=0\, .
\ee
Mathematically this follows from simple theta-function identities. For example considering the 
term proportopnal to ${\cal I}(q^\half p^\half\,a,\dots)$ in the commutator we get
\be
\frac{\theta(\frac{t}{p}a^{-2};q)}{\theta(a^{2};q)}
\frac{\theta(\frac{t}{qp}a^{-2};p)}{\theta(pa^{2};p)}-
\frac{\theta(\frac{t}{q}a^{-2};p)}{\theta(a^{2};p)}
\frac{\theta(\frac{t}{qp}a^{-2};q)}{\theta(qa^{2};q)}=0\,,
\ee etc. In more generality one can show that all the operators  ${\frak S}_{(r,s)}$ commute with each other.
Physically this result is expected since these operators compute residues in theories which enjoy S-duality and
we can employ this duality to change the order in which the residues are taken, see figure~\ref{comm}. 

\

\subsubsection*{Self-adjointness}

\begin{figure}
\begin{center}
\includegraphics[scale=0.4]{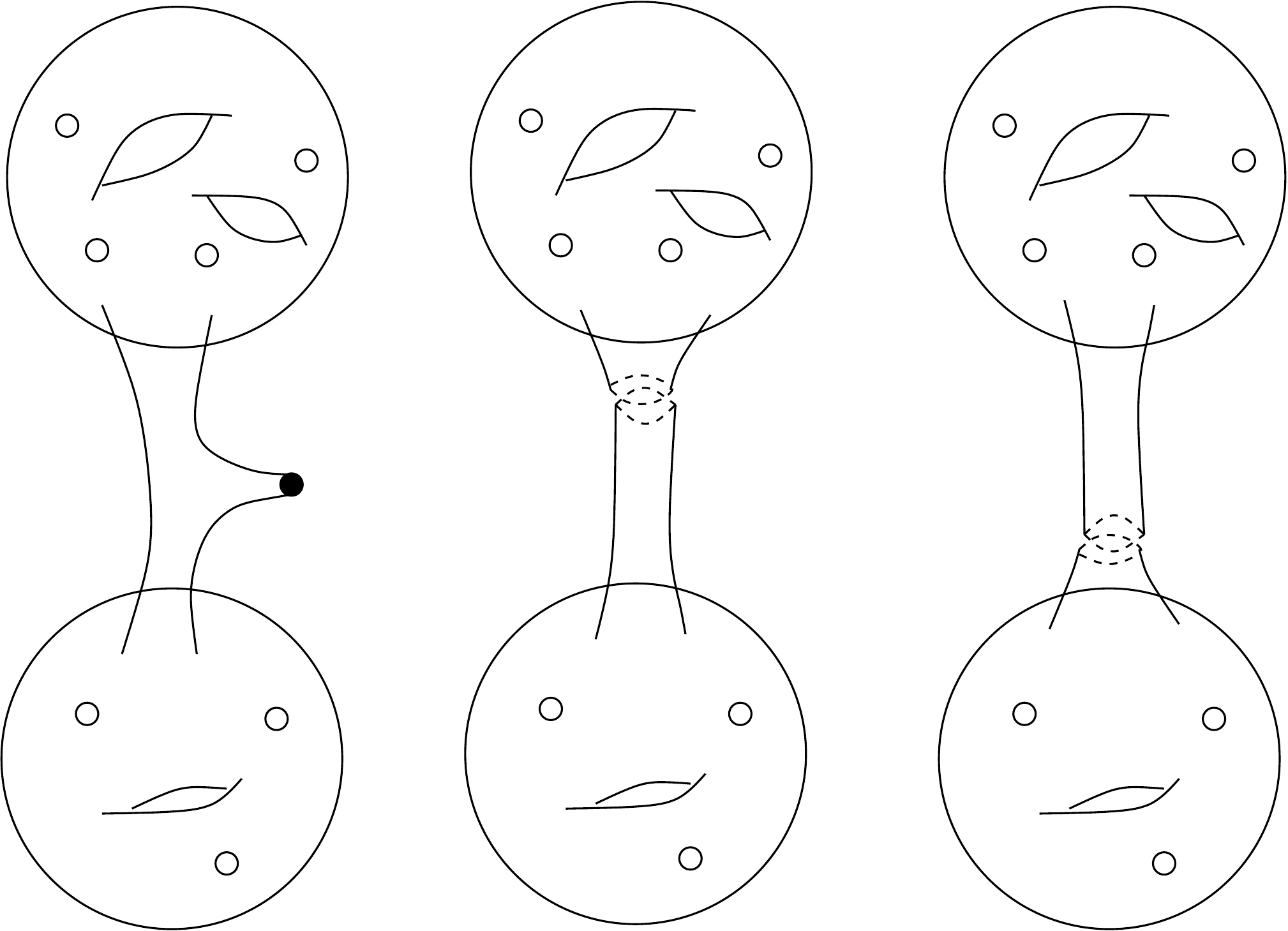}
\end{center}
\caption{Three different but equivalent ways to introduce a surface defect in a tube. First
by coupling the two edges of the tube to a free hypermultiplet and computing a $U(1)_f$ residue of the index.
Then by acting with the relevant difference operator on either of the edges and then gauging.
The equivalence of these different procedures implies that the index is self-adjoint under the natural measure.\label{figadj}}
\end{figure}

The operator ${\frak S}_{(1,0)}$ is self-adjoint under the natural propagator measure $\Delta\cdot {\cal I}_V$,
\be
\oint \frac{da}{a}\,\Delta(a)\,{\cal I}_V(a)\,
f(a)\,\left[{\frak S}_{(0,1)}\,\cdot\, g(a) \right] =
\oint \frac{da}{a}\,\Delta(a)\,{\cal I}_V(a)\,
\left[{\frak S}_{(0,1)}\,\cdot\, f(a) \right]\,g(a)\,.
\ee This can be shown simply by a change of the integration variable $a\to a\,q^{\pm\half}$,
if one assumes that the test functions $f(a)$ and $g(a)$ do not have poles in the strip $|q|^{\half}
\leq |a|\leq |q^{-\half}|$. Under this assumption we write
\be
&&\oint\frac{da}{a}{\cal I}_V(a)\Delta(a)\,f(a)\left[
\frac{\theta(\frac{t}{q}a^{-2};p)}{\theta(a^{2};p)}g(a\,q^{1/2})+
\frac{\theta(\frac{t}{q}a^{2};p)}{\theta(a^{-2};p)}g(a\,q^{-1/2})
\right]=\\
&&\oint\frac{db}{b}{\cal I}_V(bq^{-\half})\Delta(bq^{-\half})\,f(bq^{-\half})
\frac{\theta(tb^{-2};p)}{\theta(b^{2}q^{-1};p)}g(b)+
\oint\frac{db}{b}{\cal I}_V(bq^{\half})\Delta(bq^{\half})\,f(bq^{\half})
\frac{\theta(tb^{2};p)}{\theta(b^{-2}q^{-1};p)}g(b)
\,.\nonumber
\ee The integrand of the first integral satisfies
\be
&&{\cal I}_V(bq^{-\half})\Delta(bq^{-\half})\,
\frac{\theta(tb^{-2};p)}{\theta(b^{2}q^{-1};p)}=
\half(1-q b^2)(1-b^2q^{-1})PE\left[\left(\frac{q p-t}{1-p}\right)b^{-2}\right]\times\nonumber\\
&&\;PE\left[\left(\frac{q^{-1}-pt^{-1}}{1-p}\right)b^{2}\right]\,PE\left[
(-\frac{p}{1-p}-\frac{q}{1-q}+\frac{\frac{pq}{t}-t}{(1-p)(1-q)})(q^{-1}b^2+qb^{-2}-1)\right]=
\nonumber\\
&& {\cal I}_V(b)\Delta(b)\,
\frac{\theta(\frac{t}{q}b^{2};p)}{\theta(b^{-2};p)}\,,
\ee and analogously for the second term thus establishing the self-adjointness of ${\frak S}_{(0,1)}$.
This property is to be expected from how we introduced the difference operator, namely as a way
to evaluate residues in the $SU(2)_f$ flavor fugacity of ${\cal T}_{UV}$. Consider the setup
of the first row in   figure~\ref{figadj}, where two theories ${\cal T}$ and ${\cal T}'$ are connected to  a bifundamental hypermultiplet by gauging two $SU(2)$ groups.
  The index can be written as double-contour integral
for the two $SU(2)$ gauge fugacities. To extract the residue in the $SU(2)_f$ fugacity, we can follow the procedure of the previous subsection
and pinch one of the two contours for fixed values of the second integration variable. The result is 
 the second or third column in  figure~\ref{figadj}, according to the order in which we perform  the countour integrals.
Since the setup is symmetric under exchange of ${\cal T}$ and ${\cal T}'$, it follows that 
 ${\frak S}_{(r,s)}$ must be self-adjont under the propagator measure.

\

\subsubsection*{Acting with difference operators on the trinion}

In the $2d$ TQFT interpretation of the index of theories of class ${\cal S}$, the action of the
difference operator ${\frak S}_{(r,s)}$
 on a flavor fugacity corresponds to the fusion of the ``degenerate puncture'' associated to  $(r,s)$ 
  surface defect with a flavor puncture. The choice of flavor puncture on which the operator
  acts is then expected to be immaterial.
This follows directly from our construction of the operators and from the generalized S-duality enjoyed by theories of class ${\cal S}$. The difference operators compute a residue of the index at a pole 
of one of the flavor fugacities. To obtain the operators we have singled out an additional flavor fugacity by decoupling a free hypermultiplet as in figure~\ref{figdeg}.
Invariance of the index under S-duality  means that our choice of the additional flavor fugacity must be immaterial.
The crucial claim is that the difference operators must give the same result when acting on any of the three legs of the
  elementary ``trinions''. By decomposing the Riemann surface into pairs of pants, it follows  from this claim that the difference operators can be freely
  moved around the surface, indeed we 
    have already shown that they are are self-adjoint with respect to the propagator measure.

In the $A_1$ case the index of the basic building block, the index of the  trinion, is explicitly known,
so we can establish this claim by direct computation.
The $A_1$ trinion index is the index of a free tri-fundamental hypermultiplet,
\be
{\cal I}(a,b,c)=\Gamma\left(t^\half a^{\pm1}b^{\pm1}c^{\pm1};p,q\right)\,.
\ee 
 Dropping overall 
flavor-independent factors we obtain
\be\label{trinsym}
&&{\frak S}_{(0,1)}(a){\cal I}(a,b,c)\sim
\frac{\theta(\frac{t}{q}a^{-2};p)}{\theta(a^{2};p)}\Gamma(t^\half b^{\pm1}c^{\pm1}(a q^\half)^{\pm1};p,q)+
\frac{\theta(\frac{t}{q}a^{2};p)}{\theta(a^{-2};p)}\Gamma(t^\half b^{\pm1}c^{\pm1}(a q^{-\half})^{\pm1};p,q)\nonumber\\
&&\;=\Gamma\left(\sqrt{\frac{t}{q}} a^{\pm1}b^{\pm1}c^{\pm1};p,q\right)\,
\left[
\frac{\theta(\frac{t}{q}a^{-2};p)\theta(\sqrt{\frac{t}{q}}a^{}b^{\pm1}c^{\pm1};p)}{\theta(a^{2};p)}+
\frac{\theta(\frac{t}{q}a^{2};p)\theta(\sqrt{\frac{t}{q}}a^{-1}b^{\pm1}c^{\pm1};p)}{\theta(a^{-2};p)}
\right]\,.\nonumber\\
\ee In the Macdonald limit $p=0$~\cite{Gadde:2011uv} it is easy to show that the above expression is symmetric under permutations of
flavor fugacities,
\be\label{thetaind}
&&{\frak S}_{(0,1)}(a){\cal I}(a,b,c)\sim
\Gamma\left(\sqrt{\frac{t}{q}} a^{\pm1}b^{\pm1}c^{\pm1};0,q\right)\,\times\\
&&\quad\left[1+\left(\frac{t}{q}\right)^3-(\chi_{adj}(a)+\chi_{adj}(b)+\chi_{adj}(c))\left(\frac{t}{q}
+\left(\frac{t}{q}\right)^2\right)
+2\chi_f(a)\chi_f(b)\chi_f(c)\left(\frac{t}{q}\right)^{3/2}
\right]\,.\nonumber
\ee 
We can also easily check the $a \leftrightarrow b \leftrightarrow c$ symmetry  Schur limit  $q=t$~\cite{Gadde:2011uv,Gadde:2011ik},
\be
&&\frac{\theta(a^{-2};p)\theta(a^{}b^{\pm1}c^{\pm1};p)}{\theta(a^{2};p)}=
\frac{\theta(a^{2};p)\theta(a^{-1}b^{\pm1}c^{\pm1};p)}{\theta(a^{-2};p)}=
\sqrt{\theta(a^{\pm1}b^{\pm1}c^{\pm1};p)},\\
&&\Gamma\left( a^{\pm1}b^{\pm1}c^{\pm1};p,q\right)=
\frac{1}{\theta(ab^{\pm1}c^{\pm1};p)\theta(a^{-1}b^{\pm1}c^{\pm1};q)}
=\frac{1}{\sqrt{\theta(a^{\pm1}b^{\pm1}c^{\pm1};p)\theta(a^{\pm1}b^{\pm1}c^{\pm1};q)}},\nonumber
\ee we obtain
\be\label{schur0}
{\frak S}_{(0,1)}(a){\cal I}(a,b,c)\sim\frac{2}{\sqrt{\theta(a^{\pm1}b^{\pm1}c^{\pm1};q)}}\,.
\ee This expression is manifestly symmetric and also independent of $p$ as expected~\cite{Gadde:2011uv}.\footnote{The same expression can be also obtained thus from~\eqref{thetaind}
by setting $t=q$.}
For arbitrary $(p,q,t)$
 one has to verify that the combination of theta functions appearing in~\eqref{thetaind}
is symmetric under permutations of the three fugacities, an exercise we leave to our enthusiastic readers.
We have checked this claim perturbatively in $p$.

Even if the theory has no flavor punctures, we can still define an operation
that modifies its index to include surface defects. We consider any pair-of-pants decomposition of the Riemann surface, act with the difference operator
on any of the open punctures and glue the pieces back together. The result is well-defined, since the properties
just discussed guarantee that it does not depend on the specific decomposition or choice of open puncture.

\

In a $4d$ theory with surface defects, it is generally difficult to separate the $4d$ and $2d$ degrees of freedom counted
by the index. In the simple case of the theory of a free hypermultiplet  endowed with surface defect, 
we can however identify an interesting operator intrinsic to the defect.
Let us consider the ``Coulomb limit''~\cite{Gadde:2011uv}\footnote{
A related limit for ${\cal N}=4$ SYM was discussed in~\cite{Spiridonov:2010qv}.}  (see appendix~\ref{imbsec}): $t,p,q\to 0$ with $p^2/t$ and $q^2/t$ fixed. 
The index
of the trinion acted upon by ${\frak S}_{(1,0)}$~\eqref{trinsym} becomes
\be\label{coulind}
&&{\frak S}_{(0,1)}(a)\;{\cal I}(a,b,c)\sim
1+\frac{pq}{t}\,.
\ee
The non-trivial term comes entirely from the difference operator itself, and we identify it with 
a Coulomb branch operator $X$ living on the defect, obeying the constraint $X^2 = 0$ (as natural for $SU(2)$ 
gauge group).  Its $2d$ charges (see appendix~\ref{imbsec}) are
$L_0=\half$, $J_0=-1$, and $f=0$.  It is a naturally
interpreted 
as the exactly marginal twisted chiral field that couples to the relative gauge coupling, which we
mentioned at the end of section~\ref{varhiggssec}.\footnote{
For higher rank theories there is an analogous story. For instance the Coulomb index of a bi-fundamental 
free hypermultiplet with a basic surface defect is $\sum_{\ell=0}^{N-1}\left(\frac{pq}{t}\right)^\ell$.
Assuming that the Coulomb limit commutes with the residue computation this will be true for all indices.}

\

\subsubsection*{Recursion relations}

Finally, the difference operators satisfy interesting recursion relations.
Consider the computation of ${\frak S}_{(0,s)}{\frak S}_{(0,1)} {\cal I}$. 
Let ${\cal I}[{\cal T}_{UV}]$ be the index of the theory with one extra  $U(1)_f$ puncture.
To  compute ${\frak S}_{(0,s)}{\frak S}_{(0,1)} {\cal I}$ we extract 
the residue of ${\frak S}_{(0,1)} {\cal I}[{\cal T}_{UV}]$ at $a = t^{1/2} q^{s/2}$.
If we substitute the explicit expression for ${\frak S}_{(0,1)}$ we get the neat relation,
\be\label{recur}
\frac{\theta(q^{-1};p)}{\theta(t;p)}\,{\frak S}_{(0,s)}{\frak S}_{(0,1)}  = \frac{\theta(q^{-s-1};p)}{\theta(t q^s;p)}{\frak S}_{(0,s+1)}+
\frac{\theta(t^2 q^{s-1};p)}{\theta(\frac{1}{t q^{s}};p)}{\frak S}_{(0,s-1)}\,.
\ee  
This recursion relation can be formally recast as a fusion product between operators corresponding to representation of $SU(2)$. 
We can write~\eqref{recur} as
\be
{\bf{[s]}}\times {\bf{[1]}}= {\cal A}_{(s,1|s+1)}\,{\bf{[s+1]}} + {\cal A}_{(s,1|s-1)}\,{\bf{[s-1]}}\,,
\ee where ${\bf{[s]}}$ is an operator corersponding to an $s+1$-dimensional irrep of $SU(2)$. This 
is reminiscent of  the fusion relation between degenerate operator in Liouville theory,
which in the AGT correspondence are associated to surface defects~\cite{Alday:2009fs}. 

One can check  (\ref{recur}) by using the explicit expression for
${\frak S}_{(0,s)}$~\eqref{Sr0}.  One then finds the same 
theta-function identity that we encountered in verifying the symmetry under permutation 
of flavor fugacities of the trinion when acted upon with a difference operator~\eqref{trinsym}. This identity takes the 
following general form
\be
&&\theta(x b^{-1} a^{\pm 1} c^{\pm 1};p) \frac{\theta(x^2 b^2;p)}{\theta(b^{-2};p)} + \theta(x b a^{\pm 1} c^{\pm 1};p) \frac{\theta(x^2 b^{-2};p)}{\theta(b^2;p)} = \nonumber\\
&&\qquad\qquad \theta(x a^{-1} b^{\pm 1} c^{\pm 1};p)\frac{\theta(x^2 a^2;p)}{ \theta(a^{-2};p)} +  \theta(x a b^{\pm 1} c^{\pm 1};p) \frac{\theta(x^2 a^{-2};p)}{\theta(a^2;p)}\,.
\ee
As suggested in section~\ref{prescsec} it is natural to  strip off a factor 
\begin{equation}\label{stripoff}
\frac{\theta(t;p)}{\theta(q^{-1};p)}\frac{\theta(t q;p)}{\theta(q^{-2};p)} \cdots \frac{\theta(t q^{s-1};p)}{\theta(q^{-s};p)}
\end{equation}
to define an operator $\bar {\frak S}_{(0,s)}$.
Then the recursion relation takes a slightly simpler form,
\be\label{recbar}
\bar {\frak S}_{(0,s)} \bar {\frak S}_{(0,1)}  = \bar {\frak S}_{(0,s+1)}+
\frac{\theta(t^2 q^{s-1};p)}{\theta(\frac{1}{t q^{s}};p)}\frac{\theta(q^{-s};p)}{\theta(t q^{s-1};p)} \bar {\frak S}_{(0,s-1)}\,.
\ee 
Remarkably, the ratios of $\theta$-functions 
can be interpreted as  $2d$ indices $\chi_{2d}$ of free chiral fields with a specific R-charge assignment (see appendix~\ref{imbsec})
\be
\frac{\theta(\frac{pq}{t}\,x;p)}{\theta(x;p)}\equiv \chi_{2d}(x)\,.
\ee
 In terms of these $2d$ indices we can write a final form of the recursion relation,
\be
\chi_{2d}(t)\,{\frak S}_{(0,s)}{\frak S}_{(0,1)}  = \chi_{2d}(t\,q^s){\frak S}_{(0,s+1)}+
\chi_{2d}(t^{-1}q^{-s}){\frak S}_{(0,s-1)}\,.
\ee

\

\

\section{Bootstrapping the index for $A_1$ theories}\label{bootsec}

We are now ready to implement the ``bootstrap''
of the $2d$ TQFT that computes the index of class ${\cal S}$ theories of type $A_1$. In other terms, we will determine the index from consistency  conditions alone.
Our strategy is a simplified (topological) version of the ``Teschner trick'' \cite{Teschner:1995yf}. A degenerate puncture associated to a surface defect
can be fused with any of the flavor punctures, and assuming $S$-duality the result must be independent of the chosen flavor puncture.\footnote{Of course,
for the general $A_1$ theory, which has a Lagrangian description, the index is already explicitly known in terms of a matrix integral.  Since everything is explicit,
S-duality of the index for the $A_1$ quivers can be rigorously established~\cite{Gadde:2009kb} using  an identity for a certain integral
of elliptic Gamma functions~\cite{debult}. However,
the bootstrap approach   generalizes easily to higher-rank theories, whose index is a priori unknown, and even for the $A_1$ theories
it  has the virtues of directly giving the index in  the very useful ``diagonal''  form. 
}

To proceed, we make a plausible technical assumption: the difference operators  ${\frak S}_{(r,s)}$  admit
a complete set of eigenfunctions $\{ \psi_\lambda (a) \}$, 
normalizable under the propagator measure, with non-degenerate eigenvalues $E_{(r,s)}^\lambda$.
The label $\lambda$ runs over the irreducible $SU(2)$ representations.
Since the difference operators are self-adjoint in the propagator measure, an implication of this assumption is
that any two different eigenfunctions are orthogonal in the propagator measure, and can be normalized to be orthonormal.
This assumption is certainly true in the degenerations 
 limits $p=0$ or $q=0$, where the difference operators become the familiar Macdonald operators\footnote{More precisely, they are related
 to the Macdonald operators by a similarity transformation.} and the eigenfunctions are proportional to
 Macdonald polynomials. For arbitrary $(p,q,t)$, to the best of our knowledge
  the eigenfunctions  are not known in closed form: they are expected to be an ``elliptic'' deformation of the Macdonald functions.
  In appendix~\ref{eigenfuncsec} we describe an approximation scheme to determine the eigenfunctions in an expansion for small $q$ and $p$.

Expanding the index of the trinion,
\be\label{repexp}
{\cal I}_{0,3}=\sum_{\alpha,\beta,\gamma} C_{\alpha\beta\gamma}\;
\psi^\alpha(a)\psi^\beta(b)\psi^\gamma(c)\, ,
\ee
 and acting with the difference operator one any of the flavor punctures, 
we have
\be
{\frak S}_{(r,s)}\;{\cal I}_{0,3}&=&\sum_{\alpha,\beta,\gamma} C_{\alpha\beta\gamma}\,E^\alpha_{(r,s)}\,
\psi^\alpha(a)\psi^\beta(b)\psi^\gamma(c)=
\sum_{\alpha,\beta,\gamma} C_{\alpha\beta\gamma}\,E^\beta_{(r,s)}\,
\psi^\alpha(a)\psi^\beta(b)\psi^\gamma(c)\\
&=&\sum_{\alpha,\beta,\gamma} C_{\alpha\beta\gamma}\,E^\gamma_{(r,s)}\,
\psi^\alpha(a)\psi^\beta(b)\psi^\gamma(c)\,.\nonumber
\ee 
As by assumption the eigenvalues are non-degenerate,
this implies that the index is diagonal in the basis  $\{ \psi^\alpha \}$,
\be
{\cal I}_{0,3}=\sum_{\alpha} C_{\alpha}\;
\psi^\alpha(a)\psi^\alpha(b)\psi^\alpha(c)\,.
\ee 
It remains to fix
the structure constants $C_\alpha$. Using orthonormality of the eigenfuctions under the propagator measure,  
 the index of the theory associated to the four-punctured sphere is immediately given by
\be
{\cal I}_{0,4}=\sum_{\alpha} C_{\alpha}^2\;
\psi^\alpha(a)\psi^\alpha(b)\psi^\alpha(c)\psi^\alpha(d)\,.
\ee 
According to our prescription, extracting the residue at $a=t^\half$ closes a puncture,  so
\be
{\cal I}_{0,3}=2\;{\cal I}_V\;\mathrm{Res}_{a = t^{1/2}}\;\frac{1}{a}\,{\cal I}_{0,4}\,.
\ee We conclude that
\be
C_\alpha=\frac{1}{2\,{\cal I}_V\,\left(  \mathrm{Res}_{a = t^{1/2}} \;\frac{1}{a}\,\psi^\alpha(a)\right)}\,.
\ee 
 The index of general $A_1$ theory
associated to a surface with $k$ flavor punctures
is given by
\be
{\cal I}_{ \frak g, k} (a_i) =\sum_\a \left(C_\a\right)^{2\frak g-2+k} \prod_{i=1}^k\psi^\a(a_i)\,.
\ee
To introduce an $(r,s)$ surface defect, we are instructed to simply multiply by $E_{(r,s)}^\alpha$ inside the summation sign.

We can also give expressions for the eigenvalues using  the residue interpretation of the 
difference operators. For example
\be
{E_{(0,s)} }^\a=\frac{ \mathrm{Res}_{a = t^{\half}\,q^{\half s}} \;\frac{1}{a}\,\psi^\alpha(a)}{ \mathrm{Res}_{a = t^{\half}} \;\frac{1}{a}\,\psi^\alpha(a)}\,,
\ee   In particular in the Macdonald limit, $p=0$,  we have 
\be
\psi^\a(a) = \frac{1}{(t a^{\pm2};q)}\,P^\a(a;q,t)\,,\qquad
{E_{(0,s)}}^\a=\left[\prod_{i=0}^{s-1}\frac{1-t^2q^i}{1-q^{-i-1}}\right]\,
\frac{P^\a(t^\half q^{\half\,s};q,t)}{P_\a(t^\half;q,t)}\,.
\ee Here $P^\a(z;\,q,\,t)$ are Macdonald polynomials~\eqref{macpol} and we have defined as always
\be
(z;\,q)\equiv \prod_{\ell=0}^\infty (1-z\,q^\ell)\,.
\ee
It is easy to verify that then the energies satisfy the recursion relation
\be
\frac{1-q^{-1}}{1-t}\,{E_{(0,s)}}^\a{E_{(0,1)}}^\a=
\frac{1-q^{-s-1}}{1-q^s t}{E_{(0,s+1)}}^{\a}+\frac{1-t^2q^{s-1}}{1-q^{-s}t^{-1}}{E_{(0,s-1)}}^{\a}\,,
\ee which is a direct consequence of recursion relation satisfied by the operators~\eqref{recur}.

In the Macdonald limit the energies ${E_{(r,0)}}^\a$ can be 
related to modular S-matrix of the {\it refined} Chern-Simons theory~\cite{Aganagic:2011sg}.
Up to $\alpha$ independent normalization factor these energies are given by
\be
\frac{P^\a(t^\half q^{\half\,s};q,t)}{P^\a(t^\half;q,t)}=
\frac{{S_\a}^s\,{S_0}^0}{{S_0}^s\,{S_\a}^0}\,,
\ee where ${S_\a}^s$ is the modular S-matrix of the \textit{refined} Chern-Simons
theory as defined in~\cite{Aganagic:2011sg}.
In the Schur limit ($q=t$) the {\it refined} Chern-Simons theory further
reduces to the  $2d$ q-deformed Yang-Mills (in the zero-area limit), as defined in~\cite{Aganagic:2004js}. The relation of $2d$ qYM to the 
index was discussed in~\cite{Gadde:2011ik,Gadde:2011uv}.
In the $2d$ YM language the flavor punctures correspond to fixing the holonomies of the gauge fields around the punctures.
Thus, at least in the Macdonald limit, introducing a surface defect
corresponds to adding a puncture  with fixed holonomy for
 \textit{dual} variables of the $2d$ theory, {\it i.e.} the canonical momenta dual to the gauge fields~\cite{Witten:1991we,Witten:1992xu, Aganagic:2004js}. 

\

\

\section{$3d$ reduction}\label{$3d$sec}
Starting from the  $4d$ index, which is the supersymmetric partition function
on $S^3\times S^1$, we can dimensionally reduce on  the $S^1$ to obtain
the partition function of a $3d$ theory on a squashed $S^3$.
The reduction of the $4d$ index to $3d$ partition function is achieved by the following scaling of the 
fugacities,
\be\label{$3d$lim}
q=e^{i\beta \omega_1},\quad
p=e^{i\beta \omega_2},\quad
t=e^{i\beta },\quad 
a_\ell=e^{i\beta m_\ell},
\ee and taking $\beta\to0$~\cite{Dolan:2011rp,Gadde:2011ia,Imamura:2011uw}, where $a_\ell$ are $4d$ flavor fugacities and $m_\ell$ $3d$ real masses.
The $4d$ index goes to the ``ellipsoid partition function'', with squashing parameter 
\be
b=\frac{\omega_1}{\omega_2}\,.
\ee 
The $3d$ theory which arise in the infrared from the circle reduction a $4d$ theory of class ${\cal S}$ 
has a nice mirror description~\cite{Benini:2010uu} in terms takes of a star-shaped quiver: 
a central node connected to a set of linear quivers, where each linear quiver is associated to a puncture of the Riemann surface.
The wavefunction $\psi^\alpha (a)$ that we associate to a puncture in the TQFT description of $4d$ index  reduces in the $3d$ limit
 to the $S^3$ partition function of the corresponding
 linear quiver tail \cite{Nishioka:2011dq}\footnote{See also~\cite{Benvenuti:2011ga} for related work.}.

The surface operators of the $4d$ theory are expected to become line defects of the $3d$ theory.
Following the chain of dualities, one finds that
the canonical surface defects for a $4d$ theory of class ${\cal S}$, which arise from codimension four defects of the $6d$ (2,0) theory,
go to Wilson loops for the central node of the quiver in the mirror $3d$ description. We would  like to verify this fact.

For  $A_1$ theories 
the mirror description involves a $3d$ gauge theory with a $U(1)$ gauge group for each puncture, and 
a single $SU(2)$ gauge group. The matter content includes a doublet of hypermultiplets for each puncture, with charge one
under the corresponding $U(1)$ gauge group, and $\frak g$ adjoint hypermultiplets for the $SU(2)$ gauge group, where $\frak g$ is the genus of the 
Riemann surface. Each ``quiver tail'' consists of a $U(1)$ gauge theory coupled to a doublet of hypermultiplets. 
This is the so-called 
$T[SU(2)]$ theory, which is intimately related to the S-duality operation in four-dimensional ${\cal N}=4$ SYM \cite{Gaiotto:2008ak}. 
In particular, its $S^3$ partition function is related by a generalization of AGT~\cite{Hosomichi:2010vh}
to the integral kernel which implements the $S$ operation on conformal blocks on a one-punctured torus. 
This fact has a simple consequence: the  $S^3$ partition function of $T[SU(2)]$ 
satisfies a functional equation, and intertwines the action of two operators, 
which are related to the action of  't Hooft and Wilson loops of ${\cal N}=4$ SYM.

The Wilson loop operator multiplies the $T[SU(2)]$  partition function by a character of the $SU(2)$
flavor symmetry which rotates the doublet of hypermultiplets. 
In particular, the $S^3$ partition function of the full star-shaped quiver in the presence of a 
Wilson loop operator for the central $SU(2)$ node can be  obtained by acting with the 
Wilson loop operator on any of the $T[SU(2)]$ linear quiver tails partition functions. 
This can be then rewritten as the action of a 't Hooft line operator on the partition function of that quiver tail. 
We will now show that $3d$ limit of the operator which inserts a surface defect in the $4d$ theory 
exactly reproduces the 't Hooft line operator. This will verify that the insertion of the surface defect goes to the insertion of a 
Wilson line defect for the middle node of the mirror star-shaper quiver. 

Let us consider the $3d$ limit~\eqref{$3d$lim} for the operator $\bar {\frak S}_{(0,1)}$.
The prefactors
\be
\frac{\theta(\frac{t}{q}a^{-2};p)}{\theta(a^2;p)}&\to&
\frac{\sin \pi\frac{1-\omega_1-2m}{\omega_2}}{\sin 2\pi\frac{m}{\omega_2}}
\,,
\ee
while  shift operators reduce to
\be
 f(q^\half a)\quad\to\quad  f(m+\half\omega_1)\,.
\ee All in all,
\be\label{loop}
\bar {\frak S}_{(0,1)}\quad\to\quad 
\frac{\sin \pi\frac{1-\omega_1-2m}{\omega_2}}{\sin 2\pi\frac{m}{\omega_2}}
\,\Delta_{m\to m+\half \omega_1}+
\frac{\sin \pi\frac{1-\omega_1+2m}{\omega_2}}{\sin 2\pi\frac{-m}{\omega_2}}
\,\Delta_{m\to m-\half \omega_1}\,.
\ee 
(In complete analogy, the limit of the operator $ \bar {\frak S}_{(1,0)}$ is obtained by switching $\omega_1$ with $\omega_2$.)
Under an appropriate identification of the parameters this coincides with the 't Hooft line defect, as desired.
Line operators in the framework of AGT were studied in~\cite{Drukker:2009id}
for the $A_1$ case and in~\cite{Gomis:2010kv} for higher rank theories.
One can compare the operator~\eqref{loop} to equation (5.25) in~\cite{Drukker:2009id}, which displays
operator inserting a 't Hooft operator in the ${\cal N}=2^*$ theory.
The ratio of the two periods is the $b$ parameter of the underlying Liouville CFT,
and the modular properties we will discuss below are related to the change $b\to b^{-1}$.

Finally let us briefly discuss some mathematical properties of the $3d$ difference operators.
The $3d$ reduction of $\bar {\frak S}_{(1,0)}$ is nothing but the basic Macdonald difference operator,
with effective parameters
\be
t=e^{2\pi i \,\frac{1}{\omega_2} },\qquad
q=e^{2\pi i \,\frac{\omega_1}{\omega_2} },\qquad
a=e^{2\pi i \,\frac{m}{\omega_2} }\,.
\ee 
Similarly, 
for the reduction of $\bar {\frak S}_{(0,1)}$ we get the Macdonald operator with parameters
\be
t'=e^{2\pi i \,\frac{1}{\omega_1} },\qquad
q'=e^{2\pi i \,\frac{\omega_2}{\omega_1}},\qquad
a'=e^{2\pi i \,\frac{m}{\omega_1}} \,.
\ee Note that $q$ and $q'$ are related by a ``modular'' transformation,
\be
q=e^{2\pi \tau}\quad\to\quad q'=e^{-\frac{2\pi}{\tau}}\,.
\ee
Physically this follows from the fact that the (1,0) and (0,1) surface defects have support on the two 
maximal (linked) circles in $S^3$ fixed respectively by $j_{12}$ and $j_{34}$. 
 Interchanging the two types
of defects is the same as interchanging the circles.
Let us also write down the $3d$ reduction of the elliptic measure,
\be
\Delta_I=\half\frac{\Gamma(\frac{pq}ta^{\pm2};q,p)}{\Gamma(a^{\pm2};q,p)}\quad \to\quad
\half\prod_{i,j\geq 1}
\frac{\Omega_{ij}+1\pm 2m}{\Omega_{i+1\,j+1}-1\pm 2m}
\frac{\Omega_{ij}\pm 2m}{\Omega_{i+1\,j+1}\pm 2m}\, ,
\ee
where we have defined
$
\Omega_{ij}=i\omega_1+j\omega_2$.

\

\

\section{$A_{N-1}$ theories}\label{sunsec}

In this section we outline the generalization of our ``topological bootstrap'' procedure to the higher rank theories.
We first discuss, but not review in detail, the six-dimensional construction of ${\cal N}=2$ four-dimensional gauge theories and 
surface defects in them, and the correspondence with two-dimensional Toda CFT correlation functions. We refer the reader to~\cite{Gaiotto:2009hg,Drukker:2010jp,Gaiotto:2011tf}
for a complete discussion of the relevant facts.

The $(2,0)$ $6d$ SCFTs are labeled by a simply-laced Lie algebra $g$. An important class of codimension-two (to be contrasted with codimension-four) superconformal defects 
is labeled by an embedding $\rho$ of $su(2)$ in $g$, see {\it e.g.}~\cite{Drukker:2010jp,Chacaltana:2012zy}.
The properties of these defects  can be motivated, for example, upon compactification on a circle shared by all the defects. The bulk theory becomes $5d$ SYM with gauge algebra $g$.
The codimension two defect induces a singularity of the $5d$ SYM fields which breaks the gauge symmetry to $g(\rho)$ (possibly times some Abelian factors) at the defect. 
The codimension-four defect becomes a Wilson loop in the representation $R$ of $g$ or $g(\rho)$. 
In the construction of four-dimensional ${\cal N}=2$ gauge theories, one considers the twisted compactification of 
the $6d$ SCFT on a Riemann surface ${\cal C}$, with codimension two defects at points in ${\cal C}$. Codimension four defects at points in ${\cal C}$  produce surface defects in the $4d$ theory. 

Nekrasov's instanton partition function for the $4d$ theory is conjecturally the same as a conformal block for 
the $W_N$ current algebra~\cite{Alday:2009aq,Wyllard:2009hg}. The vertex operators for the $W_N$ theory come in families associated to various patterns of degenerate vectors. 
The families are labeled by the embedding $\rho$ of $su(2)$ in $g$, and an appropriate choice of Toda momentum, a vector valued in the dual of the Lie algebra $g$. 
Non-degenerate vertex operators (corresponding to $\rho=0$) are labeled by an imaginary Toda momentum. 
Semi-degenerate vertex operators are labeled by the sum of an imaginary momentum in $f(\rho)$, 
and a real momentum of the form $b \lambda + b^{-1} \lambda'$, where $\lambda$, $\lambda'$ 
are weights for $g(\rho)$. For a fully degenerate vertex operator, $f(\rho)=0$ and $g(\rho) = g$. 

The vertex operators map to a six-dimensional configuration with a codimension two defect labeled by $\rho$ 
and codimension four defects labeled by $\lambda$ and $\lambda'$, sitting on the $12$ and $34$ planes in flat space. 
In the Toda context one can change the type of a defect, by analytically continuing the continuous imaginary part of the momentum 
to some discrete choices of real momentum. 

Much of our analysis concerning the analytic structure of the index 
can be repeated for the analytic structure of the Nekrasov's partition function, {\it i.e.} for the properly normalized conformal blocks.
The conformal blocks have poles at discrete values of the Toda momenta, which correspond to special values of the mass parameters  
of the four-dimensional theory. At these values, the divergence is caused by a flat directions in the Higgs branch.
Thus the process of making a $W_N$ vertex operator more degenerate can be matched to a Higgsing process,
in the presence of vortices which become surface defects in the IR. 

Let us start from the basic example involving a bifundamental hypermultiplet. 
Our construction starts from a theory ${\cal T}_{IR}$, which for simplicity we can take to have a six-dimensional description in terms of the $A_{N-1}$ theory on a 
Riemann surface ${\cal C}$ with at least one ``full'' puncture, which carries  the $SU(N)$ flavor symmetry. \footnote{This assumption does not remove any generality from our analysis:
no matter what ${\cal T}_{IR}$ is, one can build an auxiliary six-dimensional theory in terms of a sphere with two full punctures, 
and ${\cal T}_{IR}$ added by hand at one puncture, by gauging the diagonal $SU(N)$ flavor symmetry.}
Then ${\cal T}_{UV}$ is a theory associated to ${\cal C}$ with one extra simple puncture inserted near the full puncture. The position of the simple puncture 
controls the gauge coupling of the auxiliary $SU(N)$ gauge group. The simple puncture has a $U(1)$ flavor symmetry, which we gauge in the Higgsing procedure. 

In the context of conformal blocks, the Toda momentum of a simple puncture is an imaginary multiple of a single simple weight $w_1$. 
It can be analytically continued to the real momentum of a fully degenerate puncture, but only one with $\lambda = r\, w_1$
and $\lambda' = s\, w_1$. These surface defects are naturally associated to 
 $r$-th and $s$-th symmetric powers of the fundamental representation. In the context of the superconformal index,
 they are the defects introduces by the difference operators ${\frak S}_{(r,s)}$.

Can we discuss surface defects labeled by other representations of $SU(N)$?
In the context of $5d$ SYM, Wilson loops in representations $R_1$ and $R_2$ can be brought together to a 
Wilson loop in the representation $R_1 \otimes R_2$, which can then be decomposed to a sum of Wilson loops in irreducible representations
\begin{equation}
R_1 \otimes R_2 = \sum_i N^i_{12} R_i\,.
\end{equation}
If we try to uplift such a formula to an ``OPE'' of codimension-four defects in $6d$, 
\begin{equation}
S_{R_1} \otimes S_{R_2} = \sum_i {\cal T}^i_{12} \otimes S_{R_i}\,,
\end{equation}
we need to promote the coefficients $N^i_{12}$ to $2d$ theories ${\cal T}^i_{12}$ with $N^i_{12}$
Here ``OPE'' is intended as a collision of the punctures in ${\cal C}$, not in space-time: the configuration of two codimension-four defects 
represents a single, two-parameter surface defect in four-dimension, which simplifies to a sum of defects if the parameters are tuned appropriately. 

At the level of the index, the location of the defects on ${\cal C}$ is immaterial, and the relation will become 
\begin{equation}
{\frak S}_{(R_1,0)} \otimes {\frak S}_{(R_2,0)} = \sum_i \chi[{\cal T}^i_{12}] \otimes {\frak S}_{(R_i,0)}\,,
\end{equation}
where $\chi[{\cal T}^i_{12}] $ is the $2d$ index of the ``OPE coefficient theory''. 
We have discussed such relations in the $A_1$ case (see section~\ref{propsec}).

\

\subsection*{Closing minimal punctures}

In this section we discuss the generalization of the residue
computation we performed for $A_1$ quivers to higher rank cases. 
The generalization is straightforward and we will only outline the
essential steps. 
\begin{figure}
\begin{center}
\includegraphics[scale=0.35]{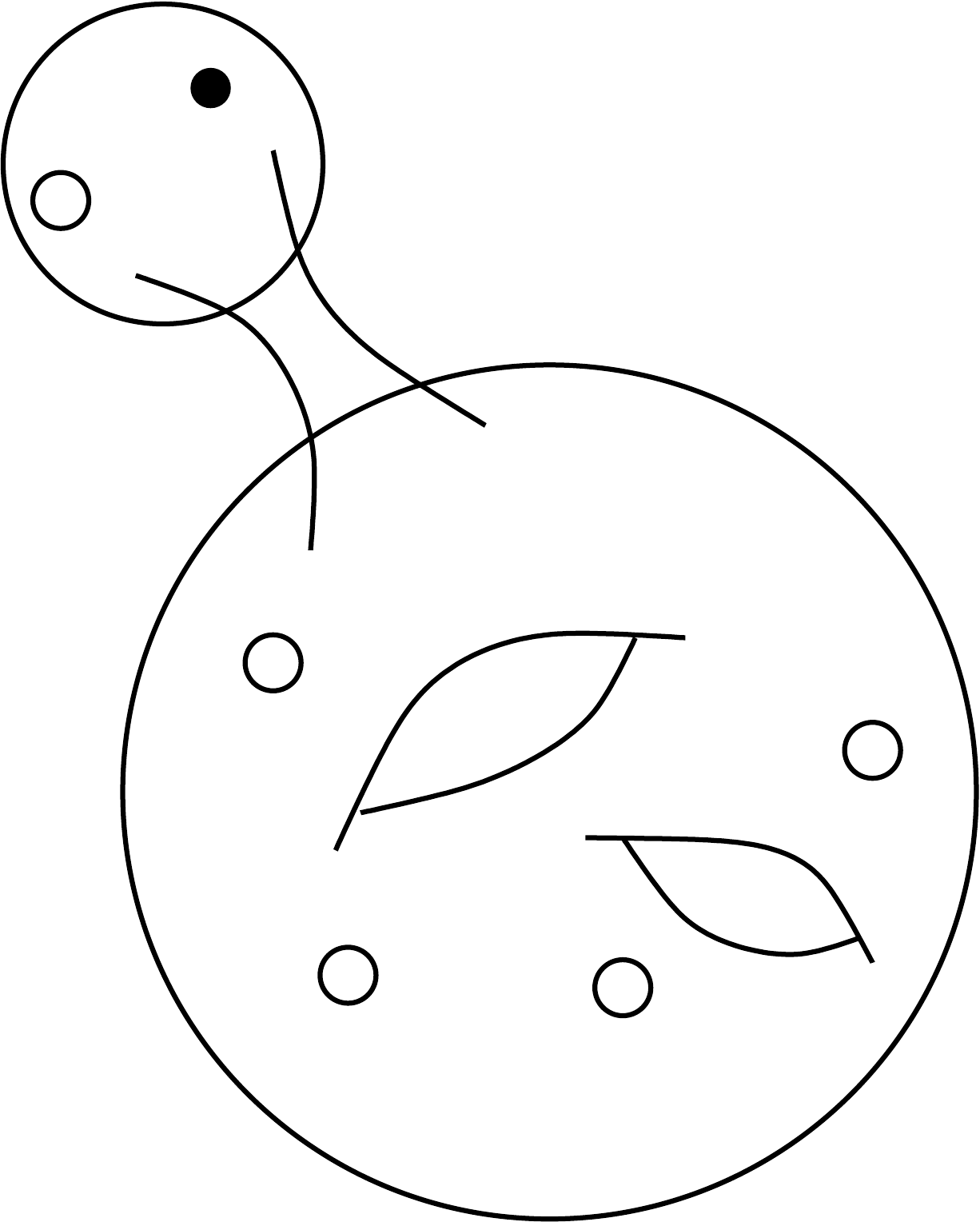}
\end{center}
\caption{To compute the residue of the index at a pole of the $U(1)$ fugacity
we consider the degeneration limit of the theory by decoupling a free-hypermultiplet.}
\end{figure}\label{figdegN}
As before, we consider a theory of class ${\cal S}$ ${\cal T}_{IR}(\equiv {\cal T}_{\cal C})$ with at least one maximal puncture 
and couple to it a free bi-fundamental hypermultiplet. The index of the free bi-fundamental hypermultiplet is 
equal to 
\be\label{hyp2}
{\cal I}_{hyp}({\mathbf b},{\mathbf c};a)=\prod_{i,j=1}^N\prod_{m,n\geq0}
\frac{1-p^{n+1}q^{m+1}t^{-\half}(a b_i c_j)^{-1} }{1-p^{n}q^{m}t^{\half}a b_i c_j }
\frac{1-p^{n+1}q^{m+1}t^{-\half}a b_i c_j }{1-p^{n}q^{m}t^{\half}(a b_i c_j)^{-1} }\,.
\ee Here $b_i$ and $c_i$ ($\prod_{j=1}^n b_j=\prod_{j=1}^n c_j=1$) are $SU(N)$ fugacities and $a$ is a $U(1)$ fugacity.
To couple this hypermultiplet to a general $A_{N-1}$ quiver
corresponding to Riemann surface ${\cal C}$ we gauge a diagonal symmetry $SU(N)$ by integrating over fugacity $c$,
\be\label{indb}
{\cal I}_{UV}=\oint\prod_{j=1}^{N-1}\frac{dc_j}{2\pi ic_j}\Delta({\mathbf c})\,{\cal I}_V({\mathbf c})\, {\cal I}_{hyp}({\mathbf b},{\mathbf c};a)\,
{\cal I}_{\cal C}({\mathbf c}^{-1},\dots)\,.
\ee
This index has poles at the following values of the $U(1)$ flavor fugacity
\be\label{poleN}
a=t^\half \, p^{r/N}\, q^{s/N}\,,
\ee with non-negative integers $r$ and $s$.
Actually, there are poles also at $a\to \exp[\frac{2\pi i}{N}\ell]\;a$ as we discussed before (see section~\ref{imppoles}). 
A way to see this is again by studying the pinchings of $c$ contour integrals.
Let us start by considering the poles of the integrand of~\eqref{indb} at 
\be
c_i=t^\half q^{m_i}p^{n_i}\frac{1}{a\,b_{\s(i)}}\,,\qquad  i=1,\dots,N\,.
\ee 
Here $\s(\cdot )$ is an element of $S(N)$.
Notice that these combined poles have a straightforward physical interpretation,
as contributions from a region in field space 
where complex bi-fundamental fields $Q_i^{\s(i)}$ receive a vev, which Higgses the $U(N)$ gauge symmetry completely. 
The $SU(N)$ gauge invariant operator is a determinant of the bifundamental fields. Thus the poles at $a^N t^{N/2} p^r q^{s}=1$. 
The residue (multiplied by $N$ for future convenience) is ${\cal I}_V^{-1}\;{\cal I}_{{\cal T}_{IR}}({\mathbf b},\dots)$.
If we gauge the $U(1)$ flavor symmetry with fugacity $a$, the contribution to the index of the 
simple pole at $a=t^\half$ is $N^{-1} {\cal I}_{{\cal T}_{IR}}({\mathbf b},\dots)$. The other poles at $a=\exp[\frac{2\pi i \ell}{N}]\,t^{\half}$ 
give the answer $N^{-1} {\cal I}_{{\cal T}_{IR}}(\exp[-\frac{2\pi i \ell}{N}]\, {\mathbf b},\dots)$, and they combine to give 
the $Z_N$ center gauging of ${\cal T}_{IR}$, as expected (see section~\ref{imppoles}).

Further specifying to $a=t^\half p^\frac{r}{N}q^{\frac{s}{N}}$ we get that the simple poles above pinch the integration contour and become double poles on the unit circle, 
because we have only $N-1$ independent $SU(N)$ flavor fugacities, if
\be
s=\sum_{i=1}^N m_i\,,\qquad
r=\sum_{i=1}^N n_i\,.
\ee This implies that the index has simple poles at~\eqref{poleN}.
The residue computation is given by the action of the following operator on the index
\begin{equation}\label{sunfin}
\boxed{
\begin{array}{l}
{\cal I}[{\cal T}_{IR},{\frak S}_{(r,s)}]={\frak S}_{(r,s)}\;{\cal I}[{\cal T}_{IR}]=\\
\\
\qquad
\displaystyle\sum_{\sum_{i=1}^N n_i=r}
\displaystyle\sum_{\sum_{i=1}^N m_i=s} {\cal I}_{\cal C}(q^{\frac{s}{N}-m_i}
p^{\frac{r}{N}-n_i}b^{}_{i})\,\times\\
\\
\qquad\qquad\displaystyle\prod_{i,j=1}^N
\left[{\displaystyle\prod_{m=0}^{m_i-1}}\frac{\theta(p^{n_j}q^{m+m_j-m_i}tb_i/b_j;p)}
{\theta(p^{-n_j}q^{m-m_j}b_j/b_i;p)}\right]
\left[\displaystyle\prod_{n=0}^{n_i-1}\frac{\theta(q^{m_j-m_i}p^{n+n_j-n_i}tb_i/b_j;q)}
{\theta(q^{m_i-m_j}p^{n-n_j}b_j/b_i;q)}\right]\,.
\end{array}
}
\end{equation}
The basic properties of these operators are the same as for the $A_1$ case.
These operators are all commuting and self-adjoint under the natural measure.

\

\

\subsection*{The index of a general theory of class ${\cal S}$ }

We are finally ready to bootstrap the index of class ${\cal S}$ theories of type $A_{N-1}$.
The difference operators ${\frak S}_{(r,s)}$ have common eigenfunctions $\{ \psi^\alpha({\mathbf a}) \}$, labelled
by irreducible $SU(N)$ representations. The eigenfunctions can be chosen to be
orthonormal under the propagator measure and we assume again that their eigenvalues are non-degenerate.
By the usual logic of introducing a degenerate puncture and colliding it with flavor punctures,
we deduce that the index of the trinion with three maximal punctures is diagonal in the $\{ \psi^\alpha({\mathbf a})\}$ basis,
\be \label{trindiag}
{\cal I}_{trin}({\mathbf b},{\mathbf c},{\mathbf d})=\sum_\alpha  C_\alpha\,\psi^\alpha({\mathbf b})\psi^\alpha({\mathbf c})\,\psi^\alpha({\mathbf d})\,.
\ee
The determination of the structure constants is somewhat more involved than in the $A_1$ case. 
Let us
outline the procedure for the case of $A_2$.

The index of the free hypermultiplet of $A_{2}$ can be expanded as
\be \label{hypdiag}
{\cal I}_{hyp}({\mathbf b},{\mathbf c};a)=\sum_\alpha  \psi^\alpha({\mathbf b})\psi^\alpha({\mathbf c})\,\phi_{(2,1)}^\alpha(a)\,.
\ee Here ${\mathbf b}$ and ${\mathbf c}$ are $SU(3)$ fugacities and $a$ is $U(1)$ fugacity. 
 The label ${(2,1)}$  denotes the auxiliary Young diagram corresponding to the $U(1)$ puncture. This equation \textit{defines}
the function $\phi_{(2,1)}^\alpha(a)$ since the left-hand side is explicitly known~\eqref{hyp2}. 
(That ${\cal I}_{hyp}({\mathbf b},{\mathbf c};a)$
must take this diagonal form follows again by consistency of the theory with an extra degenerate puncture.)
Consider now the theory associated to a sphere with
 two maximal and two minimal punctures. The index can be calculated in two different duality frames.
 In the degeneration limit of the surface where a maximal puncture collides with a minimal puncture,
the theory is obtained by gauging the diagonal $SU(3)$ of two bifundamental hypermultiplets,
\be
A=\sum_\alpha  
\phi^\alpha_{(2,1)}(b/c)\,\phi_{(2,1)}^\alpha((c\,b)^{-1})\,\prod_{\ell=1}^{2}\psi^\alpha({\mathbf a}_\ell)\,\,.
\ee Here $b$ and $c$ denote  two $U(1)$ fugacities. 
On the other hand, in the degeneration limit where two maximal punctures collide, the trinion with three maximal punctures
is coupled to a free hypermultiplet by gauging an $SU(2)$ subgroup (this is the duality discovered in~\cite{argyres-2007-0712}).
In this duality frame the index is given by
\be
A=\sum_\alpha C_\alpha\; \prod_{\ell=1}^{2}\psi^\alpha({\mathbf a}_\ell)\,
\oint \frac{dz}{2\pi i z}\Delta(z){\cal I}_V(z)\, \tilde{\cal I}_{hyp}(z,b^3)\psi^\alpha(z\, c,z^{-1}\,c,c^{-2})\,.
\ee Here $\tilde{\cal I}_{hyp}(z,\, b^3)$ is a free hypermultiplet in the fundamental of $SU(2)$, where the $SU(2)$ fugacity is parametrized by $z$;
 the two half-hypermultiplets have charges $\pm3$ 
under the $U(1)$ symmetry parametrized by $b$.
The $U(1)$ charges of the different components in the two duality frames are linearly dependent~\cite{argyres-2007-0712}, which results 
{\it e.g.} in the non-trivial power  of $b$ in the free hypermultiplet $\tilde {\cal I}_{hyp}$.
Comparing the indices in the two duality frames we obtain 
\be\label{Ceq}
C_\alpha= \frac{\phi^\alpha_{(2,1)}(b/c)\,\phi_{(2,1)}^\alpha((c\,b)^{-1})}
{\oint \frac{dz}{2\pi i z}\Delta(z){\cal I}_V(z)\,\tilde {\cal I}_{hyp}(z,\,b^3)\psi^\alpha(z\, c,z^{-1}\,c,c^{-2})}\,.
\ee
The right-hand side here is written as  a function of $b$ and $c$ but is actually independent of these
as an implication of S-duality; this can be explicitly checked~\cite{Gadde:2011uv}. 

 Finally, by gluing
elementary building blocks, we can write 
the index of a genus ${\frak g}$ theory
with $k$ maximal and $k'$ minimal punctures,
\be
\sum_\alpha \left(C_\alpha\right)^{2{\frak g}-2+k-k'}\, \prod_{\ell=1}^{k}\psi^\alpha({\mathbf a}_\ell)\,
\prod_{m=1}^{k'}\phi^\alpha_{(2,1)}(b_m)\,.
\ee 

\

This algorithm can be easily generalized to  higher rank.
The trinion with three maximal punctures is always diagonal in the $\{ \psi^\alpha({\mathbf a})\}$ basis, see (\ref{trindiag}).
The index of the free hypermultiplet has a diagonal expression analogous to (\ref{hypdiag}), in terms of two  $\psi^\alpha$ wavefunctions and 
one wavefunction $\phi^\alpha_{(2,1,\dots,1)}$ associated to the minimal puncture, which can  in fact be fixed
by the known expression of the index of the free hyper. Finally the structure constants  $C_\alpha$ are determined
by comparing two degeneration limits of the $N+1$-punctured sphere with two maximal and $N-1$ minimal punctures: in one duality frame we have a linear superconformal quiver with the two maximal punctures at the ends, and in the other one we have the $SU(N)$ trinion (the $T_N$ theory) coupled
to the superconformal tail with a maximal puncture on one end and minimal puncture on the other end.
The two duality frames are related by a generalized Argyres-Seiberg duality \cite{Gaiotto:2009we}.
The former duality frame is completely Lagrangian and the index can be computed explicitly. 
 The index in the latter
frame is written in terms of the index of the trinion.   Equating the indices in the two duality frames one determines the trinion structure
constants $C_\alpha$.
These date are sufficient to fix the index of any theory containing only maximal and minimal punctures.
More general punctures can be incorporated by a Higgsing procedure that involves more general
superconformal tails.   In appendix~\ref{Lapp} we give an example of this procedure. The final
 prescription is spelled out in~\cite{DR} and coincides with the prescription of~\cite{Gadde:2011uv}.\footnote{The prescription of~\cite{Gadde:2011uv} gave divergent results for a small subclass of 
theories of class ${\cal S}$, however following the residue logic described in this paper it was explained  in~\cite{DR}
how to resolve those singularities.}

\

\

\section{Discussion}\label{sumsec}

In summary, the superconformal index, being a protected quantity,
must be consistent with a large class of deformations of the $4d$ field theory. 
Moving along  an exactly marginal direction the index does not change, so  it must be the same
when evaluated in different  S-duality frames.  This has the non-trivial implication 
that the index of class ${\cal S}$ theories is computed by  the correlator of a $2d$ TQFT~\cite{Gadde:2009kb}.
In this paper we have considered a more interesting class of deformations: we have studied how the index is affected by 
expectation values of supersymmetric operators.
We have argued that a  pole of the index  in  flavor fugacity is associated
to a bosonic flat direction, parametrized by  the vacuum expectation value of a protected operator, 
and that the residue captures the IR physics reached at the endpoint of the flow triggered by
the vev.  We have formulated a precise prescription to evaluate the index of the IR theory.

We have focussed on a special class of poles, associated
to RG flows that introduce two-dimensional defects in the IR SCFT. 
In particular, starting from the index of a theory with  flavor symmetry ${\cal G}\times U(1)$ our residue calculus
  determines the index of the theory with smaller flavor symmetry, ${\cal G}$, but  endowed with BPS surface defects.
In the language of the $2d$ TQFT, the surface defects are associated to special ``degenerate'' punctures. The fusion
of the degenerate punctures with the  ``ordinary'' punctures associated to flavor symmetries amounts to acting
on the flavor fugacities with certain elliptic difference operators. Under a minimal set of assumptions, namely
consistency of the $2d$ TQFT structure and knowledge of the index of free theories, we were then able to  determine
the index of a general theory of class~${\cal S}$.

We can perhaps extract two general lessons for quantum field theory.
First, in a given theory, even if ultimately we are most interested in the familiar
correlators of local operators, it is fruitful to consider the larger set of observables
that includes defects of various codimension. In our problem the introduction of surface defects
was the key step.
Second, to obtain results in a given theory,
it it fruitful to enlarge the view to all the theories of the same class. 
Indeed our ``bootstrap''  is a conventional bootstrap for the $2d$ TQFT,
but for the $4d$ SCFT it is really a bootstrap  in theory space.

\  

\subsection*{Some open problems}

In closing, let us briefly mention several further research directions,  questions, and speculations arising from our work.

\begin{itemize}

\item 
 First and foremost, it would be very  valuable to have a {\it bona fide} computation of the superconformal index for Lagrangian class ${\cal S}$ theories 
 endowed with surface defects. We have in mind a first-principles calculation directly from path integral definition, using
 the methods of supersymmetric localization.

 \item 
The difference  operators that introduce surface defects are sums of terms  shifting the values of flavor fugacities in different ways, with coefficients looking as indices of $2d$ free fields, {\it i.e.} 
combinations of theta-functions  (see {\it e.g.}~\eqref{sunfin}). 

It is natural to wonder  this sum has a more direct physical interpretation. What happens
physically when the  defect puncture collides with a flavor puncture?
There is some evidence \cite{Gaiotto:2011tf} that the surface defect may degenerate to the direct sum of simpler, decoupled 
defects. Schematically, we could write 
\begin{equation}
{\frak S}_{(r,0)} \to \sum_i T_{r,i} \times \sigma_i 
\end{equation}
where $\sigma_i$ are the simpler surface defects, and the ``coefficients'' $T_i$ are further decoupled $2d$ degrees of freedom. 
We can imagine that the simple defects $\sigma_i$ give the individual shift operators in the index fugacities, while the 
$2d$ index of the $T_{r,i}$ produces the coefficients in the expansion of the difference operator. 
It would be interesting to make sense of  this conjectural structure.

Similarly, we 
have seen that the difference operators introducing the defects satisfy recursion relations. The coefficients of these
recursion relations are again combinations of theta-functions, and  could perhaps be  interpreted 
as indices of $2d$ theories.

\item The index of class ${\cal S}$ theories takes a  diagonal form in the eigenfunctions $\psi^\alpha ({\bf a} )$ of the $\frak S_{(r,s)}$ difference operators.
The eigenfunctions do not depend on the specific four-dimensional theory. It is tempting to interpret these wavefunctions
as universal modules of some intricate algebra, built out of the flavour currents and their descendants. 

\item
Finally, 
it will be important to generalize our results ${\cal N}=1$ theories. The simplest route
is to take the ${\cal N}=1$ (``Sicilian'') limit of our construction,
which corresponds to giving a mass to the chiral field in the ${\cal N}=2$
vector multiplet~\cite{Benini:2009mz}. At the level of the index this  amounts to setting~\cite{Gadde:2010en}
$p\,q=t^2 $. 
It  would be  interesting to check if the index of surface defects 
remains sensible after this mass deformation to ${\cal N}=1$.

\end{itemize}

\

\

\section*{Acknowledgments}

It is a pleasure to thank C.~Beem,  T.~Dimofte, A.~Gadde, S.~Gukov, Z.~Komargodski, J.~Maldacena, N.~Nekrasov, N.~Seiberg, and V.~Spiridonov for useful discussions. 
The research of DG was supported by Perimeter Institute for Theoretical
Physics.  Research at Perimeter Institute is supported by the
Government of Canada through Industry Canada and by the Province of
Ontario through the Ministry of Economic Development and Innovation.
The work of DG was also supported in part by 
NSF grant NSF PHY-0969448 and in part
by the Roger Dashen membership in 
the Institute for Advanced Study.
The research of LR is partially supported by NSF Grant PHY-0969919.
The research of SSR was supported in part by NSF grant PHY-0969448.  SSR would like to thank the hospitality of the HET groups at the Weizmann 
Institute, KITP, and ICTS Bangalore where part of this research was conducted. This research was also supported in part by the NSF grant PHY-1125915.
LR and SSR would like to thank the Aspen Center for Physics for hospitality during the completion of this work. The Aspen Center for Physics is partially supported by the NSF under Grant No. 1066293.

\

\

\

\appendix

\section{Embedding $2d$ in $4d$}\label{imbsec}

Let us discuss the embedding of $2d$ $(2,2)$ superalgebra in the ${\cal N}=2$ $4d$ superalgebra.
We choose to parametrize $S^3\times S^1$ by $\sum_{i=1}^4 x_i^2=1$ for $S^3$ and by an $x_0$ 
for the $S^1$.
We start from the ${\cal N}=2$ superconformal  algebra $SU(2,2|2)$, whose
 bosonic subgroup is $SU(2,2)\times SU(2)_R\times U(1)_r$.
A maximal subgroup of $SU(2,2)$ is $SU(2)_{j_1}\times SU(2)_{j_2}\times U(1)_E$.
The generators $j_{12}=j_1+j_2$ is the rotation of the $(x_1,x_2)$ plane and $j_{34}=j_2-j_1$ is the rotation
on the $(x_3,x_4)$ plane.

The two dimensional defects  wrap the $S^1$ and one of the equators:
either $x_1=x_2=0$ or $x_3=x_4=0$. 
The surface defects preserve an $SU(1,1|1)\times SU(1,1|1)\times U(1)_J$ subalgebra,
where the $U(1)_f$ commutes with all the preserved supercharges and can be considered as a flavor symmetry from the $2d$ perspective.
The bosonic subalgebra is given by two copies, left and  right, of $SU(1,1)\times U(1)_J$ and by the flavor $U(1)_f$.
The $SU(1,1)$ is a chiral half of the global conformal group in two dimensions with 
Cartan generator $L_0$ ($\bar L_0$), while  the $U(1)_J$ is generated by $J_0$ ($\bar J_0$).
We denote the generator of the flavor $U(1)_f$ by  $f$.
\begin{table}
  \begin{centering}
  \begin{tabular}{|c|c|c|c|c|}
  \hline
${\mathcal Q}$ &$SU(2)_{j_1}$&$SU(2)_{j_2}$&$SU(2)_R$&$U(1)_r$ \tabularnewline
  \hline
  \hline
$  {\mathcal Q}_{{\suup }-}$ &$-\half$& $0$& $\;\;\;\half$&$\half$ \tabularnewline
  %
  \hline
$ {\mathcal Q}_{{\suup}+}$ &$\;\;\half$& $0$& $\;\;\;\half$&$\half$ \tabularnewline
 %
   \hline
$ {\mathcal Q}_{{\sudown}-}$ &$-\half$& $0$& $-\half$&$\half$ \tabularnewline
  %
  \hline
  ${\mathcal Q}_{{\sudown}+}$  &$\;\;\half$& $0$& $-\half$&$\half$ \tabularnewline
 \hline
 %
 %
  \hline
  $\widetilde {\mathcal Q}_{{\suup}\dot{-}}$ &$0$&$-\half$&  $\;\;\;\half$&$-\half$ \tabularnewline
  \hline
$\widetilde {\mathcal Q}_{{\suup}\dot{+}}$ &$0$&$\;\;\;\half$&  $\;\;\;\half$&$-\half$  \tabularnewline
 %
\hline
$\widetilde {\mathcal Q}_{{\sudown}\dot{-}}$ &$0$&$-\half$&  $-\half$&$-\half$  \tabularnewline
    \hline
$\widetilde {\mathcal Q}_{{ \sudown}\dot{+}}$ &$0$&$\;\;\;\half$&  $-\half$&$-\half$ \tabularnewline
  \hline
  \end{tabular}
  \par  \end{centering}
  \caption{ \label{charges} For each supercharge ${\cal Q}$, we list its quantum numbers, the associated
  $\delta \equiv 2\left\{{\mathcal Q},{\mathcal Q}^\dagger\right\}$, and the other $\delta$s commuting with it.
  Here $I = \suup,\sudown$ are $SU(2)_R$ indices and
$\alpha = \pm$, $\dot \alpha = \pm$ Lorentz indices.  
 $E$  is the conformal dimension,  $(j_1, j_2)$ the Cartan generators of the $SU(2)_1 \otimes SU(2)_2$ isometry group, and $(R \, ,r)$, the Cartan generators 
  of  the  $SU(2)_R \otimes U(1)_r$ R-symmetry group.
  }\end{table}
The fermionic generators of the  two $SU(1,1|1)$ NSNS algebras are denoted by
$G_r^\pm$, $\bar G_r^\pm$ with $r = \pm \frac{1}{2}$.

There are many equivalent ways  to embed this into the $4d$ algebra
but since we are computing the index with respect to $\Q$ there are two choices which correspond to two possible 
choices of $f$,
\be\label{center}
f=j_{12}+R,\qquad f'=j_{34}+R\,.
\ee The two choices are of course isomorphic and correspond to putting the defects
on the two different equators. For concreteness we will focus on the former choice. The left and the right super-charges are then
mapped  as follows
\be
(G_{-\half}^+,G_{-\half}^-)=({\mathcal Q}_{{\suup }-},\widetilde {\mathcal Q}_{{ \sudown}\dot{+}})\,,
\qquad
(\bar G_{-\half}^+,\bar G_{-\half}^-)=(\Q,{\mathcal Q}_{{\sudown}+})\,.
\ee
The $G_\half^\pm$ and $\bar G_\half^\pm$ are the superconformal counterparts.
The map between charges
\be
&&J_0=R+r+f\,,\qquad \bar J_0=R-r+f\,,\\
&&L_0+\bar L_0=E\,, \qquad L_0-\bar L_0=j_{34}\,,\nonumber
\ee generates the canonical $(2,2)$ superalgebra
\be
&&[L_0, G_r^\pm]=-r G_r^\pm,\qquad [\bar L_0, \bar G_r^\pm]=-r \bar G_r^\pm,\\
&&[J_0, G_r^\pm]=\pm G_r^\pm,\qquad [\bar J_0, \bar G_r^\pm]=\pm \bar G_r^\pm,\nonumber\\
&&\{G_\half^\pm,G_{-\half}^\mp\}=L_0\pm \half J_0\,,\qquad
\{\bar G_\half^\pm,\bar G_{-\half}^\mp\}=\bar L_0\pm \half \bar J_0\,.\nonumber
\ee The admixture of the flavor symmetry $f$ inside $J_0$ and $\bar J_0$ is fixed 
by the last anticommutators.

\

\noindent Let us now consider our $4d$ index,
\be
{\cal I}=\Tr(-1)^F p^{j_{12}-r}\,q^{j_{34}-r}\,t^{r+R}\,.
\ee We notice that we can identify the charges as,
\be
j_{34}-r=L_0-\half J_0-(\bar L_0-\half \bar J_0),\qquad 
j_{12}-r=2f-J_0\,\qquad
r+R=J_0-f\,.
\ee Moreover only states with $\bar L_0=\half\bar J_0$ contribute to the 
index due to~\eqref{BPS}. Thus, the index can be written as follows
\be\label{indQ}
{\cal I}=
\Tr(-1)^F q^{L_0+\half J_0-f} \,p^{f}\,\left(\frac{t}{pq}\right)^{J_0-f}\,.
\ee This is an NSNS index of the $(2,2)$ theory.
Choosing the symmetry $f'$ in~\eqref{center} to be the flavor symmetry the index becomes
\be
{\cal I}=\Tr(-1)^F p^{L_0+\half J_0-f'} \,q^{f'}\,\left(\frac{t}{pq}\right)^{J_0-f'}\,,
\ee i.e. we just switch between $p$ and $q$, and $f$ with $f'$.

\

We can give a $2d$ interpretation of the prefactor that we obtained 
in the residue computation,
\be\label{ourx}
\frac{\theta(t\,p^u,\,q)}{\theta(p^{-1-u},q)}=
\frac{\theta(\frac{p\,q}{t}\;p^{-1-u},\,q)}{\theta(p^{-1-u},q)} \equiv  \chi_{2d}  (p^{-1-u}) \,,
\ee where $u$ is non-negative integer.
The numerator  can be understood as the partition function generated by two fermions with
the following charges
\be
\frac{q\,p^{-u}}{t}\qquad&:&\qquad L_0=-\half u,\qquad f=-1-u,\qquad J_0=-2-u\,,\\
t\,p^u\qquad&:&\qquad L_0=1+\half\, u,\qquad f=1+u,\qquad J_0=2+u\, ,
\ee 
and their derivatives.
The charges of the derivative which contribute to index are $L_0=1$,
$J_0=0$, $f=0$, and $\bar J_0=\bar L_0=0$. This derivative contributes a factor of $q$
 to~\eqref{indQ} and generates the infinite product in the theta-functions of~\eqref{ourx}. 
Similarly, the denominator is generated by bosons with the following charges
\be
p^{-1-u}\qquad&:&\qquad L_0=-\half-\half u,\qquad f=-1-u,\qquad J_0=-1-u\,,\\
p^{1+u}\,q\qquad&:&\qquad L_0=\frac32+\half u,\qquad f=1+u,\qquad J_0=1+u\,.
\ee
These are  unusual charge assignments. The standard charges of a free chiral field (free complex scalar 
and a complex fermions) are obtained in the formal limit $u=-1$: here the 
flavor charges are zero and the expression~\eqref{ourx} diverges and hence the limit is formal.
 We take this as further indication that these partition functions
 describe
 decoupled fields with accidental symmetries, and that they
 should be stripped off from the definition of the surface defects. 

If we redefine the charges as
\be
\hat L_0=L_0-\half\,f\,,\qquad
\hat J_0=J_0-f\,,
\ee then the new charges have natural values for the fields contributing to~\eqref{ourx}.

Note also that the combinations of theta-functions appearing in  the basic difference operator~\eqref{opdef}
are $\theta(\frac{p\,q}{t}\,b^{\pm2};q)/\theta(b^{\pm2};q)$ which is again a free chiral field, with
standard charges, also charged under flavor symmetry coupled to $b$.

\

Finally, we consider different supersymmetric limits of the index.
For concreteness we consider the index~\eqref{indQ}.
\subsection*{``Higgs'' index}

We take the limit $q\to 0$ while keeping the other fugacities finite. This is the $2d$ version 
of the $4d$ Macdonald index~\cite{Gadde:2011uv}. The index becomes
\be
{\cal I}_H=\Tr_H\;(-1)^F p^{2f-J_0}\,t^{J_0-f}\,. 
\ee Here $\Tr_H$ is summing over states satsfying 
\be
L_0=\half J_0,\qquad \bar L_0=\half\bar J_0\,.
\ee Thus this index is half-BPS and gets contributions from states annihilated
by two supercharges, one left and one right, of same R-symmetry charge.
The index of the free $2d$ chiral field~\eqref{ourx}  here becomes
\be
\frac{\theta(t\,p^u,\,q)}{\theta(p^{-1-u},q)}\qquad \to\qquad
\frac{1-t\,p^u}{1-p^{-1-u}}\,.
\ee For instance the index of the $A_1$ trinion with a surface defect in this limit is given in~\eqref{thetaind}
and gets contributions from states charged under flavor symmetry and thus captures Higgs branch states.
Note that the $2d$ $(2,2)$ algebra does not constrain the value of charge $f$ and thus there is no analogue
here of the Hall-Littlewood index of~\cite{Gadde:2011uv}.

\

\subsection*{``Coulomb'' index}

We take the limit $\{q,p,t\}\to 0$ while keeping $p/q$ and  $pq/t$ finite. This is the $2d$ version 
of the $4d$ Coulomb index~\cite{Gadde:2011uv}. The index becomes
\be
{\cal I}_C=\Tr_C\;(-1)^F \left(\frac{p}{q}\right)^{f}\,\left(\frac{t}{pq}\right)^{J_0-f}\,. 
\ee Here $\Tr_C$ is summing over states satsfying 
\be
L_0=-\half J_0,\qquad \bar L_0=\half\bar J_0\,.
\ee Thus this index is half-BPS and gets contributions from states annihilated
by two supercharges, one left and one right, of opposite R-symmetry charge.
The index of the free $2d$ chiral field~\eqref{ourx} here becomes
\be
\frac{\theta(t\,p^u,\,q)}{\theta(p^{-1-u},q)}\qquad \to\qquad
\left(\frac{p\,q}{t}\right)^{1+u}\frac{1-\frac{p\,q}{t}\left(\frac{q}{p}\right)^{1+u}}{1-\left(\frac{q}{p}\right)^{1+u}}\,.
\ee The index of the $A_1$ trinion with basic surface defect in this limit~\eqref{coulind} gets contributions from flavor singlets not counted
by the Higgs index and thus naturally captures Coulomb branch physics.

\section*{``Schur'' index}

We take the limit $p=t$ while keeping all the fugacities finite.
 This is the $2d$ version 
of the $4d$ Schur index~\cite{Gadde:2011uv}. The index becomes
\be
{\cal I}_S=\Tr\;(-1)^F q^{L_0-\half J_0}\,p^f
\ee Since $f$ commutes with all the charges in the $(2,2)$ algebra this index is actually independent of $q$.
In other words this index is simply $\Tr(-1)^F$ refined only by flavor symmetries from the $2d$ perspective.
Given such an index care has to be taken in expanding it in fugacities since there is no a-priori good expansion parameter. 
It gets contributions only from states satisfying
\be
L_0=\half J_0,\qquad \bar L_0=\half\bar J_0\,.
\ee
The index of the free $2d$ chiral field here~\eqref{ourx} becomes
\be
\frac{\theta(t\,p^u,\,q)}{\theta(p^{-1-u},q)}\qquad \to\qquad
-p^{1+u}\,,
\ee and in particular all dependence on $q$ factors out automatically as it should.
Note that here the vacuum canceled out against a fermionic state with zero $f$-charge: this is 
an artifact of the issue mentioned above.

\

\

\section{Wavefunctions of the $A_1$ elliptic RS model}\label{eigenfuncsec}

The difference operators that we defined through our residue prescription are closely
related to well-known difference operators. Consider the similarity transformation
\be
{\frak S}_{(0,1)}=\frac{\theta(t;p)}{\theta(q^{-1};p)}\,
\hat {\cal K}(b)\,H\,\hat {\cal K}^{-1}(b),\qquad
\hat {\cal K}(b)=\Gamma\left(t \,b^{\pm2};\,p,\,q\right)\, ,
\ee where
$H$ is  given by
\be\label{RSH}
H=\frac{\theta(t\,b^2;\,p)}{\theta(b^2;\,p)}\Delta_{b\to q^{1/2}b}+
\frac{\theta(t\,b^{-2};\,p)}{\theta(b^{-2};\,p)}\Delta_{b\to q^{-1/2}b}\,.
\ee This operator is known as the elliptic Ruijsenaars-Schneider (RS)
difference operator, which  is a relativistic generalization of the  Hamiltonian
of the elliptic Calogero-Moser-Sutherland model~\cite{Ruijsenaars:1986vq,Ruijsenaars:1986pp,R}.\footnote{
A relation between Calogero-Moser-type models and elliptic Gamma functions, which are the building blocks of the index of the
chiral superfield, was discussed in~\cite{SpirCM}.}
In fact we have a {\it second} set of operators obtained by interchanging $p$ with $q$. All these operators
 are commuting as we have discussed in the bulk of the paper.
 The spectrum of the RS operator is well-known  in certain degeneration limits. Taking $p=0$ the operator reduces to the Macdonald 
difference operator, whose eigenfunctions are the Macdonald polynomials. Conjugating
with $\hat {\cal K}$ the propagator measure (the Haar measure times the index of the ${\cal N}=2$ vector multiplet), 
we find 
\be
\Delta\;{\cal I}_V=\half\frac{1}{\Gamma\left(t \,b^{\pm2};\,p,\,q\right)\Gamma\left(b^{\pm2};\,p,\,q\right)}\qquad\to\qquad
\hat \Delta=\half\frac{\Gamma\left(t \,b^{\pm2};\,p,\,q\right)}{\Gamma\left(b^{\pm2};\,p,\,q\right)}\,.
\ee 
The operator $H$ is self-adjoint under this measure. Setting $p=0$ this becomes
\be
\hat \Delta=
\half\frac{\left(b^{\pm2};\,q\right)}{\left(t\,b^{\pm2};\,q\right)}\,,
\ee which is the measure under which Macdonald polynomials are orthogonal.
The Macdonald polynomials are explicitly given by
\be\label{macpol}
P^\lambda({b}|q,t)=&&(1-q^\lambda t)^{\half}\prod_{i=\lambda}^\infty\frac{\sqrt{(1-q^{1+i})(1-q^{i}t^2)}}{(1-q^it)}\,
\prod _{j=0}^{\lambda-1} \left(\frac{1-q^{j+1}}{1-t\, q^j}\right)\,\times\\
&&\sum_{i=0}^\lambda \prod_{j=0}^{i-1}\frac{1-t\,q^j}{1-q^{j+1}}
 \prod_{j=0}^{\lambda-i-1}\frac{1-t\,q^j}{1-q^{j+1}}\,b^{2i-\lambda}\,.\nonumber
\ee It is important that these eigenfunctions are non-degenerate.

For general $(p, q, t)$, although there is some discussion of  wavefunctions in the
 literature~\cite{ru1,ru2,ru3} (see also~\cite{Etingof:1994az,rains, rains2}), to the best of our knowledge
 no explicit closed form is known.
However, 
as a proof of concept that such eigenfunctions exist and are non-degenerate we can  construct
approximate wavefunction as follows. 
We solve for the eigenfunctions using the following ansatz
\be
\psi_\ell(a)=\frac{a^{\ell+1}g(a)-a^{-\ell-1}g(a^{-1})}{a-a^{-1}}\,.
\ee 
 Plugging this ansatz into the difference equation~\eqref{RSH} we obtain,
\be\label{anzeq}
&&q^{\half\ell}\[\frac{a^{\ell}g(q^\half a)\theta(t\,a^2;\,p)}{(1-q^{-1}a^{-2})\theta(a^2;\,p)}
+\frac{a^{-\ell}g(q^{\half} a^{-1})\theta(t\,a^{-2};\,p)}{(1-q^{-1}a^{2})\theta(a^{-2};\,p)}\]+\\
&&+q^{-\half\ell}\[\frac{a^{-\ell}g(q^{-\half} a^{-1})\theta(t\,a^2;\,p)}{(1-q^{}a^{2})\theta(a^2;\,p)}
+\frac{a^{\ell}g(q^{-\half} a^{})\theta(t\,a^{-2};\,p)}{(1-q^{}a^{-2})\theta(a^{-2};\,p)}\]
={\cal E}_q\,\frac{a^{\ell+1}g(a)-a^{-\ell-1}g(a^{-1})}{a-a^{-1}}\,.\nonumber
\ee
As a first approximation we neglect the term in the first line, assuming as always
that $|q|<1$, and obtain 
the simple equation
\be
\frac{g_0(q^{-\half} a^{})\theta(t\,a^{-2};\,p)}{(1-q^{}a^{-2})\theta(a^{-2};\,p)}=
q^{\half\ell}{\cal E}^{(0)}_q\,\frac{g_0(a)}{1-a^{-2}}\,.
\ee The subscript $0$ on $g_0(\cdot)$ signifies that we are making an approximation here.
 This equation is  solved using the following property of the elliptic Gamma function
\be
\Gamma(q\,z;p,q)=\theta(z;p)\, \Gamma(z;p,q)\,.
\ee The solution is 
\be
g_0(a)=(1-a^{-2})\,\frac{ \Gamma(a^{-2};p,q)}{ \Gamma(t\,a^{-2};p,q)}\,,\qquad
{\cal E}^{(0)}_q=q^{-\half\ell}\,.
\ee 
This solution is  a joint
eigenfunction of the two hamiltonians ${\frak S}_{(1,0)}$ and ${\frak S}_{(0,1)}$, {\it i.e.} it is symmetric in $p$ and $q$.
The eigenfunction  are given thus by
\be
\psi_\ell(a)\sim 
a^{\ell}\,\frac{ \Gamma(a^{-2};p,q)}{ \Gamma(t\,a^{-2};p,q)}+
a^{-\ell}\,\frac{ \Gamma(a^{2};p,q)}{ \Gamma(t\,a^{2};p,q)}\,.
\ee 
Note that this solution is correct up to order $q^{\ell/2}$ ($p^{\ell/2}$), 
since we neglected a term of the form $q^{\ell/2}g_0(q^{1/2}a)$. 
The function $g_0(q^{1/2}a)$  actually is not regular in $q$:
it has an expansion in negative powers of $q$ of arbitrary order with coefficients proportional to some power of $p$.
This expansion has terms of the form $q^{-r}p^{r}$ along with 
terms with higher powers of $p$
and same power of $q$,
\be
\left.\frac{ \Gamma(a^{-2};p,q)}{ \Gamma(t\,a^{-2};p,q)}\right|_{a\to q^{\half}a}=F_{reg}(a,q,p,t)\;
\prod_{i,j\geq 0}\frac{1-t\,\frac{p}{q}\,p^i\, q^j\, a^{-2}}{1-\frac{p}{q}\,p^i\, q^j\,a^{-2}}\,.
\ee
Here $F_{reg}(\cdot)$ is regular in $q$-expansion. If we expand in $p$ to $r$th order then the term which we neglected on the left-hand-side of~\eqref{anzeq}
contributes at order $\half\ell-r$. On the other hand the second term on the left-hand-side of~\eqref{anzeq} contributes at order $-\half\ell+r$.   
Thus, the highest order in $q$-expansion to which the eigen-function are consistent with the assumptions $r=\half \ell$.
We can make several checks of this result.
Note that ${\frak S}_{(1,0)}$ and ${H}$ can be mapped to each other by exchanging 
$t\to \frac{pq}{t}$.
 Thus,
the (exact) eigenfunctions have to satisfy
\be
 \Gamma(t\,a^{\pm 2};p,q)\psi(;p,q,t)=\psi(;p,q,\frac{pq}{t})\,.
\ee This property is obviously satisfied in the above approximation.
The approximate eigenfunctions are orthogonal under the natural measure (up to the
order they are valid),
\be
\oint\frac{da}{4\pi i a}\frac{ \Gamma(t a^{\pm 2};p,q)}{ \Gamma(a^{\pm 2};p,q)}
\, \psi_\ell(a)\,\psi_{\ell'}(a)=\delta_{\ell\ell'}\,n_\ell\,.
\ee
The normalization is given by
\be
\text{$\ell=0$  }\;&:&\;\qquad n_\ell=1+t\,,\nonumber\\
\text{$\ell\neq 0$  }\;&:&\;\qquad n_\ell=1\,.
\ee
In the special limit $p=0$ the approximate eigenfunctions coincide with Macdonald polynomials
up to order $\ell-1$ in $q$. In the limit $p=q=0$ they are identical to Hall-Littlewood
 polynomials (no approximation).
Starting from these approximate eigenfunctions  one should  
in principle be able to obtain corrections perturbatively in $q$ and $p$.

\

\

\section{Non-maximal punctures}\label{Lapp}

\begin{figure}
\begin{center}
\begin{tabular}{ccc}
\includegraphics[scale=0.4]{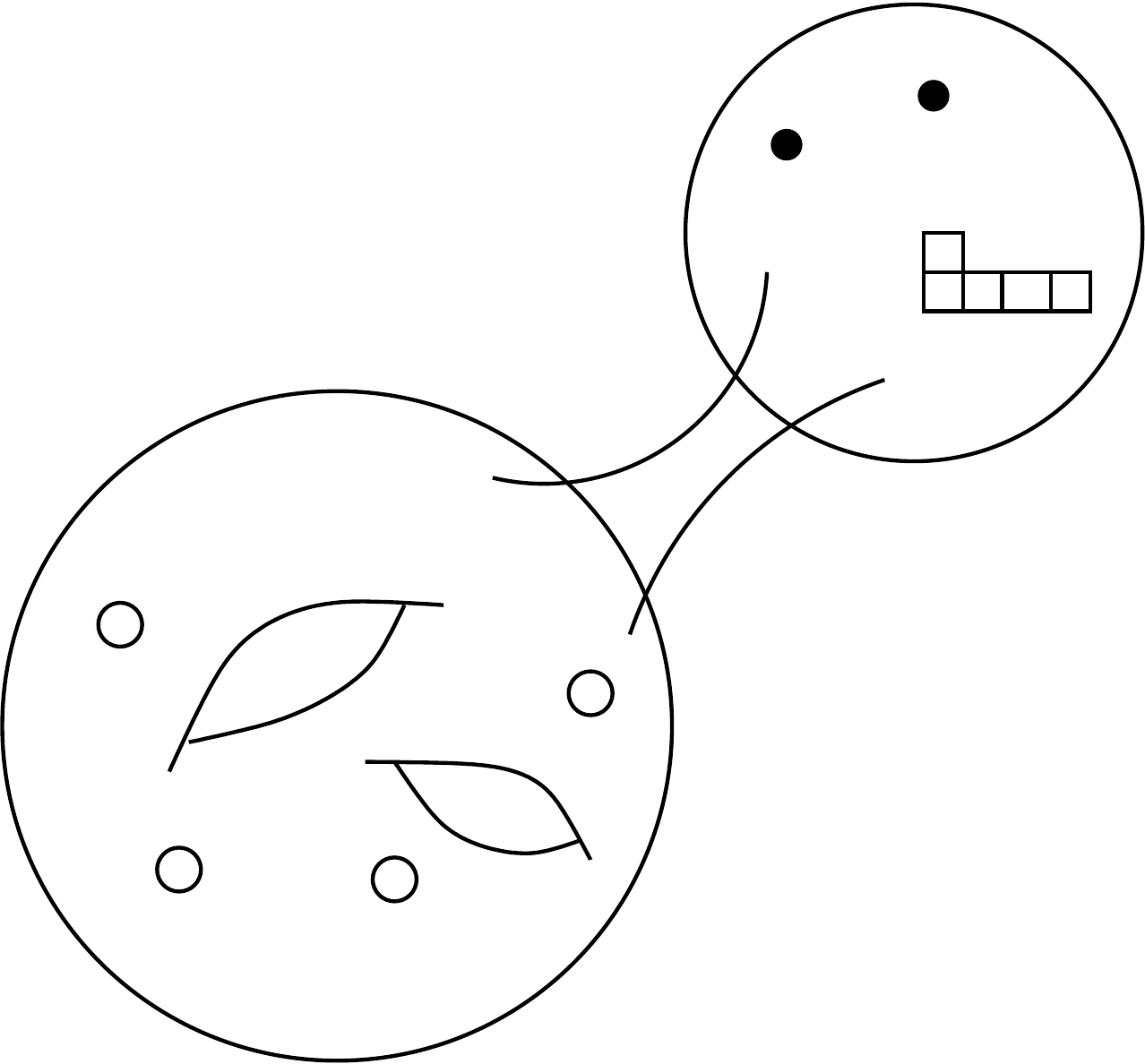}&
$\;\;\;\;$
&
\includegraphics[scale=0.8]{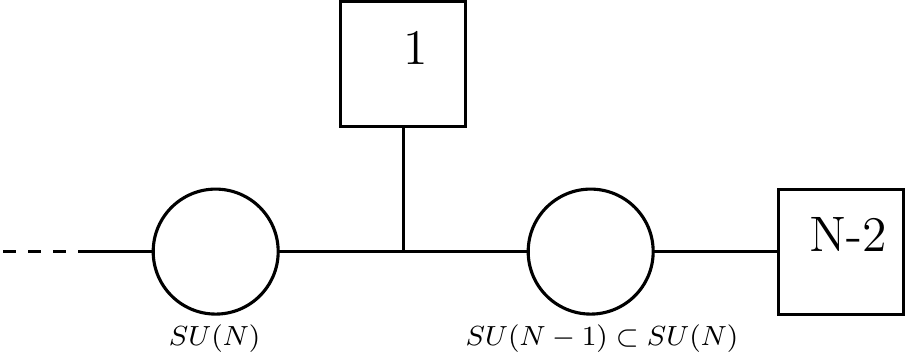}
\end{tabular}
\end{center}
\caption{Closing the full puncture down to $L$-shaped one by closing 
two minimal punctures. On the left we have a generic theory with a 
superconformal tail ending in an $L$-shaped puncture represented 
by a Riemann surface. On the right we have 
a Lagrangian representation of the superconformal tail: the squares denote
hypermultiplets and circles gauge groups.\label{Lshape}}
\end{figure}

Let us discuss the procedure of reducing the flavor symmetry of a puncture by our residue calculus.
We consider a concrete example that captures the general algorithm.
We will derive the prescription to partially-close a full puncture to obtain an $L$-shaped
puncture with two rows (see figure~\ref{Lshape}).  

First we couple a theory with a maximal puncture to superconformal tail
ending in an $L$-shaped puncture as dipicted in figure~\ref{Lshape}.
The figure on the left corresponds to a Riemann surface representation
of the theory, and on the right we have a Lagrangian description for the tail.
The flavor symmetry of the $L$-shaped puncture is $U(1)\times SU(N-2)$. 
The superconformal tail has two additional factors of $U(1)$  flavor symmetry: in the Riemann
surface picture these correspond to the two minimal punctures, and in the Lagrangian representation
the $U(1)$s are flavor symmetries of the hypermultiplets. 
To obtain the index 
with the $L$-shaped puncture only, we  have to remove the two minimal punctures 
by computing the residue of  the relevant $U(1)$ flavor fugacities at
 $t^\half$,
as we discussed in the bulk of the paper. We will implement this procedure 
in a convenient way to obtain an explicit expression for this index.
The index of this theory is given by
\be
&&\oint \prod_{i=1}^{N-1}\frac{dc_i}{2\pi i c_i}\,{\cal I}_V({\mathbf c})\Delta({\mathbf c})
{\cal I}_{\cal C}({\mathbf c}^{-1},..)\times\\
&&\qquad\oint \prod_{j=1}^{N-2}\frac{db_j}{2\pi b_j}
\,{\cal I}_V({\mathbf b})\Delta({\mathbf b})
{\cal I}_{hyp}({\mathbf b},{\mathbf d},z)\,
{\cal I}_{hyp}({\mathbf c},x,\{{\mathbf b},y\})\,.\nonumber
\ee  ${\cal I}_{\cal C}$ denotes the index of the Riemann surface to which 
we glue the superconformal tail. Here the free hypermultiplet indices are given by
\be
&&{\cal I}_{hyp}({\mathbf b},{\mathbf d},z)=\prod_{i=1}^{N-1}\prod_{j=1}^{N-2}\prod_{m,n\geq0}
\frac{1-p^{n+1}q^{m+1}t^{-\half}(d_j b_i z)^{-1} }{1-p^{n}q^{m}t^{\half}d_j b_i z }
\frac{1-p^{n+1}q^{m+1}t^{-\half}d_j b_i z}{1-p^{n}q^{m}t^{\half}(d_j b_i z)^{-1} }\,,\\
&&{\cal I}_{hyp}({\mathbf c},x,\{{\mathbf b},y\})=\prod_{i,j=1}^{N}\prod_{m,n\geq0}
\frac{1-p^{n+1}q^{m+1}t^{-\half}(x \hat b_i c_j)^{-1} }{1-p^{n}q^{m}t^{\half}x\hat b_i c_j  }
\frac{1-p^{n+1}q^{m+1}t^{-\half}x\hat b_i c_j}{1-p^{n}q^{m}t^{\half}(x\hat b_i c_j)^{-1} }\,.\nonumber
\ee We have defined $\{\hat b_1,\dots \hat b_N\}=
\{y b_1^{-1},\dots y b_{N-1}^{-1},y^{1-N}\}$. Let us look for poles which can pinch the integration
contours when a combination of the $U(1)$ fugacities associated to the $U(1)$ punctures is equal to $t^\half$.
 There are three $U(1)$ fugacities in the tail denoted above by $x$, $y$  and $z$. Let us denote
by $\alpha$ and $\beta$ the combinations of $U(1)$ fugacities corresponding to the minimal punctures and by $\gamma$
the $U(1)$ fugacity inside the $L$-shaped puncture. Then we have the following identification
of the $U(1)$ fugacities in the two descriptions: the Riemann surface and the Lagrangian,
\be
\alpha=x,\qquad 
\beta=\frac{1}{y}z^{\frac{N-2}{N}}\,,\qquad
\gamma=y^2z^{\frac2N}\,.
\ee We will consider the poles at $\alpha=\beta=t^\half$ one after the other. 
Note that if the $c$ fugacity is not gauged there are no such poles. The poles in the $c$ 
integration occur at
\be
c_j=\frac{t^\half}{x y} b_{\sigma(j)}\quad {\text {or}}\quad
c_j=t^\half y^{N-1}x^{-1}
\,,\qquad j=1\dots N-1\,;\qquad 
c_{N}=\frac{t^{\frac{1-N}{2}}}{(xy)^{1-N}}\,.
\ee  Thus if we choose
\be
x=t^\half\,,
\ee all the $c$ contours are pinched and we obtain a pole. We evaluate then the integrand 
of the $c$ contour integral at 
\be
x=t^\half;\qquad c_j=y^{-1}b_{\sigma(j)}\,,\qquad j=1\dots N-1\,;\qquad 
c_{N}=y^{N-1}\,.
\ee We obtain
\be
&&N!\oint \prod_{j=1}^{N-2}\frac{db_j}{2\pi b_j}\,
{\cal I}_V({\hat {\mathbf b}}^{-1})\Delta({\hat {\mathbf  b}}^{-1})
{\cal I}_{\cal C}({\hat {\mathbf  b}},..)\,\times
\\
&&\qquad\qquad \,{\cal I}_V({\mathbf b})\Delta({\mathbf b})
{\cal I}_{hyp}({\mathbf b},{\mathbf d},z)\,
{\cal I}'_{hyp}(\{{\mathbf b},y\})\,.\nonumber
\ee The free hypermultiplet which coupled to $c$ becomes
\be
&&{\cal I}_{hyp}({\mathbf c},x,\{{\mathbf b},y\})\to{\cal I}'_{hyp}(\{{\mathbf b},y\})=\prod_{i,j=1}^{N}\prod_{m,n\geq0}'
\frac{1-p^{n+1}q^{m+1}t^{-1}\hat b_j/\hat b_i }{1-p^{n}q^{m}t^{}\hat b_i /\hat b_j  }
\frac{1-p^{n+1}q^{m+1}\hat b_i /\hat b_j}{1-p^{n}q^{m}\hat b_j /\hat b_i }\,\nonumber\\
&&\qquad \qquad =\frac{1}{N!{\cal I}_V({\hat {\mathbf  b}}^{-1})\Delta({\hat {\mathbf  b}}^{-1}){\cal I}_V}\,.
\ee  In turn, the index becomes
\be
&&\oint \prod_{j=1}^{N-2}\frac{db_j}{2\pi b_j}\,
{\cal I}_{\cal C}({\hat {\mathbf  b}},..)\,{\cal I}_V({\mathbf b})\Delta({\mathbf b})
{\cal I}_{hyp}({\mathbf b},{\mathbf d},z)\,.\nonumber
\ee
 The poles in the $b$ 
integration contour now occur at
\be
b_j=\frac{t^\half}{z} d^{-1}_{\sigma'(j)}
\,,\qquad j=1\dots N-2\,;\qquad 
b_{N-1}=\frac{t^{\frac{2-N}{2}}}{z^{2-N}}\,.
\ee
We have $\gamma=y^2z^{\frac2N}$ and $\beta=\frac{1}{y}z^{\frac{N-2}{N}}$. Thus we set $\beta=t^\half$
\be
y\,z=t^\half \gamma,\qquad y^{N-1}=t^{-\half} \gamma^{\frac{N-2}{2}},
\qquad b_{N-1}\,y^{-1}=\frac{t^{\frac{2-N}{2}}}{yz^{2-N}}=t^{\half} \gamma^{\frac{N-2}{2}}\,.
\ee
 In particular
\be
c_j=\frac{1}{\gamma}d_j^{-1},
\qquad j=1\dots N-2\,,\qquad 
c_{N-1}=t^{\half} \gamma^{\frac{N-2}{2}},\qquad
c_{N}=t^{-\half} \gamma^{\frac{N-2}{2}}\,.
\ee If we only consider the poles in the decoupled piece (the sphere with the superconformal tail)
 there are no contour pinchings. 
However, since we know that these poles exist, say by decoupling each of the minimal 
punctures together with a maximal one as before, we deduce that  ${\cal I}_{\cal C}$ has to have 
a relevant pole.  We note that
\be
{\cal I}_V\,\Delta({\mathbf b})\,{\cal I}_V({\mathbf b})\,{\cal I}_{hyp}({\mathbf b},{\mathbf d},z)\quad\to\quad
\frac{1}
{\Gamma(t^\half(\gamma^{N/2}d_j)^{\pm1};p,q)}
\,.
\ee
Thus we deduce that the index (times the two free $U(1)$ factors)
is given by
\be\label{Lstrip}
{\cal I}_L={\cal I}'_{\cal C}(\{\frac{1}{\gamma}d_j^{-1},t^{\half} \gamma^{\frac{N-2}{2}},
t^{-\half} \gamma^{\frac{N-2}{2}}\},\dots)
\frac{1}
{\Gamma(t^\half(\gamma^{N/2}d_j)^{\pm1};p,q)}
\,.
\ee
To obtain the index of a theory with puncture corresponding to $L$-shaped Young
diagram we have to compute a certain residue of the theory with full puncture and strip
off a free hypermultiplet. This procedure can be generalized for arbitrary non-maximal
punctures.

\

Let us compare the above result with the general prescription to compute the index 
of non maximal punctures in the Macdonald case suggested in~\cite{Gadde:2011uv}.
Following the prescription of ~\cite{Gadde:2011uv} the index of the theory with $L$-shaped puncture is given by
\be\label{Lind1}
{\cal I}_L= {\hat K}_L(\gamma, {\mathbf d})\, {\hat K}(\cdot)
\sum_\lambda \psi^\lambda(\frac{1}{\gamma}{\mathbf d}^{-1},t^{\half} \gamma^{\frac{N-2}{2}},
t^{-\half} \gamma^{\frac{N-2}{2}})\, \Psi_\lambda(\cdot)\,. 
\ee Where $\Psi_\lambda(\cdot)$ and $ {\hat K}(\cdot)$ are  combination of functions 
not depending on the flavor fugacities of the $L$ shaped puncture. 
The function $\psi^\lambda$ is a Macdonald polynomial and we also have\footnote{
The plethystic exponential is defined as $PE[f(x)]=\exp(\sum_{\ell=1}^\infty\frac{1}{\ell}f(x^\ell))$ where we also have to specify what are the parameters
$x$ with respect to which the plethystics is done: in index applications the parameters $x$ are all the fugacities appearing in index definition.
}
\be\label{KL}
 {\hat K}_L(\gamma, {\mathbf d})=PE\left[\frac{t}{1-q}\sum_{i,j=1}^{N-2}d_i/d_j
+\frac{t^{3/2}}{1-q}\sum_{i=1}^{N-2}(d_i\gamma^{\half N}+d_i^{-1}\gamma^{-\half N})+
\frac{t^{2}+t}{1-q}\right]\,.
\ee On the other hand the index of the same theory but with the $L$-shaped puncture
traded with the maximal one is given by
\be
{\cal I}= {\hat K}({\mathbf a})\, {\hat K}(\cdot)
\sum_\lambda \psi^\lambda(a_1,\cdots,a_N)\, \Psi_\lambda(\cdot)\,. 
\ee
The factor $\hat{\cal K}$ for the maximal puncture has the following form
\be
\hat {\cal K}=PE\left[\frac{t}{1-q}\sum_{i,j=1}^Na_i/a_j\right]
=\prod_{i,j}\frac{1}{(t\,a_i/a_j;q)}\,.
\ee
We now want to obtain ${\cal I}_L$ from a residue of ${\cal I}$.
 ${\cal I}$ has poles whenever
\be
a_i=t\,q^\ell\, a_j\,.
\ee Since $\prod_{i=1}^Na_i=1$ a class of poles is 
\be
a_i=t^{1/2}\, q^{\ell/2}\, \frac{1}{\prod_{j\neq i}^{N-1}a_j^{1/2}}\,.
\ee Lets us consider the pole at $a_1$ with $\ell=0$ and define $b_1=\prod_{j\neq 1}^{N-1}a_j^{-1/2}$.
Then the fugacity assignment in the orthogonal functions will be
\be
\psi^\lambda(a_1,\cdots,a_N)\quad\to\quad
\psi^\lambda(t^{1/2}b_1,t^{-1/2}b_1,a_2\cdots,a_{N-1})\,,
\ee  and of course by construction $b_1^2\prod_{i=2}^{N-1}a_i=1$. 
The above assignment corresponds to two row Young diagram with one box
in the first row and $N-1$ boxes in the second and defining a puncture with flavor symmetry
$S(U(1)\times U(N-2))$. We see that it is the same as in~\eqref{Lind1} with the following
identification of fugacities,
\be
a_i=d_i\gamma^{-1}\,,\qquad b_1=\gamma^{\frac{N-2}{2}}\,.
\ee
Let us now compute the residue of the index at this pole 
\be
\underset{a_1\to t^{1/2}b_1}{Res}\,\{\frac{1}{a_1}\,\hat {\mathcal K}({\mathbf a})\}=
\frac{2}{(t;q)^2(t^2;q)(q;q)}\prod_{i,j=2}^{N-1}\frac{1}{(t\,a_i/a_j;q)}
\prod_{i=2}^{N-1}\frac{1}{(t^{3/2}(a_i/b_1)^{\pm1};q)(t^{1/2}(b_1/a_i)^{\pm1};q)}\,.\nonumber\\
\ee On the other hand we also have  from~\eqref{KL}
\be 
\hat {\mathcal K}_{L}({\mathbf a})=
\frac{1}{(t;q)(t^2;q)}\prod_{i,j=2}^{N-1}\frac{1}{(t\,a_i/a_j;q)}
\prod_{i=2}^{N-1}\frac{1}{(t^{3/2}(a_i/b_1)^{\pm1};q)}\,.
\ee The ratio of the two quantities above is simply given by
\be
\frac{\hat {\mathcal K}_{L}({\mathbf a})}{\underset{a_1\to t^{1/2}b_1}{Res}\,\{\frac{1}{a_1}\,\hat {\mathcal K}({\mathbf a})\}}=\half
{\cal I}_{V}\,
\prod_{i=2}^{N-1}(t^{1/2}(b_1/a_i)^{\pm1};q)=
\frac{{\cal I}_{V}}{2\,PE\left[\frac{t^{1/2}}{1-q}(\frac{1}{b_1}{\mathbf a}+b_1{\mathbf a}^{-1})\right]}\,.
\nonumber\\
\ee 
Thus the prescription to obtain the index of a theory with $S(U(1)\times U(N-2))$ 
puncture from the index of the theory with the maximal puncture is just to consider a certain pole of the latter discussed here 
 and multiply the residue by the index of a free $U(1)$ vector multiplet 
and divide by the index of  an apropriate free hyper-multiplet. 
We get complete agreement with~\eqref{Lstrip}. The procedure of~\cite{Gadde:2011uv} to compute the index of theories with any 
non-maximal punctures can be phrased as a computation of residues as was discussed in~\cite{DR}.

\

\

\bibliography{sdualityMAC}

\bibliographystyle{JHEP}

\end{document}